\documentstyle[12pt,epsf]{iftese}
\def\bib{\bibitem}
\def\'#1{\if#1i{\accent 19\i}\else{\accent 19 #1}\fi}

\renewcommand{\baselinestretch}{1.5}
\newtheorem{theorem}{Teorema}[chapter]

\newtheorem{corollary}[theorem]{Corol\'ario}
 
\newtheorem{definition}{Defini\c c\~ao}[chapter] 
\newtheorem{example}{Exemplo}[chapter] 
 
\newtheorem{lemma}{Lema}[chapter] 
\newtheorem{proposition}{Proposi\c c\~ao}[chapter]

\newtheorem{law}{Lei}
\newtheorem{hyp}{Hip\'otese}
\begin{document}
\pagestyle{headings}
\def\baselinestretch{1.2}
\hoffset=-1.0 true cm
\voffset=-2 true cm
\topmargin=1.0cm
\thispagestyle{empty}
\def\thefootnote{\fnsymbol{footnote}}

\thicklines
\begin{picture}(370,60)(0,0)
\setlength{\unitlength}{1pt}
\put(40,53){\line(2,3){15}}
\put(40,53){\line(5,6){19}}
\put(40,53){\line(1,1){27}}
\put(40,53){\line(6,5){33}}
\put(40,53){\line(3,2){25}}
\put(40,53){\line(2,1){19}}
\put(40,53){\line(5,-6){17}}
\put(40,53){\line(1,-1){22}}
\put(40,53){\line(6,-5){30}}
\put(40,53){\line(3,-2){22}}
\put(40,53){\line(-2,1){15}}
\put(40,53){\line(-3,1){23}}
\put(40,53){\line(-4,1){26}}
\put(40,53){\line(-6,1){36}}
\put(40,53){\line(-1,0){40}}
\put(40,53){\line(-6,-1){32}}
\put(40,53){\line(-3,-1){20}}
\put(40,53){\line(-2,-1){10}}
\put(75,45){\Huge \bf IFT}
\put(180,56){\small \bf Instituto de F\'\i sica Te\'orica}
\put(165,42){\small \bf Universidade Estadual Paulista} 
\put(-25,2){\line(1,0){433}}
\put(-25,-2){\line(1,0){433}}
\end{picture}  


\vskip .3cm
\noindent
{DISSERTA\c C\~AO DE MESTRADO}
\hfill    IFT--D.005/00\\

\vspace{3cm}
\begin{center}
{\large \bf{ M\'etodos Estoc\'asticos em Turbul\^encia Desenvolvida}}

\vspace{1.2cm}
Jo\~{a}o Paulo Viegas Carneiro
\end{center}

\vskip 3cm
\hfill Orientador
\vskip 0.4cm
\hfill {\em Gerson Francisco}
\vskip 4cm
\vfill
\begin{center}
Maio de 2000
\end{center}

\newpage

\pagenumbering{roman}

\begin{center}
{\Large \bf Agradecimentos}
\end{center}
\vskip 2.0cm
\begin{center}  
Gostaria de agradecer a todos que v\^em me ajudando, mesmo quando eu tenho pregui\c ca, a me desenvolver como ser humano: Deus, todos os Profetas, todos os Santos, Omar Ali Shah e toda a sua fam\'\i lia, minha fam\'\i lia, meus amigos e meus inimigos. Gostaria de agradecer \`a FAPESP pela bolsa, processo 98/02806-0.
\bigskip

Agrade\c co particularmente ao meu orientador, prof.{\bf Gerson Francisco}, e ao prof. {\bf Uriel Frisch} cujas cr\'\i ticas e alertas me ajudaram, e ajudam, no caminho da turbul\^encia; tamb\'em a minha m\~ae, {\bf Ana Maria}, pela revis\~ao. 
\vskip +4.0cm
Dedico este estudo a {\Large {\bf T\^ania e a Sofia}}, por serem as pessoas que s\~ao.
\

\end{center}


\newpage

\begin{center}
{\Large \bf Resumo}
\end{center}
\vskip 2.0cm

         Nesta disserta\c c\~ao, utilizamos m\'etodos da teoria de processos estoc\'asticos para a compreens\~ao da turbul\^encia em fluidos. Discutimos o modelo de Kolmogorov para turbul\^encia homog\^enea desenvolvida, resultados anal\'\i ticos recentes para a equa\c c\~ao de Burgers.


\vskip 1.0cm
\noindent
{\bf Palavras-Chaves}:Turbul\^encia; M\'etodos Estoc\'asticos; Din\^amica de Fluidos
\vskip 0.5cm
\noindent
{\bf \'Areas do conhecimento}: Turbul\^encia, Din\^amica de Fluidos, Sistemas Din\^amicos, Processos Estoc\'asticos.

\newpage

\begin{center}
{\Large \bf Abstract}
\end{center}
\vskip 2.0cm

        In this thesis, we use some methods of theory of stochastic process to approach turbulence in fluid dynamics. We discuss the Kolmogorov model for fully developed homogeneous turbulence, some recent analytical results to the 1-D Burgers' equation approach turbulence.   

\vfill \eject

\tableofcontents
\part{M\'etodos Matem\'aticos}

O objetivo destes dois primeiros cap\'\i tulos \'e fornecer o embasamento matem\'atico, por mim considerado m\'\i nimo para a leitura do corpo principal da tese (cap\'\i tulos 3,4 e 5).  Embora estes cap\'\i tulos n\~ao sejam imprescind\'\i veis, para uma pessoa com experi\^encia em \'areas correlatas, ajudam a tornar a tese mais autoconsistente e estabelecem os conceitos b\'asicos que norteiam o estudo de turbul\^encia desenvolvida na atualidade.

\chapter{Sistemas Din\^amicos}
\pagenumbering{arabic}
Neste cap\'\i tulo trataremos de algumas defini\c c\~oes e teoremas de sistemas din\^amicos, tendo como objetivo final justificar os cap\'\i tulos posteriores, que encaram o fen\^omeno de turbul\^encia desenvolvida, utilizando ferramentas de processos estoc\'asticos.

Algumas palavras devem ser ditas sobre o que entendemos como {\texttt{justificar}}. Embora o estudo matem\'atico de sistemas din\^amicos remonte ao come\c co do s\'eculo com os trabalhos vision\'arios de Poincar\'e, a aridez deste campo de pesquisa fez com que, at\'e esta d\'ecada, o foco estivesse voltado, quase que exclusivamente, ao estudo de sistemas de equa\c c\~oes diferenciais ordin\'arias. Como a passagem dos resultados obtidos em EDOs para EDPs 
permane\c ca ainda inacabada, este cap\'\i tulo servir\'a somente para um 
entendimento qualitativo\footnote{Como veremos mais adiante a teoria de Sistemas Din\^amicos n\~ao \'e suficiente para o entendimento da turbul\^encia desenvolvida.} do comportamento das solu\c c\~oes.

Apesar de o nosso interesse, neste trabalho, estar voltado para sistemas descritos por equa\c c\~oes diferenciais, existem resultados an\'alogos para mapeamentos. As refer\^encias para este cap\'\i tulo s\~ao: \cite{Eckmann,Hirsch,Guckenheimer,Lichtenberg,Palis,Ruelle1,Wiggins}.
\section{Conceitos B\'asicos}

Durante quase todo o cap\'\i tulo estaremos estudando a seguinte equa\c c\~ao:
\begin{equation}
\label{sisdyn}
\dot{x} \equiv \frac{d}{dt}x = f(x,\mu) 
\end{equation}
Ou seja, um sistema descrito por uma equa\c c\~ao diferencial (\ref{sisdyn}), onde $x\in U\subset \Re^{n}$, $t$, $\mu \in \Re_{+}$, com $U$ aberto, $t$ e $\mu$ par\^ametros, poder\'\i amos pensar que h\'a uma restri\c c\~ao, pois $f$ n\~ao depende explicitamente do tempo, portanto, estar\'\i amos considerando somente sistemas aut\^onomos. Todavia, podemos lan\c car m\~ao do seguinte artif\'\i cio: introduzimos uma vari\'avel $s$, tal que, $\frac{ds}{dt}=1$. Logo, em termos deste novo par\^ametro, o sistema \'e escrito como: 
\begin{equation}
x\prime \equiv \frac{dx}{ds}=f(x,s)\times 1. \nonumber
\end{equation}
Criamos, ent\~ao, um novo sistema $y$ dado por: 
\begin{eqnarray}
y&=&(x,s) \nonumber \\
g(y)&=&\left(f(x,s),1\right), \nonumber
\end{eqnarray} 
obtendo o seguinte sistema : 
\begin{equation}
y\prime=\frac{dy}{ds}=g(y),\hspace{+0.5cm}y\in \Re^{n}\times \Re .\nonumber
\end{equation}
Assim, um sistema n\~ao aut\^onomo pode ser transformado em aut\^onomo aumentando sua dimens\~ao.

Denotando uma solu\c c\~ao com condi\c c\~ao inicial $x_{0}(x)$ no instante $t_{0}$ por 
\begin{equation}
 x(t) = x(t,t_0,x_0),  \nonumber
\end{equation}
podemos ver que esta solu\c c\~ao define um fluxo 
\begin{eqnarray}
\phi_t&:&U\times \Re_{+} \rightarrow \Re^{n} \nonumber \\
(a,t)&\mapsto &x(t,0,a), \nonumber
\end{eqnarray}
que satisfaz a seguinte equa\c c\~ao:
\begin{equation}
\frac{d}{dt} (\phi(x,t))|_{t= \tau} = f( \phi(x, \tau)). \nonumber
\end{equation}

Neste caso, $ \{\phi _t \}_{t=0}^{\infty}$ forma um semigrupo parametrizado pelo tempo, pois o fluxo goza das seguintes propriedades:
\begin{eqnarray}
   {\phi}_0 &=& id, \nonumber \\
 \phi _t \circ \phi _s &=& \phi _{t+s}, \quad t,s \in \Re _{+}, \nonumber
\end{eqnarray}
definindo portanto, uma fam\'\i lia de difeomorfismos $ \phi _t:U \rightarrow \Re^{n}$.

De um ponto de vista geom\'etrico, $x(t,t_0,x_0)$ \'e uma curva sobre $\Re^{n}$, parametrizada pelo tempo $t$, tal que, no instante $t_0$ est\'a em $x_0\in \Re^{n}$ e sua tangente satisfaz a equa\c c\~ao diferencial (\ref{sisdyn}) em todos os tempos. Assim, uma solu\c c\~ao de equil\'\i brio pode ser vista como um ponto fixo do fluxo definido por ${\phi}_{t}$ que, por ser a mais simples poss\'\i vel, \'e a primeira a ser estudada.
\begin{definition}
\label{def:pontofixo}
Chamamos $\overline{x}( t)$ de {\bf ponto fixo} ou solu\c c\~ao de equil\'\i brio se $f\left( \overline{x}\right)=0$.
\end{definition}
Para entender esta defini\c c\~ao, basta substituir em (\ref{sisdyn}) e verificar que a derivada temporal desta solu\c c\~ao \'e nula, logo o campo n\~ao variar\'a no tempo. Como a perda de estabilidade das solu\c c\~oes(conceito que ser\'a  melhor explicado adiante) \'e o fen\^omeno que nos leva ao caos, \'e importante termos defini\c c\~oes precisas. Assim, numa primeira abordagem, poder\'\i amos exigir que solu\c c\~oes pr\'oximas permanecessem pr\'oximas para qualquer tempo, ou seja:
\begin{definition}
Uma solu\c c\~ao $\overline{x}\left( t\right) $ \'e dita {\bf L-est\'avel}(Liapunov est\'avel) se, dado $\varepsilon >0$, existe $\delta >0$, tal que para qualquer solu\c c\~ao $x\left( t\right) $, satisfazendo $\left| \overline{x}\left( t_{0}\right) -x\left( t_{0}\right) \right| <\delta $, implica $\left| \overline{x}\left( t\right) -x\left(t\right) \right| <\varepsilon $, para todo $t>t_{0}$.
\end{definition} 
\begin{definition}
Uma solu\c c\~ao $\overline{x}\left( t\right)$ \'e dita {\bf assintoticamente
est\'avel}, se \'e L-est\'avel e existe $a$ constante real positiva, tal que para toda solu\c c\~ao $x$ satisfazendo $\left| \overline{x}(t_{0})-x(t_{0})\right| <a$, temos  
${\lim }_{t\rightarrow \infty}\left| \overline{x}(t)-x(t)\right| =0$.
\end{definition}
O que induz as seguintes defini\c c\~oes:
\begin{definition} 
Chamamos os pontos fixos assintoticamente est\'aveis de {\bf sorvedouros} e os L-est\'aveis, que n\~ao s\~ao sorvedouros, de {\bf centro}.
\end{definition}
Uma aproxima\c c\~ao um pouco mais f\'\i sica sobre a estabilidade \'e pensarmos no conceito de {\it perturba\c c\~ao}.

\begin{definition} 
Seja $F \in C^r(\Re^n)$, $r,k \in {\cal Z}_{+} $, $ k \leq r$, e $\varepsilon > 0$. Ent\~ao, $G \in C^k$ \'e uma {\bf perturba\c c\~ao de tamanho} $\mathbf{ \varepsilon}$, se existe um conjunto compacto $K \in \Re^n$, tal que, $F=G \quad \forall x \in \{ \Re^n - K\}$, e para todo $(i_1, \ldots, i_n)$, com $i_1 + \cdots + i_n = i \leq k$, tivermos $\|\frac{\partial^i}{\partial^{i_1} x_1 \cdots \partial^{i_n} x_n}(F-G)\| < \varepsilon$.
\end{definition}

\begin{definition} 
Duas aplica\c c\~oes $C^r$, $F$ e $G$ s\~ao $\mathbf{C^k}$ {\bf equivalentes} ou $\mathbf{C^k}$ {\bf conjugadas} $(k \leq r)$, se existe um homeomorfismo $h \in C^k$, tal que, $h \circ F = G \circ h$. Uma equival\^encia $C^0$ \'e chamada de {\bf equival\^encia topol\'ogica}. 
\end{definition}
Seja $\phi_t(x,t)$ um fluxo cujas perturba\c c\~oes  $\varepsilon$ s\~ao ${\phi}_{t} ^{\varepsilon _i}(x,t)$. Ent\~ao, se elas s\~ao topologicamente equivalentes, existe um homeomorfismo $h$, tal que, ${\phi}_{t_i} ^{\varepsilon _i}(x,t) \circ h = h \circ {\phi}_{t_j} ^{\varepsilon _j}(x,t)$, ou seja, existe uma equival\^encia entre as \'orbitas geradas por estes dois fluxo. Todavia, $h$ n\~ao tem que necessariamente preservar a parametriza\c c\~ao temporal das \'orbitas, mas quando isto acontece, chamamos o homeomorfismo de {\bf conjuga\c c\~ao}.
\begin{definition} 
Uma aplica\c c\~ao $F \in C^r(\Re)$ \'e {\bf estruturalmente est\'avel}, se existe um $\varepsilon > 0$, tal que, para todas as perturba\c c\~oes de tamanho $\varepsilon$ de $F$ s\~ao topologicamente equivalentes.

\end{definition}

A import\^ancia destes conceitos reside no fato de que por n\~ao serem lineares muitos dos sistemas de interesse, necessitamos saber, quando fazemos lineariza\c c\~oes, quais informa\c c\~oes do sistema n\~ao linear ainda est\~ao no modelo linearizado, ou seja, quais s\~ao as propriedades locais. 
A t\'\i tulo de ilustra\c c\~ao, consideremos o seguinte exemplo: tomemos uma solu\c c\~ao arbitr\'aria $x$, de um sistema com ponto fixo $\overline{x}$ e a representemos por: 
\begin{equation}
x=\overline{x}(t)+y, \nonumber
\end{equation}
onde $y$ \'e um campo a ser determinado. Agora, substituindo esta express\~ao na equa\c c\~ao diferencial e fazendo expans\~ao em s\'erie de pot\^encia centrada no ponto fixo, vem 
\begin{equation}
\dot{x}=\dot{\overline{x}}(t)+\dot{y}=f(\overline{x}(t))+{\cal D}f(\overline{x}(t))y+{\cal O}(\left| y\right| ^{2}), \nonumber
\end{equation}
onde ${\cal D}$  \'e operador diferencial no espa\c co em quest\~ao. Mas por (def.\ref{def:pontofixo}) vem:
\begin{equation}
\dot{y}={\cal D}f(\overline{x}(t))y+{\cal O}(\left| y\right| ^{2}). \nonumber
\end{equation}

Desprezando termos de segunda ordem em $y$. Ficamos com 
\begin{equation}
\dot{y}={\cal D}f(\overline{x}(t))y, \nonumber
\end{equation}
cuja solu\c c\~ao \'e 
\begin{equation}
y\left( t,\overline{x}\right) =y_{0}\exp \left\{ \int_{t_{0}}^{t}dt{\cal
D}f(\overline{x}(t))\right\}. \nonumber
\end{equation}
Para analizarmos o sistema linearizado fazemos a seguinte defini\c c\~ao:
\begin{definition}
Sendo $x=\overline{x}$ um ponto fixo de $\dot{x}=f(x)$, $x\in \Re^{n}$, $\overline{x}$ ser\'a chamado de {\bf ponto fixo hiperb\'olico}, se nenhum dos autovalores de ${\cal D}f(\overline{x})$ tiver parte real zero.
\end{definition}
\begin{theorem}[Hartman-Grobman]
Se $\overline{x}$ \'e ponto fixo hiperb\'olico , existe um homeomorfismo $h$ definido em alguma vizinhan\c ca $U\{\overline{x}\} \subset \Re^n$, tal que $h( \phi (x,t))= e^{{\cal D}(\overline{x})}$. Note-se que este homeomorfismo preserva o sentido de \'orbita e tamb\'em pode ser escolhido de forma que a parametriza\c c\~ao temporal seja preservada.
\end{theorem}

\begin{theorem}
Se todos os autovalores de ${\cal D}f(\overline{x}\left( t\right))$ t\^em parte real negativa para um tempo suficientemente grande, a solu\c c\~ao de equil\'\i brio de $x=\overline{x}\left( t\right) $ do campo vetorial linear \'e assintoticamente est\'avel.
\end{theorem}

Assim sendo, poder\'\i amos nos perguntar se esta condi\c c\~ao n\~ao \'e excessivamente forte, ou seja, se deix\'assemos alguns auto-valores serem nulos, ser\'a que n\~ao ter\'\i amos, ainda assim, uma solu\c c\~ao est\'avel? Infelizmente o m\'etodo da lineariza\c c\~ao n\~ao nos d\'a a resposta, pois neste caso o comportamento da solu\c c\~ao \'e controlado pelos termos de ordem superior em $y$, na regi\~ao gerada por esta variedade local central. Mais adiante, no entanto, voltaremos a este problema. Para evitar as mazelas da lineariza\c c\~ao, podemos utilizar o seguinte teorema:

\begin{theorem}[Estabilidade de Liapunov]
Se $\overline{x}$ \'e um ponto fixo e $V:U\rightarrow \Re$ \ uma fun\c c\~ao $C^{1}$ definida em alguma vizinhan\c ca $U$ de $\overline{x}$, tal que:
\begin{enumerate}
\item  $V(\overline{x})=0$ e $V(x)>0$ se,  $x\neq $ $\overline{x}$;
\item  $V^\prime (x)\leq0$ em $U-\{\overline{x}\}$, ent\~ao $\overline{x}$ \'e est\'avel.

 Al\'em disso, se
\item  $\dot{V}(x)<0$ em $U-\{\overline{x}\}$, ent\~ao $V$ \'e chamada de {\bf fun\c c\~ao de Liapunov}. Se pudermos tomar $U$ como $\Re^{n}$ , ent\~ao $\overline{x}$ \'e globalmente assintoticamente est\'avel.
\end{enumerate}
\end{theorem}

Ao estudar sistemas mec\^anicos, a energia \'e sempre uma boa candidata para a fun\c c\~ao de Liapunov, j\'a que \'e sempre positiva definida e, nos sistemas dissipativos sua derivada temporal \'e negativa. Logo, para sistemas sem termos dissipativos, nos quais $\dot{V}(x)\geq 0$, este m\'etodo sempre ser\'a inconclusivo.

\begin{definition}
Seja $S\subset \Re^{n}$ um conjunto. Ent\~ao, $S$ \'e dito {\bf conjunto invariante} sob o campo vetorial $\dot{x}=f(x)$, se para qualquer $x_{0}\in S$ existir $x(t,t_0,x_{0})\in S$ para todo $t\in \Re$. Se a assertiva for v\'alida, somente para tempos positivos $S$ ser\'a chamado de {\bf conjunto positivamente invariante}.
\end{definition}
\begin{definition}
Um conjunto invariante $S\subset \Re^{n}$ \'e chamado de {\bf variedade invariante} $C^{r}(r\geq 1)$, se $S$ tiver a estrutura de uma variedade diferenci\'avel $C^{r}.$
\end{definition}

Com base nestas defini\c c\~oes, voltemos a olhar o sistema linearizado.

\begin{definition}
Seja $\overline{x}$ um ponto fixo, e o seu sistema linear associado $\dot{y}=Ay,\hspace{+0.5cm}y\in \Re^n$, se representamos $\Re^{n}$ como uma soma direta de tr\^es subespa\c cos denotados por $E^{s}$, $E^{u}$ e $E^{c}$, tais que 
\begin{eqnarray}
E^{s}&=&span\{e_{1},...e_{s}\}, \\
E^{u}&=&span\{e_{s+1},...e_{s + u }\}, \\
E^{c}&=&span\{e_{s+\mu +1},...,e_{s+\mu +c}\},
\end{eqnarray}
sendo $\{e_{1},...e_{s}\}$, $\{e_{s+1},...e_{s+u}\}$, $\{e_{s+u+1},...,e_{s+u+c}\}$ os conjuntos de autovetores de $A(\overline{x})$ com parte real negativa, com parte real positiva e com parte real nula, respectivamente. Al\'em disso se,
\begin{equation}
\phi _{t}\left( E^{\alpha }\right) =E^{\alpha },\quad \alpha =s,u,c
\end{equation}
para todos os tempos. Ent\~ao chamaremos $E^s$, $E^u$ e $E^c$ de {\bf subespa\c cos est\'aveis, inst\'aveis e centrais}, respectivamente.
\end{definition}
 Observe que $E^s$, $E^u$ e $E^c$ s\~ao subespa\c cos invariantes do fluxo linearizado. Mas, se considerarmos o fluxo n\~ao linear, eles ser\~ao invariantes somente em $\overline{x}$. Isto devido a contribui\c c\~ao dos termos de ordem maior em $y$, que promovem uma distor\c c\~ao nos autovalores de $A$ conforme nos dist\^anciamos de $\overline{x}$. 
Por exemplo, seja a solu\c c\~ao do sistema aut\^onomo $x=y+\overline{x}$, $\overline{x}$ ponto fixo. Logo, podemos reescrever a EDO do sistema como 
\begin{equation}
\dot{y}={\cal D}f(\overline{x})y+R(y), \hspace{+0.5cm}y\in \Re^{n},
\end{equation}
sendo $R(y)=O(|y|^{2})$. Como existe uma transforma\c c\~ao linear $T$, que leva a forma linearizada desta equa\c c\~ao em uma forma bloco-diagonal 
\begin{equation}
\left( 
\begin{array}{c}
\dot{u} \\ 
\dot{v} \\ 
\dot{w}
\end{array}
\right) =\left( 
\begin{array}{ccc}
A_{s} & 0 & 0 \\ 
0 & A_{u} & 0 \\ 
0 & 0 & A_{c}
\end{array}
\right) \left( 
\begin{array}{c}
u \\ 
v \\ 
w
\end{array}
\right),
\end{equation}
onde $T^{-1}y\equiv (u,v,w)\in \Re^{s}\otimes \Re^{u}\otimes \Re^{c},$ $s+u+c=n$, $A_{s}$ \'e uma matriz $s\times s$ tendo autovalores com parte real negativa, $A_{u}$ \'e uma matriz $n\times n$ com autovalores de parte real positiva e $A_{c}$ \'e uma matriz $c\times c$ com autovalores com parte real nula, aplicando esta transforma\c c\~ao linear para as coordenadas do problema n\~ao linearizado, teremos 
\begin{eqnarray}
\dot{u} &=&A_{s}u+R_{s}(u,v,w)  \nonumber \\
\dot{v} &=&A_{u}v+R_{u}(u,v,w)   \label{sd1} \\
\dot{w} &=&A_{c}w+R_{c}(u,v,w),  \nonumber
\end{eqnarray}
logo, devemos estudar n\~ao os subespa\c cos invariantes sobre o ponto fixo, mas
variedades, pois assim, em qualquer ordem de aproxima\c c\~ao teremos as
propriedades desejadas. Estes fatos nos conduzem ao seguinte teorema:
\begin{theorem}
Suponhamos (\ref{sd1}) $C^{r}$ com $r\geq 2.$ Ent\~ao, o ponto fixo $%
(u,v,w)=0$ do sistema possui uma variedade est\'avel local $W_{loc}^{s}(0) \in C^{r}$, de dimens\~ao $s$; uma variedade inst\'avel local $W_{loc}^{u}(0) \subset C^{r}$, de dimens\~ao $u$; uma variedade central local$W_{loc}^{c}(0) \subset C^{r}$, de dimens\~ao $c$. H\'a uma interse\c c\~ao entre estas variedades em $(u,v,w)=0$. Estas variedades s\~ao tangentes ao campo vetorial linearizado na origem($E^\alpha , \alpha = s,u,c$) e, portanto, podem ser por eles representadas localmente. Em particular,
\begin{eqnarray}
W_{loc}^{s}(0)&=&\{(u,v,w){\cal |}v=h_{v}^{s}(u),w=h_{w}^{s}(u),{\cal D}h_{v}^{s}(0)={\cal D}h_{w}^{s}(0)=0 \} \nonumber \\
W_{loc}^{u}(0)&=&\{(u,v,w){\cal |}u=h_{u}^{u}(v),w=h_{w}^{u}(v),{\cal D}h_{u}^{u}(0)={\cal D}h_{w}^{u}(0)=0 \}\\
W_{loc}^{c}(0)&=&\{(u,v,w){\cal |}v=h_{v}^{c}(w),u=h_{u}^{c}(w),{\cal D}h_{v}^{c}(0)={\cal D}h_{v}^{c}(0)=0\}, \nonumber
\end{eqnarray}
\end{theorem}
com $|u|$, $|v|$ e $|w|$ suficientemente pequenos nas express\~oes para $W_{loc}^{s}(0)$, $W_{loc}^{u}(0)$ e $W_{loc}^{c}(0)$, respectivamente. Vale lembrar que  $h_{v}^{s}(u),$\ $h_{w}^{s}(u),h_{u}^{u}(v),h_{w}^{u}(v),h_{v}^{c}(w),h_{u}^{c}(w)$ s\~ao fun\c c\~oes $C^{r}$. Al\'em disso, $W_{loc}^{s}(0)$ e $W_{loc}^{u}(0)$ t\^em propriedades assint\'oticas de $E^{s}$ e $E^{u}$, respectivamente.

Passemos ao estudo de um outro tipo de solu\c c\~ao, chamada de solu\c c\~ao peri\'odica.
\begin{definition}
Uma solu\c{c}\~{a}o de um sistema aut\^onomo atrav\'es do ponto $x_{0}$ ser\'a chamada de {\bf peri\'odica} de per\'\i odo $T$, se existir $T=\min_a \{a;x(t,x_0)=x(t+a,x_0), \quad a > 0\}$, tal que $x(t,x_0)=x(t+T,x_0)$ para todo $t \in \Re$.
\end{definition}
Considerando um campo vetorial bidimensional sujeito \`a equa\c c\~ao abaixo 
\begin{eqnarray}
\dot{x} &=&f(x,y)  \label{sd2} \\
\dot{y} &=&g(x,y),  \nonumber
\end{eqnarray}
com $(x,y)\in \Re^{2}$, ent\~ao \'e verdade que :
\begin{theorem}[Crit\'erio de Bendixson]
Se numa regi\~ao simplesmente conexa $D\subset \Re^{2}$ a express\~ao $\frac{\partial f}{\partial x}+\frac{\partial g}{\partial y}$ n\~ao for identicamente zero e n\~ao mudar de sinal, ent\~ao (\ref{sd2}) n\~ao possui \'orbita fechada, inteiramente contida em $D$.
\end{theorem}
\begin{theorem}
Seja $B(x,y)$ cont\'\i nua com derivada cont\'\i nua em uma regi\~ao simplesmente conexa $D\subset \Re^{2}.$ Se $\frac{\partial (Bf)}{\partial x}+\frac{\partial (Bg)}{\partial y}$ n\~ao for identicamente zero e n\~ao mudar de sinal em $D$, ent\~ao (\ref{sd2}) n\~ao possui \'orbita fechada inteiramente contida em $D$.
\end{theorem}

Existe um conjunto de resultados chamados de Teoria de \'Indices para a an\'alise de fluxos bidimensionais, que pode ser resumida da seguinte forma:

Seja $\Gamma $ qualquer \'orbita fechada num plano que n\~ao contenha nenhum ponto fixo do campo. Movendo-se, por exemplo, no sentido hor\'ario(positivo) por $\Gamma$, os vetores que representam o campo rotacionam-se, sendo essa rota\c c\~ao de $2\pi k$ para uma volta completa em $\Gamma $, e $k$ um n\'{u}mero inteiro. Este $k$ \'e chamado de \'\i ndice de $\Gamma$. O \'\i ndice de um circuito fechado, n\~ao contendo ponto fixo, pode ser calculado pela integra\c c\~ao do \^angulo dos vetores em cada ponto do circuito $\Gamma $, medido em rela\c c\~ao a algum sistema de coordenadas escolhido. Assim, o valor de $k$ \'e dado por
\begin{equation}
k=\frac{1}{2\pi }\oint_{\Gamma }d\phi =\frac{1}{2\pi }\oint_{\Gamma }d\left(\tan ^{-1}\frac{g(x,y)}{f(x,y)}\right) =\frac{1}{2\pi }\oint_{\Gamma }\frac{fdg-gdf}{f^{2}+g^{2}}.
\end{equation}
E a classifica\c c\~ao do ponto em quest\~ao \'e dada pelo seguinte teorema:
\bigskip

\begin{theorem}[Teorema dos \'Indices]
As seguintes propriedades s\~ao v\'alidas: \ \ \ 

\begin{enumerate}
\item  O \'\i ndice de um sorvedouro, uma fonte ou um centro \'e +1;

\item  O \'\i ndice de um ponto de sela hiperb\'olico \'e -1;

\item  O \'\i ndice de uma \'orbita fechada \'e +1;

\item  O \'\i ndice de uma curva fechada, n\~ao contendo qualquer ponto
fixo, \'e 0;

\item  O \'\i ndice de uma curva fechada \'e igual \`a soma dos
\'\i ndices dos pontos fixos no seu interior.
\end{enumerate}

\begin{corollary}
Dentro de qualquer \'orbita fechada $\gamma $ deve haver, no m\'\i nimo, um ponto fixo. Se existe um, ent\~ao deve haver uma fonte, um sorvedouro ou um centro. Se todos os pontos fixos em $\gamma $ s\~ao hiperb\'olicos, ent\~ao deve haver um n\'{u}mero \'\i mpar $2n+1$, dos quais $n+1$ s\~ao fontes, sorvedouros ou centros.
\end{corollary}
\end{theorem}

Para discutir o comportamento assint\'otico das solu\c c\~oes, cabem as seguintes defini\c c\~oes:

\begin{definition}
Um ponto $x_{\omega }\in \Re^{n}$ \'e chamado $\mathbf{\omega}${\bf-limite} de $x\in \Re^{n} $, se existe uma seq\"u\^encia $\left\{ t_{i}\right\} ,t_{i}\uparrow \infty $, tal que, $\lim \phi \left( x,t_{i}\right) =x_{\omega }$, sendo o conjunto de todos os pontos $\omega $-limite denotado por $\omega \left(x\right) $.
\end{definition}

\begin{definition}
Um ponto $x_{\alpha }\in \Re^{n}$ \'e chamado $\mathbf{\alpha}${\bf-limite} de $x\in \Re^{n} $, se existe uma seq\"u\^encia $\left\{ t_{i}\right\} ,t_{i}\uparrow-\infty $, tal que, $\lim \phi \left( x,t_{i}\right) =x_{\alpha }$, sendo o conjunto de todos os pontos $\alpha $-limite denotado por $\alpha \left(x\right) $.
\end{definition}

\begin{definition}
Um ponto $x_{0}$ \'e dito {\bf n\~ao-errante} se, para todo aberto $U$, tal que $x_{0}\in U$ existe pelo menos um $t\neq 0$ tal que $\left( \phi \left(U,t\right) \cap U\right) \neq \emptyset $.
\end{definition}
Da defini\c c\~ao anterior, segue que pontos fixos e \'orbitas peri\'odicas s\~ao \ pontos n\~ao-errantes. 

Um conceito que se mostrou \'util nos estudos de Sistemas Din\^anicos, ao
contr\'ario dos conceitos de $\omega , \alpha$-limite, foi a id\'eia de conjunto
hiperb\'olico que conjuntamente com o conceito de conjunto n\~ao-errante formam
a base para definir atratores estranhos e portanto para formar um cen\'ario de transi\c c\~ao \`a turbul\^encia.
\begin{definition}
 Dizemos que um conjunto compacto e invariante $\Lambda$ \'e  {\bf hiperb\'olico} para o fluxo $\phi _t: M \rightarrow M$, se  existe $C>0$, $0< \lambda <1$, e seu fibrado tangente $T_{\Lambda} M$ pode ser continuamente decomposto como a seguinte soma direta: $T_{\Lambda} M = E^s \oplus E^u$, onde, para todo $x \in \Lambda$ e todo $t$, temos:
 \begin{eqnarray}
 \forall v \in E^u _x, \Rightarrow {\parallel \phi_{-t}v \parallel}  \leq C\lambda^{t}{\parallel v \parallel}; \nonumber \\
 \forall v \in E^s _x, \Rightarrow {\parallel \phi_{t}v \parallel} \leq C\lambda^{t}{\parallel v \parallel}. \nonumber  
 \end{eqnarray}
\end{definition}
Ou seja, o espa\c co tangente de cada ponto de um conjunto hiperb\'olico pode ser decomposto como a soma direta de um subespa\c co est\'avel com um inst\'avel.
\begin{definition}
Um conjunto invariante fechado $A\subset \Re^{n}$ \'e chamado {\bf conjunto atrator}, se existe alguma vizinhan\c ca $U$ de $A$, tal que 
\begin{equation}
\forall x\in U,\quad \forall t\in \Re_{+},\quad \phi(x,t) \in U,\quad \lim_{t\rightarrow \infty }\phi (x,t)=A.
\end{equation}
\end{definition}
Segue da defini\c c\~ao que o conjunto $A=\cap _{t>0}\phi (t,M)$ \'e um conjunto atrator.

\begin{definition}
Se $B=\cup _{t\leq 0}\phi \left( U,t\right) $, onde $U$ \ \'e tal que $\lim_{t\rightarrow \infty }\phi (U,t)=A$, onde $A$ \'e o conjunto atrator. Ent\~ao, $B$ \'e chamado de {\bf bacia de atra\c c\~ao} de $A$.
\end{definition}

\begin{definition}
Um conjunto fechado e conexo $M$ \'e chamado de {\bf regi\~ao de aprisionamento}, se $\phi (t,M)\subset M,\forall t\geq 0$.
\end{definition}

\begin{definition}
Um conjunto invariante fechado $A$ \'e {\bf topologicamente transitivo} se, para quaisquer dois abertos $U,V\subset A,\quad \exists t\in \Re$, tal que $\phi (t,U) \cap V \neq \emptyset $.
\end{definition}

\begin{definition}
Um conjunto $A$ \'e chamado de {\bf atrator}, se $A$ \'e um conjunto atrator, e al\'em disto, for topologicamente transitivo.
\end{definition}

\begin{definition}
Chamamos um fluxo $\phi_t(x,t)$ de {\bf fluxo tipo Axioma A} se o conjunto dos pontos n\~ao-errantes $\Omega$ de $\phi_t$ goza das seguintes propriedades: compacto, hiperb\'olico e os pontos fixos e as \'orbitas peri\'odicas s\~ao densos em $\Omega$.
\end{definition}

\begin{definition}
Chamamos um conjunto $U$ de {\bf tipo Axioma A} se goza das seguintes propriedades: invariante, hiperb\'olico e os pontos fixos e as \'orbitas peri\'odicas s\~ao densos em $U$ e $\phi$ \'e topologicamente transitivo sobre $U$.
\end{definition}

\begin{definition}
Chamamos um conjunto $\Lambda$ tipo Axioma A de {\bf atrator tipo Axioma A}, se existe $U$, tal que $ \Lambda = \cap _{t \geq 0} \phi_t (U)$.
\end{definition}

\begin{proposition}
Seja $\phi $ um fluxo e $M$ um conjunto compacto invariante por $\phi $. Ent\~ao, para todo $p\in M$ vale:
\begin{enumerate}
\item  $\omega (p)\neq \emptyset $;

\item  $\omega (p)$ \'e fechado;

\item  $\omega (p)$ \'e invariante sobre o fluxo;

\item  $\omega (p)$ \'e uma uni\~ao de \'orbitas fechadas;

\item  $\omega (p)$ \'e convexo.
\end{enumerate}
\end{proposition}

\begin{lemma}
Se $\omega \left( p\right) $ n\~ao tem pontos fixos, ent\~ao $\omega \left( p\right) $ \'e uma \'orbita fechada.
\end{lemma}

\begin{lemma}
Sejam os pontos fixos $p_{1}\neq $ $p_{2}\in $ $\omega\left( p\right) $, dentro de uma regi\~ao de aprisionamento. Ent\~ao existe, no m\'aximo, uma \'orbita $\gamma \subset \omega\left( p\right) $, tal que $\alpha \left( \gamma \right) =p_{1}$ e $\omega \left( \gamma \right) $ $=p_{2}$.
\end{lemma}
Para fluxos bidimensionais, podemos classificar completamente toda a fam\'\i lia de conjuntos $\omega(x)$ associada aos pontos de um conjunto invariante, pelo seguinte teorema:
\begin{theorem}[Poincar\'e-Bendixson]
Seja $p\in M\subset \Re^{2}$ , sendo $M$ uma regi\~ao positivamente invariante para o campo vetorial contendo um n\'{u}mero finito de pontos fixos. Ent\~ao, uma das afirma\c c\~oes abaixo, mutuamente excludentes, \'e verdadeira:
\begin{enumerate}
\item  $\omega \left( p\right) $ \'e um ponto fixo;

\item  $\omega \left( p\right) $ \'e uma \'orbita fechada;

\item  $\omega \left( p\right) $ consiste em um n\'{u}mero infinito de pontos fixos $\left\{ p_{i}\right\} _{i=1}^{\infty }$ e \'orbitas $\gamma$ com $\alpha \left( \gamma \right) =p_{i}$ e $\omega \left( \gamma \right)=p_{j}$.

Ou seja, os atratores de um fluxo bidimensional s\'o podem ser de tr\^es tipos:

\begin{enumerate}
\item  pontos fixos;

\item  \'orbitas fechadas;

\item  uni\~ao de ponto fixos e \'orbitas fechadas, geradas pelas intersec\c{c}\~oes das variedades est\'aveis e inst\'aveis destes pontos fixos, ou seja, \'orbitas de pontos fixos hiperb\'olicos.
\end{enumerate}
\end{enumerate}
\end{theorem}

\begin{theorem}[Hopf]
Se o sistema $\dot{x}=f_{\mu }\left( x\right) $, $x\in \Re^{n}$, $\mu \in \Re$ possui um ponto fixo $\left( x_{0},\mu _{0}\right) $, tal que as seguintes propriedades s\~ao verificadas:
\begin{enumerate}
\item[H1]  $D_{x}f_{\mu _{0}}\left( x_{0}\right) $ tem um par de autovalores imagin\'arios, e nenhum outro autovalor com parte real nula.
\end{enumerate}
Ent\~ao (H1), implica que existe uma curva suave de pontos fixos $\left( x_{\mu },\mu \right) $ com $x_{\mu _{0}}=x_{0}$, sendo que os autovalores $\lambda _{\mu} $ e $\lambda _{\mu }^{\ast }$ de $D_{x}f_{\mu _{0}}\left( x_{\mu }\right) $ variam suavemente com $\mu $. Se, al\'em disto, valer
\begin{enumerate}
\item[H2] $\frac{d({\cal R}e \lambda_{\mu})}{d\mu }|_{\mu_0} = d \neq 0$,
\end{enumerate}
ent\~ao, existe, uma \'unica variedade central tridimensional passando por $\left( x_{0},\mu _{0}\right) \in \Re^n \times \Re $ e um sistema de coordenadas suaves, tal que a expans\~ao de Taylor at\'e terceira ordem sobre a variedade central \'e dada por
\begin{eqnarray}
  \dot{x} = \left( d\mu + a( x^2 + y^2) \right)x - (\omega + c\mu + b(x^2 + y^2))y, \nonumber \\
  \dot{y} = \left( d\mu + a( x^2 + y^2) \right)y - (\omega + c\mu + b(x^2 + y^2))x. \nonumber
\end{eqnarray}
Portanto temos tr\^es casos:
\begin{enumerate}

\item[Se $ a \neq 0$,] existe uma superf\'\i cie de solu\c c\~oes peri\'odicas na variedade central que possui tangenciamento quadr\'atico com os autoespa\c cos de $\lambda _{ \mu _0}, $ e $\lambda _{\mu _0 }^{\ast }$, dado(com corre\c c\~ao at\'e segunda ordem) pelo parabol\'oide $ \mu = - \frac{a}{d}(x^2 + y^2)$.

\item[Se $a < 0$,] ent\~ao estas solu\c c\~oes peri\'odicas s\~ao ciclos limites est\'aveis.

\item[Se $a > 0$,] ent\~ao estas solu\c c\~oes s\~ao repulsivas.

\end{enumerate}

\end{theorem}

\begin{theorem}[Peixoto]
Um campo vetorial $C^{r}$ sobre uma variedade $M^2$ bidimensional diferenci\'avel compacta \'e estruturalmente est\'avel, se e somente se,
\begin{enumerate}
\item  O n\'umero de pontos fixos e \'orbitas peri\'odicas \'e finito, sendo todos hiperb\'olicos;
\item  N\~ao existe \'orbita conectando pontos de sela;
\item  O conjunto n\~ao-errante \'e formado por pontos fixos e \'orbitas peri\'odicas.
\end{enumerate}
Al\'em do mais, se $M^2$ \'e orient\'avel, o conjunto dos campos estruturalmente est\'aveis \'e aberto e denso em $H^r(M^2)$\footnote{O espa\c co de Hilbert dos campos $C^r$ sobre $M^2$.}.
\end{theorem}

\begin{definition}
Seja $\Sigma _{T}\left( \tilde{x}\right) $ uma superf\'\i cie transversal a $\dot{x}\left( t,x_{0}\right) $, \'orbita peri\'odica de per\'\i odo $T$, em um ponto $\tilde{x}=x\left( t_{1},x_{0}\right) $ para algum $t_{1}$. Chamamos $\Sigma _{T}\left( \tilde{x}\right) $ de {\bf sec\c c\~ao de Poincar\'e}.
\end{definition}
\begin{definition}
Chamamos de {\bf aplica\c c\~ao de Poincar\'e}, a aplica\c c\~ao $P:V\subset \Sigma _{T}\left( \tilde{x}\right) \rightarrow \Sigma _{T}\left( \tilde{x}\right) $, tal que para todo $x\in V,~P\left( x\right) =\phi \left(x,T\right) $, onde $\Sigma _{T}\left( \tilde{x}\right) $ \'e a sec\c c\~ao de Poincar\'e do fluxo $\phi$, e $V \subset \Sigma _{T}\left( \tilde{x}\right) $ um subconjunto aberto.
\end{definition}
\bigskip 
Para ilustrar estas defini\c c\~oes, vamos a um exemplo simples. Seja um sistema peri\'odico no tempo, i.e., $f\left( x,t\right) =f\left(x,t+ 2\pi \right) $, podendo ser facilmente reduzido a 
\begin{eqnarray}
\dot{x} &=&f\left( x,\theta \right) \\
\dot{\theta} &=&\omega \hspace{+0.5cm }\left( x,\theta \right) \in
\Re^{n}\times S^{1}.
\end{eqnarray}
Portanto, seu fluxo ser\'a $\phi ( x,t) =( x(t),\theta(t)) $, com $\theta( t) =\omega(t) +\theta _{0}$, $(mod2\pi)$. Definimos uma se\c{c}\~ao transversal $\Sigma _{\tau }\left( \theta _{0}\right) =\left\{\left( x,\theta \right) \in \Re^{n}\times S^{1};~\theta =\theta _{0}\in \left[0,2\pi \right] \right\} $, que \'e a se\c c\~ao de Poincar\'e, sendo que neste exemplo, em particular, ela \'e global, exceto o caso $\omega =0$. Da defini\c c\~ao de aplica\c c\~ao de Poincar\'e segue
\begin{equation}
P:x\left( \frac{\theta -\theta  _{0}}{\omega }\right) \mapsto x\left( \frac{\theta -\theta_{0} +2\pi }{\omega }\right),
\end{equation}
ou seja, a aplica\c c\~ao de Poincar\'e \'e uma geratriz arbitr\'aria do cilindro infinito $\Re \times S^1$.
\begin{definition}
Seja $A$ um espa\c co m\'etrico compacto, $N\left( r,A\right) $ a menor cobertura de $A$ com abertos de raio $r$. Chamamos de {\bf dimens\~ao de capacidade}, e denotamos por $\dim _{K}A$, a quantidade 
\begin{equation}
\dim _{K}A=-\lim_{r\rightarrow 0}\sup \frac{\log N\left( r,A\right) }{\log r}.
\end{equation}
\end{definition}

\begin{definition}
Seja $A$ conjunto n\~ao-vazio, $\sigma $ uma cobertura finita, cujo k-\'esimo conjunto $\sigma _{k}$ tem di\^ametro $d_{k}=\dim \sigma_{k}<r $. Dado $\alpha $ positivo, e $m_{r}^{\alpha }\left( A\right)=\inf_{\sigma }\Sigma _{k}\left( d_{k}\right) ^{\alpha }$, definimos a {\bf dimens\~ao Hausdorff} de $A$, como 
\begin{equation}
\dim _{H}A=\sup \left\{ \alpha ;\lim_{r\rightarrow 0}m_{r}^{\alpha }\left(A\right) >0\right\}.
\end{equation}
\end{definition}

\begin{theorem}[Sinai-Ruelle-Bowen]
Se $\phi _t : \Re^n \rightarrow \Re^n$ \'e um fluxo classe $C^2$ e possui um atrator hiperb\'olico $A$. Ent\~ao, existe uma \'unica medida $\mu$, com suporte em $A$. Ent\~ao, dada uma $g$ cont\'\i nua q.t.p.[$\lambda$] \footnote{ Abrevia\c c\~ao para quase por toda parte em rela\c c\~ao \`a medida de Lebesgue.} vale:
\begin{equation}
\label{SRB}
 \lim _{T \rightarrow \infty} \frac{1}{T} \int _{t_0} ^{T} g( \phi _t(x)) dt = \int _A g d\mu \quad .
\end{equation} 
\end{theorem}

\section{Um Mecanismo Matem\'atico para a Transi\c c\~ao \`a Turbul\^encia}

O primeiro cen\'ario para a transi\c{c}\~ao \`a turbul\^encia que existiu, foi o proposto Landau \cite{Landau}. Neste cen\'ario, o fluxo vai passando por bifurca\c c\~oes de Hopf sucessivas, com o aumento do N\'umero de Reynolds. Ou seja, o fluxo migraria de um n-t\'oro para um (n+1)-t\'oro, numa s\'erie infinita! Embora este cen\'ario concorde com a intui\c c\~ao, ele n\~ao \'e correto, pois experimentalmente e em simula\c c\~oes num\'ericas \cite{Frisch,Lichtenberg}, ap\'os o 3-t\'oro, o fluxo j\'a se torna turbulento, e n\~ao somente aperi\'odico\footnote{Onde consideramos um n\'umero de Reynolds suficientemente grande($10^6$).}\cite{Ruelle}.

Tentemos agora fazer uma reconcilia\c c\~ao dos conceitos expostos acima com a situa\c c\~ao encontrada na equa\c c\~ao de Navier-Stokes bidimensional, j\'a que para a tridimensional nem mesmo temos provas de exist\^encia\cite{Temam}. Identificaremos elementos da Teoria de Sistemas Din\^amicos na equa\c c\~ao de Navier-Stokes e ordenaremos os resultados acima, mostrando como estes constr\~oem quadros de transi\c c\~ao \`a Turbul\^encia.

Escrevendo a equa\c c\~ao de Navier-Stokes como 
\begin{equation}
\frac{d}{dt}x=X_{Re}(x),
\end{equation}
para $Re=0$, o sistema tem uma solu\c c\~ao $x=0$. Consideremos agora, o jacobiano desta solu\c c\~ao 
\begin{equation}
J_{k}^{l}=\frac{\partial X_{0}^{j}}{\partial x^{k}}\left( 0\right)
\end{equation}
que tem parte real negativa, ou seja, $x=0$ \'e um ponto fixo atrativo. Assim, o determinante do jacobiano \'e n\~ao nulo. Ent\~ao, pelo Teorema da Fun\c c\~ao Impl\'\i cita, existe $\xi _{Re}\left(t\right) $, chamada solu\c c\~ao estacion\'aria, dependendo continuamente de $Re$, tal que 
\begin{equation}
X_{Re}\left( \xi _{Re}\right) =0,
\end{equation}
ou seja, para todo $Re$ existir\'a uma solu\c c\~ao de ponto fixo(solu\c c\~ao estacion\'aria). Mas, por continuidade, o jacobiano de $\xi _{Re}$, denotado por $J_{k}^{l}\left[ \xi _{Re}\right]$, para um n\'{u}mero de Reynolds suficientemente pequeno, deve ter parte real negativa.

Ao aumentarmos o N\'{u}mero de Reynolds, a parte real dos autovalores pode se anular, fazendo com que os pontos fixos se unam ou desapare\c cam. Ou que sucessivos pares de autovalores complexos possam cruzar o eixo imagin\'ario. Logo pelo teorema de Hopf o que era um ponto fixo(solu\c c\~ao estacion\'aria), passa a ser uma \'orbita peri\'odica(solu\c c\~ao peri\'odica), sendo sua amplitude proporcional ao N\'{u}mero de Reynolds. Todavia, \`a medida que aumentamos $Re$, podemos ter uma outra bifurca\c c\~ao de Hopf ou transforma\c c\~ao de uma \'orbita peri\'odica em uma quase-peri\'odica. Sendo assim, pelo Teorema de Peixoto, a vizinhan\c ca desta \'orbita est\'a em um conjunto que \'e o complemento de um subconjunto denso e aberto do espa\c co de Hilbert dos campos $C^{r}$ sobre o t\'oro. Este conjunto, em diversas situa\c c\~oes, \'e um atrator, sendo este conjunto que defi ne o chamado cen\'ario de transi\c c\~ao \`a Turbul\^encia.

O mais conhecido cen\'ario, por ser o mais antigo(1971,1978), \'e o chamado Cen\'ario de Ruelle-Takens, embasado no seguinte teorema:
\begin{theorem}[Newhouse-Ruelle-Takens]
Seja $v$ um campo vetorial constante sobre um t\'oro $T^{n}=\Re^{n}/Z^{n}$. Se $n\geq 3$, toda vizinhan\c ca $C^{2}$ de $v $ cont\'em um campo vetorial $\tilde{v}$ com um atrator estranho tipo Axioma A. Se $n\geq 4$, podemos tomar $C^{\infty }$ em vez de $C^{2}$.
\end{theorem}

Como o teorema exige $n\geq 3$, ent\~ao o sistema deve passar primeiro por tr\^es bifurca\c c\~oes de Hopf antes de encontrar o atrator estranho\footnote{ Um atrator tipo Axioma A \'e chamado de estranho, se apresenta sensibilidade em rela\c c\~ao as condi\c c\~oes iniciais, i.e., a dist\^ancia de duas \'orbitas cresce exponencialmente no tempo.} tipo Axioma A.

O cen\'ario de Feigenbaum consiste em um processo de duplica\c c\~ao infinito de per\'\i odos da \'orbita em um intervalo finito de varia\c c\~ao do N\'{u}mero de Reynolds, sendo este atrator aperi\'odico, ou seja, composto por \'orbitas est\'aveis de per\'\i odo $2^{\infty }$. Assim, o sistema migra para a fase turbulenta por interm\'edio de bifurca\c c\~oes, como as do tipo \textit{pitchfork} at\'e que o N\'umero de Reynolds atinja seu valor cr\'\i tico e o sistema se estabele\c ca em um atrator aperi\'odico e erg\'odico.

Como fica patente pela exposi\c c\~ao acima, a aproxima\c c\~ao, atrav\'es da Teoria de Sistemas Din\^amicos, n\~ao nos traz informa\c{c}\~oes sobre a estrutura espacial do fluxo. Por exemplo, n\~ao nos diz quais s\~ao os eventos f\'\i sicos associados \`a mudan\c ca das propriedades do espa\c co de Hilbert das solu\c c\~oes, exist\^encia das regi\~oes inerciais, leis de pot\^encia. Logo, se desejarmos caracterizar a estrutura espacial das solu\c c\~oes assint\'oticas, \'e muito \'util lan\c car m\~ao sobre a hip\'otese de ergodicidade, via Teorema Sinai-Ruelle-Bowen. Partindo desta hip\'otese, que estes atratores s\~ao erg\'odigos, podemos trocar os c\'alculos espa\c co-temporais por m\'edias sobre o atrator e vice-versa, ou seja, estaremos fazendo uma an\'alise estat\'\i stica da Turbul\^encia(ver Ruelle em\cite{Sirovich}).

\chapter{Processos Estoc\'asticos}
Este cap\'\i tulo vai no mesmo esp\'\i rito do anterior, assim que, para as pessoas sem contato pr\'evio com o assunto, aconselhamos as seguintes refer\^encias: \cite{Ash,Feller, Honerkamp,Lamperti,Yaglom}.
\section{Conceitos B\'asicos}
\begin{definition}
Uma tripla $(\Omega, {\cal F}, P)$ \'e chamada de {\bf espa\c co de probabilidade}, se $ \Omega$ \'e um conjunto n\~ao-vazio, ${\cal F}$ uma $\sigma$-\'algebra minimal sobre os borelianos de $ \Omega$, e $P: \Omega \rightarrow E$ uma medida de probabilidade sobre o espa\c co linear $E$.
\end{definition}

\begin{definition}
 Uma {\bf vari\'avel aleat\'oria} $x$ sobre o espa\c co de probabilidade$(\Omega, {\cal F}, P)$ \'e uma fun\c c\~ao $x: \Omega \rightarrow E$ mensur\'avel em rela\c c\~ao a ${\cal F}$.
\end{definition}

\begin{definition}
Um {\bf Processo Estoc\'astico} \'e uma cole\c c\~ao de vari\'aveis aleat\'orias $\ X$ definidas em um mesmo espa\c co de probabilidade ($\Omega,{\cal F},P$), indexados por elementos de um conjunto de par\^ametros $T$. Quando $T=Z,Z^{+}$, tamb\'em chamamos o processo de {\bf seq\"u\^encia aleat\'oria}, se $T=\Re^{n}$ chamamos de {\bf campo aleat\'orio} ou fun\c c\~ao aleat\'oria.
\end{definition}

\begin{definition}
Se $X$ \'e um processo estoc\'astico, ent\~ao para $t\in T,\omega \in \Omega, x_{t}(\omega)\equiv x(t,\cdot ):\Omega \rightarrow E$, \'e mensur\'avel em rela\c c\~ao a ${\cal F} $; por outro lado $x_{\omega}(t) \equiv x(\cdot,\omega ):T\rightarrow E$ \'e cont\'\i nua em rela\c c\~ao \`a $T$, sendo chamada de {\bf trajet\'oria} ou {\bf realiza\c c\~ao do processo}.
\end{definition}
\begin{definition}
Seja $\{t_{i}\}_{i=1}^{n}\in T$, uma seq\"u\^encia nos \'\i ndices e $C\in {\mathcal B}(E^{n})$ um boreliano no espa\c co linear $E^{n}.$ Definimos a {\bf distribui\c c\~ao finita} por $P_{t_{i}...t_{n}}(C)=P(\{\omega \in \Omega;(x(t_{1},\omega ),...,x(t_{n},\omega ))\in C\})$ . 
\end{definition}
Assim, a distribui\c c\~ao finita \'e uma medida de probabilidade sobre ${\mathcal B}(E^{n})$, j\'a que \ $x(t_{i},\cdot )$ \'e ${\cal F}$-mensur\'avel para cada $t_{i}\in T$. H\'a um teorema de Kolmogorov que garante a rec\'\i proca da afirma\c c\~ao anterior.

\begin{theorem}[Kolmogorov]
Dado um conjunto $T$ e uma fam\'\i lia de medidas $\{P_{t_{1}...t_{n}}\}$ sobre um espa\c co linear $E$ satisfazendo:
\begin{enumerate}
\item $ P_{t_{1}...t_{j}...t_{n}}(C)=P_{t_{j}t_{2}...t_{1}...t_{n}}(C)$, para todo $C\in {\mathcal B}(E^{n}),1\leq j\leq n$, ou seja, invari\^ancia por permuta\c c\~ao em $T$.
\item $P_{t_{1}...t_{j}...t_{n}}(y_{1},...y_{j},\infty,\ldots,\infty )=P_{t_{1}...t_{j}}(y_{1},...y_{j})$.
\end{enumerate}
Ent\~ao, existe, pelo menos um processo aleat\'orio $X$, definido sobre um espa\c co $(\Omega ,{\cal F} ,P)$, tal que as medidas $\{P_{t_{1}...t_{n}}\}$ s\~ao fun\c c\~oes de distribui\c c\~ao finita do processo $X$.
\end{theorem}

Portanto, n\~ao precisamos nos preocupar, em primeira an\'alise, com a medida no
espa\c{c}o de probabilidade, pois as distribui\c c\~oes induzidas no espa\c co linear, no nosso caso o $\Re^n$, j\'a nos d\~ao informa\c c\~oes suficientes sobre o processo estoc\'astico em quest\~ao.

\begin{definition}
 Definimos o n-\'esimo {\bf momento} da vari\'avel aleat\'oria $x$ por:
 \begin{equation}
  \langle x^n \rangle \equiv E(x^n) = \int _{ \Omega} x^n(\omega)dP(\omega)= \int _E x^n (w)d\lambda(w) . \nonumber
 \end{equation}
\end{definition}
 Logo, para $E=\Re$ teremos:
 \begin{equation}
   \label{eq:momento}
   \langle x^n \rangle = \int _{- \infty} ^{+ \infty} x^n(k) \left[ \frac{d}{dk} P_x(k) \right]dk ,\nonumber
 \end{equation}
pois, $P_x(k)$ \'e a distribui\c c\~ao finita, n\~ao a densidade de probabilidade. Vale lembrar aqui que a derivada tomada em (\ref{eq:momento}), \'e a derivada no sentido do teorema de Radon-Nikodym, ou seja, $\frac{d}{dk} P_x(k)$ \'e uma distribui\c c\~ao, e n\~ao \'e necessariamente absolutamente cont\'\i nua em rela\c c\~ao \`a medida de Lebesgue.
\begin{definition}
Se um processo \'e tal que para todo elemento de $T$, $\ E(\left|x_{t}\right| ^{2})<\infty $, ent\~ao o chamamos de {\bf processo de segunda ordem}.
\end{definition}

Observe que podemos construir classes de fun\c c\~oes iguais em quase todos os pontos e quadrado integr\'avel\footnote{Fun\c c\~oes que diferem sobre um conjunto de medida nula s\~ao identificadas e geram uma classe. As classes, produzidas sob esta identifica\c c\~ao, geram um espa\c co que goza da propriedade de todos os seus elementos(classes de equival\^encia), serem quadrado-integr\'aveis.}. Essas classes, vistas como elementos de um conjunto, podem ser consideradas como um espa\c co de Hilbert $L_{2}(\Omega ,{\cal F} ,P)$, quando definimos a norma e o produto interno pelas rela\c c\~oes: 
\begin{equation}
(x,y) \equiv E(xy^{\ast }),\qquad \left\| x\right\| \equiv \sqrt{E(\left| x\right|^{2})}. \nonumber
\end{equation}
\begin{definition}
Definimos a {\bf covari\^ancia}, chamada em turbul\^encia de {\bf correla\c c\~ao}, por $K(s,t)=E(x_{s}x_{t}^{\ast })=(x_{s},x_{t}^{\ast })$ com $s,t\in T$.
\end{definition}
Agora que j\'a temos os conceitos b\'asicos, passemos para a pr\'oxima etapa, que \'e a constru\c c\~ao do c\'alculo para processos estoc\'asticos. 

A abordagem escolhida \'e a que nos parece mais intuitiva, por ser mais parecida formalmente com o c\'alculo usual. Adotaremos, doravante, $T=\Re$, e assumiremos que todos os processos estoc\'asticos s\~ao de segunda ordem, ou seja, tem covari\^ancia finita. Com isto esperamos facilitar o entendimento. Mas, cabe ressaltar que existem trabalhos, nos quais s\~ao considerados processos mais gerais\cite{Chavanis,Min} para a descri\c c\~ao dos v\'ortices do campo de velocidade de um fluxo turbulento.

\section{C\'alculo Diferencial}

Diremos que $\lim_{t\rightarrow t_{0}}x_{t}=y,se$\ $y\in L_{2}$\ e $\lim_{t\rightarrow t_{0}}E(\left| x_{t}-y\right| ^{2})=0$, ou seja, diremos que o limite existe se ele converge na m\'edia quadr\'atica.

Assim definimos 
\begin{equation}
   \frac{d}{dt}x_{t}={x \prime}_{t},  \nonumber
\end{equation}
se ${x \prime}_t \in L_2$ e
\begin{equation}
  \lim_{h\rightarrow 0}\left\| \frac{x_{t+h}-x_{t}}{h}-x_{t}^{,}\right\| =0.
\end{equation}
Obviamente diremos que $x_{t}$ \'e diferenci\'avel, se a derivada existe, neste sentido.

Como os processos de segunda ordem s\~ao caracterizados, a menos de uma isometria no espa\c co de Hilbert, pela covari\^ancia, a proposi\c c\~ao abaixo \'e muito \'util.
  
\begin{proposition}
Dado um processo estoc\'astico de segunda ordem $X$ valem as seguintes propriedades:
\begin{enumerate}
\item  Se $\lim_{t\rightarrow t_{0}}x_{t}$ \ existe$\Longleftrightarrow \lim_{s,t\rightarrow t_{0}}K(s,t)$ existe.
\item  Se $\{x_{t}\}$ \'e cont\'\i nuo em $t=t_{0}\Longleftrightarrow K(s,t)$ \'e cont\'\i nuo em $(t_{0},t_{0}).$
\item  ${x \prime}_{t}$ existe, se somente se, 
\begin{equation}
\label{eq:propK}
\lim_{h,k\rightarrow 0}\frac{1}{hk}\left\{ K(t+h,t+k)-K(t+h,t)-K(t,t+k)+K(t,t)\right\}
\end{equation}
existe.
\item  $\{x_{t}\}$ \'e continuamente diferenci\'avel em $(a,b)\Longleftrightarrow \frac{\partial ^{2}}{\partial s\partial t}K(s,t)$ existe e \'e cont\'\i nua $\forall s,t\in (a,b)$.
\end{enumerate}
\end{proposition}
\begin{definition}
Um processo \'e dito {\bf estacion\'ario} se suas distribui\c c\~oes finitas s\~ao invariantes sob transla\c c\~ao \footnote{Normalmente, quando a fam\'\i lia de par\^ametros $T=\Re^{n}$ representa um processo no espa\c co f\'\i sico, o chamamos de {\bf homog\^eneo}; se representa o tempo chamamos de {\bf estacion\'ario}.}, ou seja, se para todo $n$ fixo e toda seq\"u\^encia $\{t_k\}_{k=1}^{n}\in \Re$, e $h\in \Re$ tivermos $P_{t_{i}...t_{n}}(C)=$ $P_{t_{i}+h...t_{n}+h}(C),C\in {\mathcal B}(E^{n})$.
\end{definition}
\begin{definition}
Um processo $X$ \'e dito {\bf estacion\'ario num sentido amplo}, se a fun\c c\~ao covari\^ancia \'e invariante sob transla\c c\~ao, ou seja, $K(s,t)=K(s+h,t+h)$.
\end{definition}
\begin{definition}
Dizemos que o processo \'e de {\bf incrementos estacion\'arios}, se para todo $n$ fixo e toda seq\"u\^encia $\{t_{k}\}_{k=1}^{n}\in \Re$, e $h\in \Re$ \ a distribui\c c\~ao conjunta dos incrementos $(x_{t_{2}}-x_{t_{1}}),...,(x_{t_{n}}-x_{t_{n-1}})$ \'e invariante pela troca \ $t_{i}\rightarrow t_{i}+h.$
\end{definition}
\begin{definition}
Chamamos um processo $W$ de {\bf movimento browniano}, \footnote{Os matem\'aticos chamam-no de {\bf processo de Wiener}} se for de incrementos estacion\'arios, independentes, cont\'\i nuo na m\'edia e de quadrado integr\'avel.
\end{definition}

\begin{theorem}
Todo movimento browniano $W$ \ com $W_{0}=0$ , centrado na m\'edia, goza das seguintes propriedades:
\begin{enumerate}
\item $E(W_{t})={\mu}t,{\mu}=E(W_{1})$.
\item $E(W_{t}^{2})=\sigma ^{2}t,\sigma ^{2}=E(W_{1}^{2})$.
\item $K(s,t)=\sigma ^{2}\min (s,t)$.
\item $W$ \'e cont\'\i nuo, mas\ n\~ao diferenci\'avel({\it em rela\c c\~ao ao tempo}).
\end{enumerate}
\end{theorem} 
\textbf{Prova}
\smallskip
\begin{enumerate}
\item Observe que: $E(W_{t+s})=E(W_{t})+E(W_{t+s}-W_{t})=E(W_{t})+E(W_{s})$, como W \'e cont\'\i nuo na m\'edia, a expectativa deve ser cont\'\i nua, como $E(W_{0})=0$ teremos $1.)$.

\item Por analogia com o item $1.)$, vem: $E(W_{t+s}^{2})=E(W_{t}^{2})+E\left[(W_{t+s}-W_{t})^{2}\right] $ $=$ $E(W_{t}^{2})$ $+E\left[ (W_{s}-W_{t-t})^{2}\right] = E(W_{t}^{2})+E(W_{s}^{2})$, pelas mesmas raz\~oes de $1.)$ segue o resultado $2.)$.

\item Por ser $K(s,t)=E(W_{s}W_{t})=\frac{1}{2}\left\{E(W_{t}^{2})+E(W_{s}^{2})-E(\left| W_{t}-W_{s}\right| ^{2})\right\}=$ \\
$\frac{\sigma ^{2}}{2}\left\{ t+s-\left| t-s\right| \right\}.$

\item Por (\ref{eq:propK}), a derivada do processo s\'o existe se o limite equivalente de covari\^ancia existe. Como \'e f\'acil ver por $3.)$, este limite n\~ao existe, logo $W$ \ n\~ao \'e diferenci\'avel.

\end{enumerate}

\section{C\'alculo Integral}
 
Seja um processo estoc\'astico $X$, assumindo valores em um espa\c co de Banach $V$ arbitr\'ario, $P$ uma parti\c c\~ao de $[a,b]\in \Re$, $\delta (P)=\max\{(t_{i+1}-t_{i});t_{i+1},t_{i}\in \Re \}$, o maior intervalo da parti\c c\~ao $P$, $\zeta _{i}\in \lbrack t_{i},t_{i+1}]$, um ponto do i-\'esimo intervalo, onde  $0\leq i\leq n-1$.
\begin{definition}
 Definimos {\bf soma de Riemann} da fun\c c\~ao $\{x_{t}\}$ o vetor: 
\begin{equation}
S_{P}=\sum_{i=0}^{n-1}x_{\zeta _{i}}(t_{i+1}-t_{i}).
\end{equation}
\end{definition}

\begin{definition}
Se o limite da soma de Riemann, quando $\delta (P)\rightarrow 0$, existe, dizemos que $\{x_{t}\}$ \'e {\bf Riemann integr\'avel} em $[a,b]$ e escrevemos: 
\begin{equation}
\lim_{\delta (P)\rightarrow 0}S_{P}=\int_a^{b}x_{t}dt.
\end{equation}
\end{definition}

\begin{proposition}
A $\int_{a}^{b}x_{t}dt$ existe, se e somente se,$\int_{a}^{b}ds\int_{a}^{b}dtK(s,t)$ existe no sentido de Riemann e vale a igualdade: 
\begin{equation}
\int_{a}^{b}ds\int_{a}^{b}dtK(s,t)=\left\| \int_{a}^{b}x_{t}dt\right\| ^{2}.
\end{equation}
\end{proposition}

\begin{proposition}
Seja $y$ quadrado-integr\'avel. Ent\~ao, o produto interno e a integra\c c\~ao comutam, ou seja, vale a seguinte igualdade:
\begin{equation}
  \left(y, \int_{a}^{b}dtx_{t} \right) = \int_{a}^{b} dt \left(y, x_{t}\right).
\end{equation}
\end{proposition}

Fica claro que podemos estender v\'arios resultados cl\'assicos para esta integral, tais como o Teorema Fundamental do C\'alculo, integrais impr\'oprias e integra\c c\~ao por partes. Existe outra formula\c c\~ao mais sofisticada, que se assemelha a de Lebesque, que nos permite fazer integrais sobre processos estoc\'asticos, embora seja mais abstrata.

\subsection{Integra\c c\~ao estoc\'astica}

Como j\'a vimos, as vari\'aveis aleat\'orias induzem uma medida sobre a \'algebra gerada pelos subconjuntos de seu contradom\'\i nio, chamada fun\c c\~ao de distribui\c c\~ao finita. Donde, surge a seguinte pergunta: Quais as condi\c c\~oes que um processo estoc\'astico deve satisfazer para induzir uma medida no espa\c co das fun\c c\~oes de $T\rightarrow V $, com $T$ conjunto de par\^ametros e $V$ um espa\c co de Banach?

Buscamos definir $\int_{T}fdx_{t}, \forall f\in L_{2}(T,dF)$ tal que, $dF$ seja uma medida que dependa de um processo esoc\'astico. Ou seja, desejamos construir uma integral estoc\'astica que nos permita saber o valor m\'edio de fun\c c\~oes que dependem de processos estoc\'asticos, por exemplo: valores esperados e representa\c c\~ao espectral de processos estoc\'asticos.

\begin{definition}
Dizemos que uma fun\c c\~ao aleat\'oria possui {\bf incrementos ortogonais}, se $\forall (a,b),(c,d)\in T;(a,b)\cap(c,d)=\emptyset$ tivermos $(x_{b}-x_{a},x_{d}-x_{c})=0$
\end{definition}

\begin{theorem}
Seja $\{x_t\}$ uma fun\c c\~ao aleat\'oria de segunda ordem com incrementos ortogonais. Ent\~ao, existe  $F:T\rightarrow \Re$, fun\c c\~ao n\~ao decrescente real, tal que, $F$ \'e uma medida sobre a reta induzida pelo processo $\{x_t\}$. Sendo $F$ definida por: 
\begin{eqnarray}
F(t) &=&E\left\{ \left| x_{t}-x_{t_{0}}\right| ^{2}\right\} ,t>t_{0} \nonumber \\
F(t) &=&-E\left\{ \left| x_{t}-x_{t_{0}}\right| ^{2}\right\} ,t\leq t_{0} \label{eq:intest}.
\end{eqnarray}
\end{theorem}
Por (\ref{eq:intest}) fica claro que: ${x_t}$ \'e cont\'\i nuo em $t$, se somente  se, $F(t)$ \'e cont\'\i nua em $t$ . Agora estamos aptos a construir a integra\c c\~ao estoc\'astica. 
\begin{definition}
Seja $f(t)=\sum\limits_{i=0}^{n-1}c_{i}I_{(t_{i},t_{i+1})}(t)$ uma fun\c c\~ao escada onde cada $t_{i}$ \'e ponto de continuidade de $F$. Definimos a {\bf integral estoc\'astica} de $f$ pelo processo $\{ x_{t}\} $ por: 
\begin{equation}
\int_{T}fdx_{t}=\sum\limits_{i=0}^{n-1}c_{i}(x_{t_{i+1}}-x_{t_{i}}).
\end{equation}
\end{definition}

O ponto importante \'e que a correspond\^encia $\int_{T}fdx_{t}\leftrightarrow f$ \'e uma isometria entre o espa\c co das vari\'aveis aleat\'orias da forma dada pela equa\c c\~ao acima e o conjunto das fun\c c\~oes- escada consideradas como elementos de
$L_{2}(T,dF)$. Mostremos esta afirma\c c\~ao por interm\'edio de um c\'alculo simples:
\begin{eqnarray}
\left\| \int_{T}fdx_{t}\right\| ^{2}&=&E\left\{\sum\limits_{i=0}^{n-1}\sum\limits_{j=0}^{n-1}c_{i}c_{j}^{\ast }\left(x_{t_{i+1}}-x_{t_{i}}\right) \left( x_{t_{j+1}}^{\ast }-x_{tj}^{\ast}\right) \right\} = \nonumber \\
\sum\limits_{i=0}^{n-1}\left| c_{i}\right| ^{2}E\left\{ \left|x_{t_{i+1}}-x_{t_{i}}\right| ^{2}\right\} &=&\sum\limits_{i=0}^{n-1}\left|c_{i}\right| ^{2}\left[ F(t_{i+1})-F(t_{i})\right] =\int \left| f(t)\right|^{2}dF(t). \nonumber
\end{eqnarray}

Como para toda $f$, Borel mensur\'avel em $T$, existe uma seq\"u\^encia crescente de fun\c c\~oes- escada\footnote{O que est\'a por tr\'as desta argumenta\c c\~ao \'e que o conjunto das fun\c c\~oes escadas \'e denso neste espa\c co.} $\{f_n\}$, tal que, $\lim f_{n}=f$, a seq\"u\^encia de fun\c c\~oes-escada \'e uma seq\"u\^encia de Cauchy. Ent\~ao, usando o Teorema da Completeza \ para o espa\c co $L_{2}(T,dF)$, temos assegurada a rela\c c\~ao $f\in L_{2}(T,dF)$. Mas pela isometria, $\left\{ \int_{T}f_{n}dx_{t}\right\} $ \'e uma seq\"u\^encia de Cauchy em $L_{2}(\Omega ,{\cal F} ,P)$. Em seguida, aplicando o Teorema da Converg\^encia Mon\'otona vem $\lim \int_{T}f_{n}dx_{t}=\int_{T}fdx_{t}$. Vale a pena notar que a integra\c c\~ao n\~ao precisa ser sobre $T$, mas pode ser sobre qualquer conjunto de Borel em $T$. Desta forma, a integra\c c\~ao estoc\'astica gera um novo processo estoc\'astico com incrementos ortogonais.

\part{Turbul\^encia Desenvolvida}
\chapter{Teoria de Kolmogorov}
Os modelos para a turbul\^encia s\~ao devidos \`a inexist\^encia de condi\c c\~oes que fechem o sistema, ou seja, o c\'alculo da correla\c c\~ao de segunda ordem envolve a de terceira ordem, o c\'alculo da correla\c c\~ao de terceira ordem envolve a de quarta ordem, e assim sucessivamente, levando a um sistema indeterminado, infinito, e acoplado de equa\c c\~oes diferenciais parciais. Um m\'etodo para atacar esta situa\c c\~ao \'e fazer uma hip\'otese que feche o sistema em alguma ordem.

O modelo de Kolmogorov\cite{Kolmogorov,Kolmogorov1} foi sugerido em 1941 para explicar quais as condi\c c\~oes suficientes para obtermos a chamada Lei dos 2/3\footnote{A Lei dos 2/3 diz que a correla\c c\~ao tem um comportamento assint\'otico tipo $r^{2/3}$ para um fluxo isotr\'opico e homog\^eneo} e nos d\'a tamb\'em uma express\~ao para a correla\c c\~ao de terceira ordem. Buscando um melhor entendimento, seguiremos por etapas. A obten\c c\~ao da rela\c c\~ao de K\'arman-Howarth\cite{Karman} \'e um primeiro passo para o entendimento do problema, pois gera o cen\'ario onde as diversas teorias entrecruzam-se e buscam respaldo te\'orico e experimental \cite{Batchelor, Yaglom}. Depois explicaremos brevemente as leis experimentais; passaremos para as hip\'oteses de Kolmogorov e suas implica\c c\~oes lim\'\i trofes; exporemos os argumentos de Landau\cite{Landau} e Kraichnan\cite{Frisch}, contra a universalidade da constante de Kolmogorov.

\section{Simetria e Tensores de Correla\c c\~ao}

Consideremos, primeiramente, o caso da correla\c c\~ao de segunda ordem das componentes da velocidade entre os pontos $P\left( x_{1},x_{2},x_{3}\right)$ e $P^{\prime}\left({x^{\prime}}_1,{x^{\prime}}_2,{x^{\prime}}_3\right)$ definida por: 
\begin{equation}
E\{ u_i u^{\prime} _k\} =E\{ u\cdot u\} R_{ik}.  \label{eq:stc1}
\end{equation}
Calculando a correla\c c\~ao entre a componente da velocidade $u\left(
P,t\right) $, baseada em $P$ numa dire\c c\~ao arbitr\'aria $a_{i}$ e a
componente da velocidade $u^{\prime} \equiv u\left( P^{\prime} ,t\right) $,
baseada em $P^{\prime} $ numa dire\c c\~ao arbitr\'aria $b_{k}$, temos: 
\begin{equation}
E\left\{ u\cdot u\right\} R\left( a,b\right) =E\left\{ u_{i}u^{\prime} _{k}\right\} a_{i}b_{k},  \nonumber
\end{equation}
mas por (\ref{eq:stc1}) podemos reescrever a rela\c c\~ao acima como 
\begin{equation}
R\left( a,b\right) =R_{ik}a_{i}b_{k}.  \label{eq:stc2}
\end{equation}
Se impusermos que $R(a,b)$ deva ser invariante sob rota\c c\~oes de corpo r\'\i gido\footnote{Isto quer dizer que as reflex\~oes s\~ao deixadas de fora.} e transla\c{c}\~oes, para executarmos a invari\^ancia translacional, basta termos uma depend\^encia funcional somente na dist\^ancia relativa, ou seja, em $r_i=(x^{\prime} _i -x_i) ,\quad a_i,b_k$. Por conseguinte, a pergunta que se imp\~oe \'e: Qual a fun\c c\~ao mais geral $R\left( r;a,b\right) $ da forma de (\ref{eq:stc2}) que \'e invariante sob rota\c c\~oes arbitr\'arias dos vetores $r_{i},a_{i},b_{k}$ ?

A princ\'\i pio poder\'\i amos pensar que qualquer contra\c c\~ao dos tr\^es vetores($r,a,b$) pudesse aparecer na fun\c c\~ao para $R\left(r;a,b\right)$. Entretanto, por (\ref{eq:stc2}), $R\left( r;a,b\right)$ deve ser uma fun\c c\~ao bilinear em $a$ e $b$, i.e., n\~ao devem participar termos como $a\cdot a$ e $b\cdot b$. Portanto $R$ pode ter no m\'aximo a seguinte forma: 
\begin{equation}
R\left( r;a,b\right) =A(r\cdot a)(r\cdot b)+B(a\cdot b)+C(rab),  \nonumber
\end{equation}
onde $A,B,C$ fun\c c\~oes de $r^{2}$.

Comparando (\ref{eq:stc2}) com a equa\c c\~ao acima, \'e f\'acil concluir que: 
\begin{equation}
R_{ik}=A(r^{2})r_{i}r_{k}+B(r^{2})\delta _{ik}+C(r^{2})\epsilon _{ikl}r_{l}.
\label{eq:stc:3}
\end{equation}

Agora voltemos nossa aten\c c\~ao para o tensor de  correla\c c\~ao de terceira ordem
a dois pontos, que \'e definida por:
\begin{equation}
E\left\{ u_{i}u_{j}u^{\prime} _{k}\right\} =\left[ E\left\{ u\cdot u\right\}
\right] ^{3/2}T_{ijk}(r).  \label{eq:stc4}
\end{equation}

Exigindo homogeneidade(invari\^ancia translacional) e invari\^ancia rotacional do escalar

\begin{equation}
  T\left(r;a,b,c\right)=T_{ijk}a_{i}b_{j}c_{k},
\end{equation}
onde $a_{i}$ e $b_{j}$ s\~ao vetores baseados em $P$ e $c_{k}$ baseado em $P^{\prime} $. 

As contra\c c\~oes poss\'\i veis, j\'a que $T\left( r;a,b,c\right) $ tem que ser linear em $a,b,c$, s\~ao: 
\begin{equation}
 r\cdot r, \quad r\cdot a,\quad r\cdot b,\quad r\cdot c,\quad a\cdot b,\quad a\cdot c,\quad b\cdot c,\quad (rabc) ,\quad (abc). 
\end{equation}
Logo sua forma \'e: 
\begin{eqnarray}
T(r;a,b,c) =T_{1}( r\cdot a) ( r\cdot b)
\left( r\cdot c\right) +T_{2}\left( r\cdot a\right) \left( b\cdot c\right) +
 \\
+T_{3}\left( r\cdot b\right) \left( a\cdot c\right) +T_{4}\left( r\cdot
c\right) \left( a\cdot b\right) +T_{5}\left( rabc\right) +T_{6}\left(
abc\right).  \nonumber
\end{eqnarray}
Por (\ref{eq:stc4}) e pela equa\c c\~ao acima vem: 
\begin{equation}
T_{ijk}(r)=T_{1}r_{i}r_{j}r_{k}+T_{2}r_{i}\delta _{jk}+T_{3}r_{j}\delta
_{ik}+T_{4}r_{k}\delta _{ij}+T_{5}\epsilon _{ijkl}r_{l}+T_{6}\epsilon _{ijk},
\label{eq:stc5}
\end{equation}
onde os ${T_i}\prime s$ s\~ao fun\c c\~ao de $r^{2}$.

\section{A Rela\c c\~ao de K\'arm\'an-Howarth}

Consideremos a equa\c c\~ao de Navier-Stokes, tal que, o campo de velocidades e a for\c ca sejam estatisticamente isotr\'opicos e homog\^eneos. 
\begin{eqnarray}
\partial _{t}u_{i}+u_{j}\partial _{j}u_{i}&=&-\partial _{i}p+f_{i}+\nu \partial _{j}\partial _{j}u_{i} \nonumber \\
\partial _{i}u_{i} &=&0 . \nonumber
\end{eqnarray}
Multiplicando a equa\c c\~ao acima por $u_{k}(x+r,t)=u^{\prime} _{k}$ e tomando a m\'edia no espa\c co amostral\footnote{Note que a correla\c c\~ao entre velocidade e press\~ao \'e nula, j\'a que \'e imposta a condi\c c\~ao de isotropia e incompressibilidade.}, vem: 
\label{NS}
\begin{eqnarray}
E\left\{ u^{\prime} _{k}\partial _{t}u_{i}\right\}+E\left\{ u^{\prime} _{k}u_{j}\partial _{j}u_{i}\right\}=-E\left\{ u^{\prime} _{k}\partial _{i}p\right\} +E\left\{ u^{\prime} _{k}f_{i}\right\}+\nu E\left\{ u^{\prime} _{k}\partial _{j}\partial_{j}u_{i}\right\}  \nonumber \\
E\left\{ u_{i}\partial _{t}u^{\prime} _{k}\right\} +E\left\{ u_{i}u^{\prime} _{j}\partial ^{\prime} _{j}u^{\prime} _{k}\right\}=-E\left\{ u_{i}\partial ^{\prime} _{k}p^{\prime} \right\} +E\left\{ u_{i}f^{\prime} _{k}\right\} +\nu E\left\{ u_{i}\partial ^{\prime} _{j}\partial ^{\prime}_{j}u^{\prime} _{k}\right\}.  \nonumber
\end{eqnarray}
Somando as equa\c c\~oes acima, e lembrando que para processos estacion\'arios a covari\^ancia depende somente da diferen\c ca dos argumentos, a saber $r$, $\partial _{i}=-\partial _{r_{i}}=$ $-\partial^{\prime} _{i}$ e os operadores envolvidos comutam com a m\'edia, teremos: 
\begin{eqnarray}
\frac{1}{2}\partial _{t}E\left\{ u_{i}u^{\prime} _{k}\right\} +\frac{1}{2}\partial _{r_{j}}\left[ E\left\{ u_{i}u^{\prime} _{j}u^{\prime} _{k}\right\} -  E\left\{ u^{\prime} _{k}u_{j}u_{i}\right\} \right]  \nonumber \\
=E\left\{ u_{i}f^{\prime} _{i}\right\}  +  \nu \partial _{r_{j}}\partial_{r_{j}}E\left\{ u_{i} {u^{\prime}}_{k}\right\},  \nonumber
\end{eqnarray}

podendo tamb\'em ser escrita sob a forma de:
\begin{equation}
\frac{1}{2}\partial _{t}R_{ik} -\frac{1}{2}\partial _{r_{j}}\left[ T_{jki} +T_{ijk} \right] = S_{ik} + \nu \partial _{r_{j}}R_{ik},  \label{RKH}
\end{equation}
onde $S_{ik}=E\left\{ u_{i}f^{\prime} _{i}\right\}$. 

A equa\c c\~ao (\ref{RKH}) \'e chamada de Rela\c c\~ao de K\'arm\'an-Howarth ({\bf RKH}). Tomando $r=0$ na \textbf{RKH}, esta transforma-se em uma equa\c c\~ao para a varia\c c\~ao de energia do fluxo m\'edio. O primeiro termo do lado direito diz que a for\c ca injeta energia diretamente no fluxo m\'edio; o segundo termo, que a \'unica forma de dissipa\c c\~ao \'e atrav\'es da viscosidade. \'E interessante chamar a aten\c c\~ao para o fato \ da \textbf{RKH} n\~ao ser um sistema fechado, j\'a que para conhecermos a correla\c c\~ao de segunda ordem devemos conhecer a de terceira ordem.

\subsection{An\'alise do Caso Isotr\'opico}

A hip\'otese de isotropia\footnote{Nesta sec\c c\~ao, exigimos que o termo de for\c ca fosse escrito como o gradiente de uma fun\c c\~ao, e portanto, encorporado \`a press\~ao e posteriormente elimidado da RKH, via homogeneidade e isotropia.} traz uma grande simplifica\c c\~ao na obten\c c\~ao do espectro da covari\^ancia, que passa a ser equivalente a do espectro unidimensional, e tamb\'em um ganho no entendimento do mecanismo da turbul\^encia.

Reescrevendo as correla\c c\~oes na forma de (\ref{eq:stc:3}) e (\ref{eq:stc5}), e lembrando que no caso isotr\'opico $R_{ik}$ e $T_{ijk}$ t\^em que ser invariantes sob reflex\~ao, os termos $C,T_{5},T_{6}$ devem ser nulos. Pela incompressibilidade do fluxo vem: 
\begin{eqnarray}
\partial _{r_{k}}R_{ik} &=&\partial _{r_{i}}R_{ik}=0  \label{eq2} \\
\partial _{r_{k}}T_{ijk} &=&0 . \nonumber
\end{eqnarray}

Reescevendo as fun\c c\~oes de K\'arm\'an como:
\begin{eqnarray}
A &=&\frac{f-g}{r^{2}};B=g  \label{eq3} \\
T_{1} &=&\frac{k-h-2q}{r^{3}};T_{4}=\frac{h}{r};T_{2}=T_{3}=\frac{q}{r},
\nonumber
\end{eqnarray}

teremos por (\ref{eq2}) e (\ref{eq3}) que: 
\begin{eqnarray}
g&=&f+\frac{r}{2}\frac{\partial }{\partial r}f  \label{eq3.1} \\
0&=&\frac{r_{i}r_{j}}{\left| r\right| ^{2}}\left[ \frac{\partial }{\partial r}\left( k-h\right) +\frac{2}{r}(k- h-2q) \right] +\frac{2\delta_{ij}}{r}\left[ h+q+\frac{1}{2}\frac{\partial }{\partial r}h\right].
\nonumber
\end{eqnarray}
Substituindo (\ref{eq3.1}) em (\ref{eq:stc:3}) e (\ref{eq:stc5}) vem: 
\begin{eqnarray}
R_{ik}(r)& =\delta _{ik}f(r^{2})+\frac{1}{2}\left[ \delta _{ik}-\frac{r_{i}r_{k}}{\left| r\right| }\right] \frac{\partial }{\partial r}f(r^{2})\label{eq3.2} \\
T_{ijk}(r)& =-\frac{1}{2}\left[ \frac{r_i r_j r_k}{\left| r\right| ^{3}}\left( \frac{\partial }{\partial r}-1\right)+\delta _{ij}\frac{r_{k}}{\left| r\right| }-\left( \delta_{jk}r_{i}+\delta _{ik}r_{j}\right) \left( \frac{1}{\left| r\right| }+\frac{1}{2}\frac{\partial }{\partial r}\right) \right] k(r^{2}).
\nonumber
\end{eqnarray}
Definindo as transformadas de Fourier unidimensionais por: 
\begin{eqnarray}
E\left\{ u\cdot u\right\} f(r,t)& =\frac{1}{2}\int_{-\infty }^{\infty}dk\exp (ikr)F(k,t)  \nonumber \\
F(k,t)& =\frac{E\left\{ u.u\right\} }{\pi }\int_{-\infty }^{\infty }dr\exp
\left( -ikr\right) f(r,t) \label{eq5}\\
E\left\{ u\cdot u\right\} g(r,t)& =\frac{1}{2}\int_{-\infty }^{\infty
}dk\exp (ikr)G(k,t)  \nonumber \\
G(k,t)& =\frac{E\left\{ u\cdot u\right\} }{\pi }\int_{-\infty }^{\infty
}dr\exp \left( -ikr\right) g(r,t),
\end{eqnarray}
substituindo em (\ref{eq3.1}) vem: 
\begin{equation}
G=\frac{1}{2}\left[ F-k\frac{\partial }{\partial k}F\right],  \label{eq6}
\end{equation}
levando-nos \`a seguinte representa\c c\~ao tridimensional da correla\c c\~ao: 
\begin{eqnarray}
R_{ij}(r,t) &=&\int_{-\infty }^{\infty }dk^{3}\exp \left( ik\cdot r\right)
\Phi _{ij}(k,t)  \nonumber \\
\Phi _{ij}(k,t) &=&\frac{1}{\left( 2\pi \right) ^{3}}\int_{-\infty }^{\infty
}dr^{3}\exp \left( -ik\cdot r\right) R_{ij}(r,t),  \nonumber
\end{eqnarray}
onde, obviamente, $R_{ii}(r,t)=R_{ii}(-r,t)$. Definindo $E(k,t)=2\pi k^{2}\Phi _{ii}(k,t)$, podemos escrever a soma das componentes diagonais da correla\c c\~ao como: 
\begin{equation}
R_{ii}(r,t)=\int_{-\infty }^{\infty }dkE(k,t)\frac{\sin (kr)}{kr},
\label{eq8}
\end{equation}
igualando (\ref{eq3.2}) e (\ref{eq8}) obtemos: 
\begin{equation}
k\frac{\partial }{\partial k}\left[ F+2G\right] =-2E  .\label{eq9}
\end{equation}
Substituindo (\ref{eq6}) na express\~ao acima vem: 
\begin{eqnarray}
E(k,t) &=&{\cal D}_{3}F(k,t)  \label{eq10} \\
{\cal D}_{3} &\equiv &\frac{1}{2}\left( k^{2}\frac{\partial ^{2}}{\partial k^{2}}-k\frac{\partial }{\partial k}\right) .  \nonumber
\end{eqnarray}

A equa\c c\~ao acima nos diz que, dado o espectro da fun\c c\~ao de correla\c c\~ao de segunda ordem unidimensional, podemos obter o espectro tridimensional da correla\c c\~ao de segunda ordem. Analogamente, por (\ref{eq3.2}), podemos expressar a correla\c c\~ao de terceira ordem, por meio da fun\c c\~ao de correla\c c\~ao unidimensional\footnote{Estas fun\c c\~oes tamb\'em s\~ao chamadas de \emph{fun\c c\~oes de K\'arm\'an-Howarth} ou \emph{correla\c c\~oes de K\'arm\'an-Howath}.} $k=\frac{E\left\{ u^{\prime} _{1}\left( u_{1}\right)^{2}\right\}}{ \left[ E\left\{ u\cdot u\right\} \right] ^{3/2}}$. Importante frisar que esta possibilidade existe em virtude da incompressibilidade do fluxo, que nos permitiu escrever (\ref{eq2}).

\subsection{Transformada de Fourier da RKH}

Olhemos para a RKH, desta vez no espa\c co de Fourier, para entender por interm\'edio de quais mecanismos a turbul\^encia pode ser gerada, mantida e amortecida. A (\ref{RKH}) para a dire\c c\~ao paralela aos pontos em
quest\~ao \'e: 
\begin{equation}
\frac{\partial }{\partial t}\left( \left\| u\right\| ^{2}f\right) +2\left\|u\right\| ^{2/3}\left( \frac{\partial }{\partial r}h+4\frac{h}{r}\right)=2\nu \left\| u\right\| ^{2}\left( \frac{\partial ^{2}}{\partial r^{2}}f+\frac{4}{r}\frac{\partial }{\partial r}f\right),  \label{eq22}
\end{equation}
lembrando que: $f \propto (u_{1}u_{1}^{^{\prime}})$ e $h\propto (u_{2}^{2}u_{1}^{^{\prime}})$. Substituindo (\ref{eq5}) vem:
\label{eq26}
\begin{eqnarray}
\frac{\pi}{2}\frac{\partial }{\partial t}F+4k^{2}H_{1}-8 \int\limits_{k_{0}}^{k}sH_{1}(s)ds =- 2\nu k^{2}F+4\int\limits_{k_{0}}^{k}sF(s)ds  \label{eq26a} \\
H_{1}(k)=\frac{\left\| u\right\| ^{2/3}}{k\pi }\int\limits_{-\infty}^{\infty }h(r)e^{ikr}dr  .\label{eq26b}
\end{eqnarray}
Por (\ref{eq10}) obtemos uma equa\c c\~ao para o espectro da energia, aplicando o operador ${\cal D}_{3}$ em (\ref{eq26a}). 
\label{eq27}
\begin{eqnarray}
\frac{\partial }{\partial t}E &=&-W-2\nu k^{2}E  \label{eq27a} \\
W &=&4k^{2}{\cal D}_{3}\left[ H_{1}\left( k\right) \right].  \label{eq27b}
\end{eqnarray}

A equa\c c\~ao (\ref{eq27a}) nos diz que a perda de energia dos turbilh\~oes de escala $1/k$ \'e causado pela dissipa\c c\~ao direta $2\nu k^{2}E$ e pela transfer\^encia de uma escala para outra $W$. Como $W$ \'e essencialmente a transformada de Fourier da correla\c c\~ao tripla, esta \'e a respons\'avel pelo transporte de energia de uma escala para outra, sendo a correla\c c\~ao tripla heran\c ca do termo n\~ao linear da equa\c c\~ao. Este fato corrobora o modelo fenomenol\'ogico de Prandtl, onde o mecanismo respons\'avel pelo transporte de energia \'e a colis\~ao dos turbilh\~oes. 

Nos modelos fenomenol\'ogicos \cite{Chandrasekhar,Chandrasekhar1,Heisenberg,Hinze,Stanisic} a fun\c c\~ao $W$ \'e modelada com o intuito de ajustar os resultados experimentais. Mas, como estes modelos assumem independ\^encia estat\'\i stica entre os turbilh\~oes da regi\~ao energ\'etica e da regi\~ao dissipativa, o que \'e uma aproxima\c c\~ao grosseira; e n\~ao levam em considera\c c\~ao que os turbilh\~oes s\~ao bem localizados tanto espacialmente como temporalmente; n\~ao s\~ao capazes de prover um entendimento da din\^amica da turbul\^encia, muito embora, d\^eem resultados satisfat\'orios na regi\~ao inercial.   

\section{Leis Experimentais e Escalas em Turbul\^encia}

O objetivo desta sec\c c\~ao \'e identificar escalas de movimento nos fluidos e comentar brevemente as leis experimentais da turbul\^encia.

Consideremos um fluxo com uma dimens\~ao caracter\'\i stica $L$, velocidade $U$ e viscosidade cinem\'atica $\nu$. Geralmente tomamos $U$ como a raiz quadrada da m\'edia espacial e amostral e $L$ como sendo $1/k_{0}$, onde $k_{0}$ \'e ponto de m\'aximo global da transformada de Fourier da covari\^ancia da velocidade a dois pontos. Neste caso, $L$ \'e chamado de \emph{escala energ\'etica}\footnote{Esta \'e uma tradu\c c\~ao totalmente livre do termo: \emph{energy-containing scale} corrente na literatura em l\'\i ngua inglesa.}, pois os turbilh\~oes que contribuem com uma parcela maior, geralmente, est\~ao nessa vizinhan\c ca. Chamamos de \emph{tempo caracter\'\i stico de advec\c c\~ao}\footnote{H\'a aqui um abuso de linguagem pois o termo {\it advec\c c\~ao} refere-se ao movimento de um escalar passivo sobre fluido, caso este n\~ao considerado na defini\c c\~ao.} a 
\begin{equation}
t_a=\frac{L}{U},
\end{equation}
pois este d\'a a ordem de grandeza do tempo gasto por um turbilh\~ao para percorrer uma dist\^ancia $L$ no fluxo. Definimos tamb\'em \emph{tempo de difus\~ao } 
\begin{equation}
t_{d}=\frac{L^{2}}{\nu },
\end{equation}
que nos d\'a a escala de tempo dos processos de difus\~ao viscosa no fluido.

A exist\^encia destes tempos caracter\'\i sticos nos sugere a interpreta\c c\~ao do N\'umero de Reynolds como uma raz\~ao entre eles. 
\begin{equation}
Re=\frac{t_{d}}{t_a}.
\end{equation}

Como  $Re$ \'e o par\^ametro que controla a turbul\^encia, fica patente que a concorr\^encia entre a difus\~ao, advec\c c\~ao e for\c ca \'e o que gera a turbul\^encia, pois a difus\~ao tende a diminuir o valor dos expoentes de Liapunov do fluxo, enquanto que o termo advectivo acopla os modos normais e a for\c ca compensa a perda, fazendo com que o sistema n\~ao tenda para um ponto fixo. O que qualitativamente concorda com a an\'alise feita para EDO com \'orbitas homocl\'\i nicas\cite{Palis}, onde a intersec\c c\~ao entre as variedades inst\'aveis e est\'aveis, em uma configura\c c\~ao tipo ferradura de Smale, \'e o mecanismo gerador de caos.

De um ponto de vista mais f\'\i sico, esperar\'\i amos que os fen\^omenos difusivos fossem de pequena import\^ancia para a an\'alise din\^amica da turbul\^encia. Todavia, o termo viscoso \'e respons\'avel pela dissipa\c c\~ao de energia cin\'etica do fluido em forma de calor, fen\^omeno este que \'e predominante nos turbilh\~oes de menor escala. Mas, como veremos mais adiante neste cap\'\i tulo, a concorr\^encia entre estes dois fen\^omenos (advec\c c\~ao e difus\~ao) gera tr\^es escalas de movimento, a saber : energ\'etica, inercial e dissipativa, sendo os fen\^omenos mais interessantes, turbul\^encia e intermit\^encia, est\~ao ligados a escala inercial, ou seja, a escala de transi\c c\~ao entre a energ\'etica e a dissipativa. Logo, o entendimento da turbul\^encia \'e a explica\c c\~ao da exist\^encia destas escalas.

\subsection{Escalas em Turbul\^encia}

\'E uma constata\c c\~ao visual que em um fluxo podemos identificar estruturas de diversos tamanhos, por exemplo, nos fen\^omenos atmosf\'ericos somos capazes de distinguir tornados, pequenos redemoinhos, deslocamentos de centros de baixa press\~ao. Isto \'e poss\'\i vel porque por\c c\~oes do fluxo compartilham propriedasdes din\^amicas iguais ou muito parecidas e, de alguma forma, diferente das propriedades da vizinhan\c ca(fen\^omeno este ligado ao aumento abrupto do gradiente do campo de velocidade). Uma primeira aproxima\c c\~ao \'e definir \emph{turbilh\~ao }ou \emph{redemoinho} como \ qualquer regi\~ao do fluxo, cuja velocidade n\~ao varia apreciavelmente. Mas agora defrontamo-nos com o problema de quantificar este {\it apreciavelmente}. Podemos utilizar a transformada de Fourier para esta an\'alise.

Se $u(x,t)$ \'e um campo de velocidade e $\widetilde{u}(k,t)$ sua transformada de Fourier definimos como \emph{Turbilh\~ao de escala menor que} $l=1/k$ a fun\c c\~ao 
\begin{equation}
u^{\leq }(x,t,k)=\int\limits_{_{\left| s\right| \leq k}}\widetilde{u}(s,t)e^{is.x}d^{3}s.
\end{equation}
Definimos, tamb\'em, \emph{Turbilh\~ao de escala }$l=1/k$ a fun\c c\~ao 
\begin{equation}
u(x,t,k)=\int\limits_{_{\left| s\right| =k}}\widetilde{u}(s,t)e^{is.x}ds^{3}.
\end{equation}
Definimos \emph{escala longitudinal integral }ou simplesmente\emph{\ escala integral} por 
\begin{equation}
L_{i}(t)=\frac{\int_{0}^{\infty }R_{11}(r,0,0,t)dr}{R_{11}(0,0,0,t)}=\int_{0}^{\infty }f(r)dr.
\end{equation}
\'E f\'acil ver, pela defini\c c\~ao e pelo comportamento de $f$, que o valor da escala integral \'e determinado principalmente pela regi\~ao energ\'etica do espectro.

Definimos como \emph{regi\~ao dissipativa}, a regi\~ao do espectro dos pequenos turbilh\~oes, ou seja, dos turbilh\~oes que s\~ao respons\'aveis pela dissipa\c c\~ao da energia do fluxo em forma de calor. Dentro desta vis\~ao, podemos, ainda, diferenciar duas regi\~oes: primeiro uma grande regi\~ao, chamada de \emph{regi\~ao estacion\'aria}, definida como o conjunto de todos os turbilh\~oes, cuja energia $E(k,t)$ tem depend\^encia temporal fraca ou nula. Esta regi\~ao, geralmente, abarca a regi\~ao dissipativa e a chamada \emph{regi\~ao inercial}, que \'e a parte da regi\~ao estacion\'aria na qual a dissipa\c c\~ao \'e pequena, ou seja, \'e dominada pelas for\c cas inerciais(termo n\~ao-linear da NS). Esta divis\~ao tem origem em algumas leis experimentais e s\~ao amplamente utilizadas em modelos fenomenol\'ogicos ou semi-fenomenol\'ogicos\cite{Stanisic}. 

\subsection{Leis Experimentais}

\begin{law}[Lei dos Dois-Ter\c cos]

Em um fluxo turbulento, com n\'umero de Reynolds muito alto, a m\'edia amostral do quadrado do incremento da velocidade entre dois pontos
\begin{equation}
E\left\{ \left[ u_{i}\left( x+r\right) -u_{i}\left( x\right) \right]
^{2}\right\}
\end{equation}
 \'e proporcional a $r^{2/3}.$
\end{law}

Observemos que se um processo estacion\'ario ou de incrementos estacion\'arios \'e tal que a transformada de Fourier da m\'edia
espectral do quadrado do incremento \'e uma pot\^encia $k^{-n},1<n<3$, ent\~ao, a m\'edia do quadrado do incremento \'e proporcional a $r^{n-1} $. Ou seja, se $F_{ii}$ \'e a transformada de Fourier da correla\c c\~ao do processo $u$, vem: 
\begin{equation}
E\left\{ u_{i}(x+r)u_{i}(x)\right\} =\int e^{ikr}F_{ii}(k)dk.  \label{le1}
\end{equation}
Ent\~ao, 
\begin{equation}
E\left\{ \left[ u_{i}(x+r)-u_{i}(x)\right] ^{2}\right\} =2\int \left(1-e^{ikr}\right) F_{ii}(k)dk.  \label{le2}
\end{equation}
Fazendo $F_{ii}=C\left| k\right| ^{-n},\: C>0$ observamos que (\ref{le1}) diverge para qualquer valor de $n$, portanto processos estoc\'asticos de segunda ordem n\~ao podem ter espectro em lei de pot\^encia. N\~ao obstante, considerando somente processos de incrementos estacion\'arios, nos quais vale (\ref{le2}), observamos que para $1<n<3$ n\~ao h\'a diverg\^encia.
\begin{eqnarray}
E\left\{ \left[ u_{i}(x+r)-u_{i}(x)\right] ^{2}\right\} =\widetilde{C}r^{n-1}
\label{le3a} \\
\widetilde{C}=2C\int \left( 1-e^{ix}\right) \left| x\right| ^{-n}dx.
\label{le3b}
\end{eqnarray}
Portanto, nos casos em que vale a Lei dos Dois-ter\c cos no espa\c co f\'\i sico, vale tamb\'em no espa\c co das freq\"u\^encias a Lei dos Cinco-Ter\c cos, que vem a ser a mesma coisa.

A regi\~ao de $r$ que respeita a Lei dos Dois-Ter\c cos\footnote{Em \cite{Batchelor, Frisch} h\'a uma s\'erie de exemplos experimentais e, particularmante em \cite{Frisch}, uma simula\c c\~ao.} \'e denominada de \emph{regi\~ao inercial}. Encontramos esta regi\~ao no centro do espectro, ocupando, geralmente, tr\^es ordens de grandeza, sendo seguida, \`a direita(n\'umero de onda maior), pela regi\~ao de dissipa\c c\~ao, e \`a esquerda(n\'umero de onda menor), pela regi\~ao energ\'etica.

\begin{law}[Lei da Dissipa\c c\~ao finita de Energia]

Se, em um experimento sobre um fluxo turbulento, todos os par\^ametros de controle s\~ao mantidos constantes, exceto a viscosidade, que poder\'a ser diminu\'\i da tanto quanto poss\'\i vel, a dissipa\c c\~ao de energia por unidade de massa tem um comportamento consistente com a exist\^encia de um limite finito e positivo.
\end{law}

Diferentes experimentos e simula\c c\~oes foram feitos para verifica\c c\~ao desta lei\cite{Frisch,Landahl}, sendo que a dispers\~ao estat\'\i stica varia entre 20-40\%, dependendo da geometria e do n\'umero de Reynolds em que foram realizados.

Portando, devemos encarar esta lei com cautela, j\'a que n\~ao existe nenhum resultado conhecido que obrigue a sua exist\^encia. Todavia, podemos escrever a dissipa\c c\~ao do fluxo como 
\begin{equation}
\left\langle \varepsilon \right\rangle =2\nu \int_{0}^{\infty}k^{2}E(k,t)dk.
\end{equation}

Como a regi\~ao dissipativa do espectro \'e a principal respons\'avel pelo valor de $\left\langle \varepsilon \right\rangle $, a forma detalhada de $E(k,t)$ na regi\~ao dos grandes turbilh\~oes n\~ao influencia significativamente a integral acima. Isto implica que a transfer\^encia de energia ocorre predominantemente entre modos vizinhos, como nos mostra o modelo {\it DIA} de Kraichnan\cite{Stanisic}, e  tamb\'em que a regi\~ao dissipativa \'e estatisticamente independente da regi\~ao energ\'etica. Portanto, o mecanismo de transf\^erencia de energia \'e tal que os pequenos turbilh\~oes perdem toda a informa\c c\~ao sobre as condi\c c\~oes iniciais, gerando uma regi\~ao cujas propriedades s\~ao independentes das condi\c c\~oes iniciais e, al\'em do mais estacion\'arias. Este tipo de argumenta\c c\~ao nos leva diretamente ao modelo do Kolmogorov.

\section{As Hip\'oteses de Kolmogorov e suas implica\c c\~oes}

Na teoria de Kolmogorov consideramos um campo de velocidades incompress\'\i vel, tal que seus incrementos sejam homog\^eneos, isotr\'opicos, e tenham segunda derivada cont\'\i nua. Esta defini\c c\~ao tem um aspecto generalizador, pois n\~ao exige que o campo de velocidades tenha sua sec\c c\~ao espacial estacion\'aria. O incremento do campo, contando tamb\'em sua depend\^encia temporal, deve ter incrementos estacion\'arios. Por outro lado, a inser\c c\~ao da parte temporal do campo na defini\c c\~ao n\~ao permite que este varie livremente no tempo.

\subsection{Descri\c c\~ao via Incrementos}

Seja a transforma\c c\~ao de coordenadas 
\begin{equation}
y=x^{\prime}-x-u\left( x,t\right) \left[ t^{\prime}-t\right]  \label{rgt}
\end{equation}
e o incremento do campo das velocidades 
\begin{equation}
w_{i}(y)=u_{i}(x+y)-u_{i}(x)  \label{indekolm}
\end{equation}
pela homogeneidade, vem: 
\begin{equation}
E\left\{ w_{i}(y)\right\} =0.  \label{zeromodeB}
\end{equation}
Definimos um tensor de rank 2, sim\'etrico que nos d\^e a parte sim\'etrica da correla\c c\~ao do incremento por: 
\begin{equation}
B_{ij}(y,y^{\prime})=\frac{1}{2}E\left\{w_{i}(y)w_{j}(y^{\prime})+w_{i}(y^{\prime})w_{j}(y)\right\},  \label{eq11}
\end{equation}
que pode ser reescrita como: 
\begin{equation}
B_{ij}(y,y^{\prime})=\frac{1}{2}\left\{B_{ij}(y,y^{\prime})+B_{ij}(y^{\prime},y^{\prime})-B_{ij}(y-y^{\prime},y-y^{\prime})\right\}.
\label{eq12}
\end{equation}
Pelos mesmos argumentos utilizados para a forma da covari\^ancia, no caso do campo de velocidades ser homog\^eneo e isotr\'opico, vem: 
\label{eq13}
\begin{eqnarray}
E\left\{ (w_{i}w_{j})(y)\right\} &=&\frac{y_{i}y_{j}}{y^{2}}\left[B_{dd}(y)-B_{nn}(y)\right] +\delta _{ij}B_{nn}(y)  \label{eq13a} \\
B_{kk}(y) &=&E\left\{ \left[ u_{k}(y+x)-u_{k}(x)\right] ^{2}\right\} , \: k=n,\: d.
\label{eq13b}
\end{eqnarray}
onde o  \'\i ndice \textbf{d} denota a dire\c c\~ao radial e o \'\i ndice \textbf{n} a dire\c c\~ao normal entre os pontos considerados. Pela homogeneidade e isotropia podemos reescrever, sem perda de generalidade: 
\begin{equation}
B_{kk}(y)=E\left\{ \omega _{k}(y,0,0)^{2}\right\}.  \label{eq14}
\end{equation}
Podemos tamb\'em escrever a correla\c c\~ao tripla
\label{eq16}
\begin{eqnarray}
B_{ijk}(a,b,c) &=&E\left\{ \omega _{i}(a)\omega _{j}(b)\omega _{k}(c)\right\}
\label{eq16a} \\
B_{ddd}(y) &=&E\left\{ \left[ u_{d}(x+y)-u_{d}(x)\right] ^{3}\right\}.
\label{eq16b}
\end{eqnarray}
\'E f\'acil ver, pela constru\c c\~ao dos $B^{\prime} s$, que suas rela\c c\~oes com as fun\c c\~oes de K\'arm\'an-Howarth s\~ao dadas pelas f\'ormulas abaixo: 
\label{eq17}
\begin{eqnarray}
B_{dd}(r) &=&2\left[ 1-f(r)\right] E\left\{ u_{d}^{2}\right\}  \label{eq17a}
\\
B_{nn}(r) &=&2\left[ 1-g(r)\right] E\left\{ u_{n}^{2}\right\}  \label{eq17b}
\\
B_{ddd}(r) &=&6k(r)E\left\{ u_{d}^{2}\right\} ^{3/2}.  \label{eq17c}
\end{eqnarray}

Se $u(x,t)$ descreve um fluido, ent\~ao existem suas derivadas de primeira e segunda ordem, pois a equa\c c\~ao de Navier-Stokes \'e de segunda ordem. Por conseguinte, existem as derivadas at\'e terceira ordem da covari\^ancia, donde ressaltam as seguintes propriedades:
\label{eq15}
\begin{eqnarray}
B_{kk}(0) &=&0  \label{eq15a} \\
\partial _{r}B_{kk}(0) &=&0  \label{eq15b} \\
\partial _{r}^{2}B_{dd}(0) &=&2E\left\{ \left[ \partial _{1}\omega _{1}(y,0,0)\right] ^{2}\right\} =2a^{2}  \label{eq15c} \\
\partial _{r}^{2}B_{nn}(0) &=&2E\left\{ \left[ \partial _{1}\omega _{2}(y,0,0)\right] ^{2}\right\} =2b^{2},  \label{eq15d}
\end{eqnarray}
mas, pela incompressibilidade, vem que $b^{2}=2a^{2}$.

\subsection{As Hip\'oteses de Similaridade de Kolmogorov}

\begin{hyp}[Primeira Hip\'otese de Similaridade]
Para uma turbul\^encia homo-g\^enea e isotr\'opica, as distribui\c c\~oes finitas $F_{n}$ dos incrementos da velocidade s\~ao unicamente determinadas pelas quantidades: $\nu $ viscosidade e por$\left\langle \varepsilon \right\rangle $ dissipa\c c\~ao m\'edia por unidade de massa.
\end{hyp}

Se a fam\'\i lia $F_{n}$ depende somente de $\nu $ e\emph{\ }$\left\langle \varepsilon \right\rangle $, ent\~ao as correla\c c\~oes ir\~ao depender somente delas tamb\'em. Todavia, as correla\c c\~oes dependem da dist\^ancia $r$. Portanto, deve existir\textbf{\ }$\eta $, chamado de \emph{comprimento de Kolmogorov}, com dimens\~ao de dist\^ancia, formado por $\nu $ e $\left\langle \varepsilon \right\rangle $, tal que: $B_{kk}(r)=\beta _{kk}(r/\eta )\overline{u}^{2},k=d,n$, onde $\beta _{kk}$ \'e uma fun\c c\~ao universal e $\overline{u}$ \'e uma constante com dimens\~ao de velocidade que s\'o pode depender de $\nu $ e\emph{\ }$\left\langle \varepsilon \right\rangle $. Fazendo uma an\'alise dimensioal simples, chegamos a: 
\label{eq18}
\begin{eqnarray}
\eta &=&\left[ \nu ^{3}/\left\langle \varepsilon \right\rangle \right]
^{1/4}  \label{eq18a} \\
\overline{u} &=&\left[ \nu \left\langle \varepsilon \right\rangle %
\right] ^{1/4}.  \label{eq18b}
\end{eqnarray}

A turbul\^encia \'e isotr\'opica, logo $B_{kk}(r)=B_{kk}(-r)$, e por (\ref{eq15}) a fun\c c\~ao \'e par. Retendo o primeiro termo da
expans\~ao em s\'erie de pot\^encia e utilizando novamente (\ref{eq17}%
):
\label{eq19}
\begin{eqnarray}
\beta _{kk}(r/\eta ) &\cong &\left( \frac{r}{\eta }\right) ^{2},k=d,n
\label{eq19a} \\
B_{dd}(r) &\cong &\frac{\left\langle \varepsilon \right\rangle }{\nu }%
r^{2}  \label{eq19b} \\
B_{dd}(r) &\cong &2\frac{\left\langle \varepsilon \right\rangle }{\nu }%
r^{2}.  \label{eq19c}
\end{eqnarray}

\begin{hyp}[Segunda Hip\'otese de Similaridade]

Para uma turbul\^encia homo-g\^enea e isotr\'opica, quando $\left| y\right| ,\left| y^{,}\right| ,\left| y^{,}-y\right| \gg \eta $, as ditribui\c c\~oes finitas do incremento da velocidade s\~ao unicamente determinadas por $\left\langle \varepsilon \right\rangle $.
\end{hyp}

Tomando $r$ suficientemente grande, as fun\c c\~oes universais $\beta
_{kk} $ $e$ $\beta _{ddd}$ devem ter uma forma tal, que as fun\c c\~oes de
correla\c c\~ao $B_{kk}$, $B_{ddd}$ n\~ao apresentem
depend\^encia em $\nu $. Portanto, pela Primeira Hip\'otese de
Similaridade e pela (\ref{eq18}), teremos: 
\
\begin{equation}
\beta _{kk}\cong C\left( \frac{r}{\eta }\right) ^{2/3}.  \label{eq20}
\end{equation}
Utilizando (\ref{eq17}) e fazendo um procedimento an\'alogo para a correla%
\c c\~ao tripla: 
\label{eq21}
\begin{eqnarray}
B_{dd} &\cong &C\left( \left\langle \varepsilon \right\rangle r\right) ^{2/3}
\label{eq21a} \\
B_{nn} &\cong &\frac{4}{3}C\left( \left\langle \varepsilon \right\rangle
r\right) ^{2/3}  \label{eq21b} \\
B_{ddd} &\cong &\widetilde{C}\left\langle \varepsilon \right\rangle r,
\label{eq21c}
\end{eqnarray}
onde $C,\widetilde{C}$ s\~ao constantes adimensionais a serem determinadas.

\subsection{As Hip\'oteses de Kolmogorov e a Rela\c c\~ao de K\'arm\'an-Howarth}

Substituindo (\ref{eq17}) em (\ref{eq22}), obtemos a \textbf{RKH} para as
correla\c c\~oes escalares de Kolmogorov. 
\begin{equation}
3\frac{\partial }{\partial t}\left[ 2\left\| u\right\| ^{2}-B_{dd}\right]
+\left( \frac{\partial }{\partial r}B_{ddd}+4\frac{B_{ddd}}{r}\right)
=6\nu \left( \frac{\partial ^{2}}{\partial r^{2}}B_{dd}+\frac{4}{r}%
\frac{\partial }{\partial r}B_{dd}\right),  \label{eq23}
\end{equation}
lembrando que a turbul\^encia \'e estacion\'aria e $\left\langle
\varepsilon \right\rangle =\frac{1}{2}\frac{\partial }{\partial t}\left\|
u\right\| ^{2}$, reescrevemos a equa\c c\~ao acima como: 
\begin{equation}
4\left\langle \varepsilon \right\rangle =\left[ \frac{\partial }{\partial r}+%
\frac{4}{r}\right] \left( 6\nu \frac{\partial }{\partial r}%
B_{dd}-B_{ddd}\right),  \label{eq24}
\end{equation}
integrando e usando as condi\c c\~oes iniciais vem: 
\begin{equation}
\frac{4}{5}\left\langle \varepsilon \right\rangle r=6\nu \frac{\partial 
}{\partial r}B_{dd}-B_{ddd}.  \label{eq25}
\end{equation}

Observemos que, se tomarmos $r$ da ordem do comprimento de Kolmogorov, esperamos que $B_{ddd}\approx r^{3}$ e por (\ref{eq25}) temos que $\frac{\partial }{\partial r}B_{dd}\rfloor _{r=0}\approx \frac{2\left\langle \varepsilon \right\rangle }{15\nu }r$, ou seja $B_{dd}\cong \frac{\left\langle \varepsilon \right\rangle }{15\nu }r^{2}$. Tomando $r$ muito maior que o comprimento de Kolmogorov, esperamos pelas Hip\'oteses de Similaridade, i.e. (\ref{eq21}), que $\frac{\partial }{\partial r}B_{dd}\ll B_{ddd}$, portanto $B_{ddd}=\frac{4}{5}\left\langle \varepsilon \right\rangle r$.

\section{Obje\c c\~oes \`a Universalidade de Kolmogorov}

Landau foi o primeiro que suspeitou da universalidade das constantes, baseado no seguinte argumento:

Suponhamos, a princ\'\i pio, que o valor instant\^aneo de $\left[ \omega_{d}(r)\right] ^{2}$ seja expresso como uma fun\c c\~ao universal da dissipa\c c\~ao $\varepsilon \left( t\right) $. Quando tomarmos a m\'edia desta express\~ao, \ a varia\c c\~ao de $\varepsilon $ sobre tempos da ordem do per\'\i odo dos turbilh\~oes grandes(tempo carcter\'\i stico de advec\c c\~ao) influencia o valor da m\'edia. Todavia, esta varia\c c\~ao \'e diferente para cada fluxo, pois a regi\~ao energ\'etica \'e permanente e caracterizada pelas condi\c c\~oes de contorno do problema. Logo, o resultado n\~ao pode ser universal.

Este argumento tamb\'em \'e utilizado por Kraichnan\cite{Kraichnan}, s\'o que de uma
maneira mais detalhada e quantitativa. Cuja linha mestra \'e:

A varia\c c\~ao temporal de $\varepsilon \left( t\right) $ \'e determinada pela convec\c c\~ao da escala dissipativa, pelas escalas maiores(pertencentes \`a regi\~ao inercial) e por suas distor\c c\~oes
internas, sendo que a pr\'opria convec\c c\~ao pode causar distor\c c%
\~oes. E por seu turno, as distor\c c\~oes internas e o per\'\i odo de
convec\c c\~ao s\~ao caracter\'\i sticas de cada fluxo, assim as
constantes para as correla\c c\~oes n\~ao t\^em um car\'ater universal.
Ent\~ao, se $\varepsilon $ independe dos detalhes do fluxo, a regi\~ao
inercial deve ser estatisticamente independente da escala de dissipa\c c%
\~ao. Ou seja, o mecanismo de transfer\^encia de energia dos
turbilh\~oes deve apagar detalhes das informa\c c\~oes \cite{Heisenberg} contidas nos
turbilh\~oes maiores e ter um car\'ater local.

Todavia, Kraichnan mostrou que, se considerarmos inicialmente a escala dissipativa estatisticamente independente da regi\~ao inercial e deixarmos que a \'unica forma de intera\c c\~ao entre elas seja o termo n\~ao-linear estas escalas passar\~ao a ser estatisticamente dependentes e a ter momentos depententes do tempo para instantes diferentes.

Para ilustrarmos os par\'agrafos anteriores, calculemos o exemplo de Kraichnan :

\begin{example}[Kraichnan]
Tomemos um campo de velocidade $u+v$ tal que: $v$ seja constante no espa\c co e no tempo; tenha uma distribui\c c\~ao gaussiana e isotr\'opica; $u$ seja vari\'avel no espa\c co, muito menor que $v$, e em $t=0$ tenha ditribui\c c\~ao gaussiana, homog\^enea, isotr\'opica e independente de $v$. Escrevendo a equa\c c\~ao de Navier-Stokes sem o termo viscoso, oomo o termo viscoso \'e respons\'avel pela dissipa\c c\~ao e o fen\^omeno em quest\~ao \'e a intera\c c\~ao, este somente atrapalharia. Assim, cada componente de Fourier da velocidade obedecer\'a \`a seguinte equa\c c\~ao: 
\begin{eqnarray}
\frac{\partial }{\partial t}u(k,t) &=&i(k\cdot v)u(k,t)  \nonumber \\
u(k,t) &=&u(k,0)e^{-i(k\cdot v)t}.  \label{eq28}
\end{eqnarray}
Calculando a correla\c c\~ao temporal 
\begin{eqnarray}
T(k;t,s)=\frac{E\left\{ u(k,t)u^{\ast }(k,s)\right\} }{\left[ E\left\{
u(k,s)u^{\ast }(k,s)\right\} E\left\{ u(k,t)u^{\ast }(k,t)\right\} \right]
^{1/2}} \nonumber \\
=E\left\{ e^{-iv\cdot k\left( t-s\right) }\right\} =e^{-\frac{1}{2}%
v_{0}^{2}k^{2}\left( t-s\right) ^{2}},  \nonumber
\end{eqnarray}
onde $v_{0}$ \'e a vari\^ancia de $v$. Mas, por hip\'otese, $u$ e $v$
s\~ao independentes no instante inicial, logo os valores simult\^aneos
dos dois campos s\~ao estatisticamente independentes em qualquer
instante posterior. Decorre de (\ref{eq28}) que 
\begin{equation}
E\left\{ v_{i}u_{j}(k,t)u_{m}^{\ast }(k,s)\right\} =-iv_{0}^{2}k_{i}\left(
t-s\right) e^{-\frac{1}{2}v_{0}^{2}k^{2}\left( t-s\right) ^{2}}E\left\{
u_{j}(k,0)u_{m}^{\ast }(k,0)\right\}
\end{equation}
\end{example}
Poder\'\i amos generalizar o resultado acima e mostrar que momentos mais gerais somente se anulam para tempos iguais, tal qual o exemplo acima.

\section{Observa\c c\~oes}

Muito embora a Lei dos Dois-Ter\c cos se aplique a uma grande variedade de
fluxos, como nos mostrou Kolmogorov, esta n\~ao parece estar completamente de
acordo com os resultados experimentais\cite{Frisch}, j\'a que estes parecem
indicar a exist\^encia de termos de corre\c c\~ao, ou seja, $B_{nn} \sim r^{2/3
+ \zeta}$. Mas ainda n\~ao temos dados, com precis\~ao suficiente, para saber se
$\zeta$ \'e causado pela inaplicabilidade das Hip\'otese de Kolmogorov ou pelas
t\'ecnicas intrusivas utilizadas na maior parte dos experimentos\cite{Landahl}.
Poder\'\i amos pensar em realizar simula\c c\~oes para termos comprova\c c\~oes, mas para este fim seria necess\'ario um tempo de m\'aquina impens\'avel.

Existem, hoje, evid\^encias da exist\^encia de estruturas coerentes\cite{Chorin1,Holmes} nas solu\c c\~oes da equa\c c\~ao de Navier-Stokes, como veremos mais adiante para o caso da equa\c c\~ao de Burgers, estas estruturas s\~ao respons\'aveis pelas deforma\c c\~oes da distribui\c c\~ao de probabilidade. Assim, $\zeta$ pode ser o resultado da exist\^encia destas estruturas na regi\~ao inercial. Contudo, a quantifica\c c\~ao desta influ\^encia n\~ao \'e trivial devido a intera\c c\~ao entre turbilh\~oes de diversos tamanhos. Mas as simula\c c\~oes v\^em avan\c cando muito nos \'ultimos anos, e no futuro pr\'oximo teremos uma resposta final\footnote{Come\c car\'a, no pr\'oximo ano, um esfor\c co conjunto entre o Observatoire de la C\^ote D'Azur(Prof. Frisch) e o Laborat\'orio de Los Alamos(Prof. Kraichnan) para executar uma simula\c c\~ao cuja malha ser\'a da ordem de $2048^3 \sim 10^{11}$ pontos.} para esta quest\~ao secular.

\chapter{Formula\c c\~ao Funcional}

A an\'alise da Turbul\^encia via funcionais, mais precisamente, por meio do funcional caracter\'\i stico da distribui\c c\~ao de probabilidade do campo de velocidades, come\c cou a ser aventado por Kolmogorov. Todavia, sua primeira formula\c c\~ao sistem\'atica deve-se a Hopf\cite{Hopf,Hopf1}, que introduziu uma equa\c c\~ao integro-diferencial para o funcional caracter\'\i stico. Mas esta abordagem \'e incompleta, pois somente os momentos espaciais, ou seja, momentos calculados no mesmo instante de tempo, s\~ao pass\'\i veis de serem calculados. Lewis e Kraichnan\cite{Lewis} generalizaram a abordagem de Hopf de tal forma que os momentos espa\c co-temporais podem ser calculados. Como o processo dedutivo das duas formula\c c\~{o}es \'e an\'alogo, utilizaremos como base o artigo de Lewis e Kraichnan. Cabe ressaltar aqui que, embora a equa\c c\~ao de Hopf n\~ao nos forne\c ca uma descri\c c\~ao completa do fen\^{o}meno estat\'\i stico, ela est\'a em melhor posi\c c\~ao para ser comparada com os resultados can\^{o}nicos do modelo de Kolmogorov j\'a que, quando tomamos o limite de tempos iguais para os momentos espa\c{c}o-temporais, estes tendem, for\c cosamente, para zero se a separa\c c\~ao espacial permanece finita.

\section{Funcional Caracter\'\i stico}

Definimos o valor m\'edio de um funcional $F$ sobre o espa\c co de solu\c{%
c}\~{o}es da equa\c c\~ao de Navier-Stokes para uma determinada condi\c c%
\~ao de contorno, como: 
\begin{equation}
\left\langle F\right\rangle =\int dP(u)F(u),
\end{equation}
onde $P\left( u\right) $ \'e a medida de probabilidade neste espa\c co\footnote{Observe que este funcional \'e essencialmente diferente do encontrado em Mec\^anica Qu\^antica, onde as integrais para as probabilidades de transi\c c\~ao podem ser em alguns casos discretizadas e depois tomado o limite do cont\'\i nuo e assim calculadas diretamente.}.

Agora seja $y=(y_{1},y_{2},y_{3})=y(x,t)$ um campo real arbitr\'ario, que
se anula no infinito espacial, em particular, podemos tomar $y$ como tendo
suporte compacto. Definimos um funcional de $y$, que \'e Funcional
Caracter\'\i stico de $P$, como sendo: 
\begin{equation}
F[y]=\left\langle \exp i\left\{ \int_{\Re ^{+}}dt\int_{\Re^{3}}d^{3}xy_{i}(x,t)u_{i}(x,t)\right\} \right\rangle =\left\langle
e^{i(y,u)}\right\rangle.  \label{eqt1}
\end{equation}
Calculando a n-\'esima derivada funcional parcial de $F$ para $y=0$ teremos a seguinte igualdade: 
\begin{equation}
i^{-n}\frac{\delta ^{n}F[y]}{\delta y_{\alpha _{1}}(x_{1},t_{1})...\delta
y_{\alpha _{n}}(x_{n},t_{n})}\mid _{y=0}=\left\langle u_{\alpha
_{1}}(x_{1},t_{1})...u_{\alpha _{n}}(x_{n},t_{n})\right\rangle .
\end{equation}
Por outro lado, caso desejemos obter a fun\c c\~ao caracter\'\i stica da
densidade de probabilidade a n-pontos basta tomarmos 
\begin{equation}
y_{i}(x,t)=\sum_{j=1}^{n}h_{i\left( j\right) }\delta _{i\alpha _{j}}\delta
(t-t_{j})\delta ^{3}(x-x_{j}),
\end{equation}
onde em $h_{i\left( j\right) }$, o primeiro \'\i ndice indica a componente do
campo e o segundo o ponto referido. Substituindo na defini\c c\~ao do
Funcional Caracter\'\i stico temos 
\begin{equation}
F[y]=\left\langle \exp i\left\{ \sum_{j=1}^{n}h_{\alpha _{j}\left( j\right)
}u_{\alpha _{j}}(x_{j},t_{j})\right\} \right\rangle =\widetilde{p}(h_{\left(
1\right) },...,h_{\left( n\right) }),
\end{equation}
logo, a densidade de probabilidade \'e 
\begin{eqnarray}
p\{u_{\alpha _{1}}(x_{1},t_{1}),...,u_{\alpha _{n}}(x_{n},t_{n})\}= \\
(2\pi )^{-n}\int dh_{\left( 1\right) }...dh_{\left( n\right) }\widetilde{p}%
(h_{\left( 1\right) },...,h_{\left( n\right) })\exp
\{-i\sum_{j=1}^{n}u_{\alpha _{j}}h_{\left( j\right) }\} .
\end{eqnarray}

\section{Equa\c {c}\~ao Para o Funcional Caracter\'{\i}stico}

Para obtermos uma equa\c {c}\~ao diferencial para $F$, basta observarmos
que 
\begin{equation}
\frac{\delta F[y]}{\delta y_\alpha (x_1,t_1)}=i\left\langle u_\alpha
(x_1,t_1)e^{i(y,u)}\right\rangle
\end{equation}

\begin{equation}
\frac{\delta ^{n}F[y]}{\delta y_{\alpha }(x_{1},t_{1})\delta y_{\beta
}(x_{2},t_{2})}=-\left\langle e^{i(y,u)}u_{\alpha }(x_{1},t_{1})u_{\beta
}(x_{2},t_{2})\right\rangle
\end{equation}
\begin{equation}
\frac{\partial }{\partial x_{\beta }}\frac{\partial }{\partial x_{\beta }}%
\frac{\delta F[y]}{\delta y_{\alpha }(x,t)}=i\left\langle \frac{\partial }{%
\partial x_{\beta }}\frac{\partial }{\partial x_{\beta }}u_{\alpha
}(x,t)e^{i(y,u)}\right\rangle
\end{equation}
\begin{equation}
\frac{\partial }{\partial x_{\beta }}\frac{\delta ^{n}F[y]}{\delta y_{\alpha
}(x,t)\delta y_{\beta }(x,t)}=-\left\langle e^{i(y,u)}u_{\beta }(x,t)\frac{%
\partial }{\partial x_{\beta }}u_{\alpha }(x,t)\right\rangle .
\end{equation}
Para obtermos a \'ultima equa\c c\~ao, utilizamos a condi\c c\~ao de
incompressibilidade. Assim, tamb\'em teremos 
\begin{equation}
\frac{\partial }{\partial t}\frac{\delta F[y]}{\delta y_{\alpha }(x,t)}%
=i\left\langle e^{i(y,u)}\frac{\partial }{\partial t}u_{\alpha
}(x,t)\right\rangle .
\end{equation}
Substituindo, no lado direito da equa\c c\~ao (4.12), a equa\c c\~ao de
Navier-Stokes com uma for\c ca determin\'\i stica e utilizando as rela\c c%
\~oes acima vem: 
\begin{eqnarray}
\frac{\partial }{\partial t}\frac{\delta F[y]}{\delta y_{\alpha }(x,t)}-i%
\frac{\partial }{\partial x_{\beta }}\frac{\delta ^{2}F[y]}{\delta y_{\alpha
}(x,t)\delta y_{\beta }(x,t)}=  \label{eqt2} \\
\nu \frac{\partial }{\partial x_{\beta }}\frac{\partial }{\partial x_{\beta }%
}\frac{\delta F[y]}{\delta y_{\alpha }(x,t)}+if_{\alpha }F[y]-\frac{\partial 
}{\partial x_{\alpha }}\Pi . \nonumber
\end{eqnarray}
onde $\Pi =i\left\langle p(x,t)e^{i(y,u)}\right\rangle $. Para eliminarmos o
campo de press\~ao, ao inv\'es de recorrermos \`a utiliza\c c\~ao
de operadores n\~ao locais, preferimos introduzir um campo de teste $%
l(x,t)=(l_{1},l_{2},l_{3})$, tal que este se anule suficientemente
r\'apido no infinito espacial e seja solenoidal. Fazendo uma integra\c c%
\~ao por partes vem : 
\begin{equation}
\int d^{3}xl_{i}\left( \frac{\partial }{\partial x_{i}}\Pi \right) =-\int
d^{3}x\Pi \left( \frac{\partial }{\partial x_{i}}l_{i}\right) =0 .
\end{equation}
Assim, integrando o campo $l$ conjuntamente com a equa\c c\~ao para $F$, obtemos 
\begin{eqnarray}
\int dtd^{3}xl_{\alpha }\left[ \frac{\partial }{\partial t}\frac{\delta F[y]%
}{\delta y_{\alpha }(x,t)}-i\frac{\partial }{\partial x_{\beta }}\frac{%
\delta ^{2}F[y]}{\delta y_{\alpha }(x,t)\delta y_{\beta }(x,t)}\right] =
\label{eqt3} \\
\int dtd^{3}xl_{\alpha }\left[ \nu \frac{\partial }{\partial x_{\beta }}%
\frac{\partial }{\partial x_{\beta }}\frac{\delta F[y]}{\delta y_{\alpha
}(x,t)}+if_{\alpha }F[y]\right] . \nonumber
\end{eqnarray}
Lembrando que esta equa\c c\~ao precisa ser satisfeita para qualquer campo de
teste solenoidal e pode ser tomada como substituta da equa\c c\~ao de
Navier-Stokes, com a condi\c c\~ao inicial reescrita como: 
\begin{equation}
F_{0}[y]=\left\langle \exp \left\{ i\int d^{3}xy_{i}(x,0)u_{i}(x,0)\right\}
\right\rangle .
\end{equation}
A condi\c c\~ao de incompressibilidade pode ser escrita como: 
\begin{equation}
\frac{\partial }{\partial x_{\alpha }}\frac{\delta F[y]}{\delta y_{\alpha
}(x,t)}=0 ,
\end{equation}
que \'e uma conseq\"u\^encia direta da defini\c c\~ao de $F$. Cabe
lembrar tamb\'em que $F$ tem que ser normalizado e a probabilidade deve ser n\~ao negativa, implicando as seguintes rela\c c\~oes 
\begin{equation}
F[0]=1\qquad F^{\ast }[y]=F[-y]\qquad F[y]\leq 1
\end{equation}

\begin{equation}
\left\langle \left| \sum_{j=1}^{n}z_{j}e^{i(y_{j},u)}\right|
^{2}\right\rangle =\sum_{i=1}^{n}\sum_{j=1}^{n}z_{i}^{\ast
}z_{j}F[y_{j}-y_{i}]\geq 0 .
\end{equation}
Esta \'ultima propriedade \'e chamada de defini\c c\~ao positiva do
funcional. Podemos estabelecer, tamb\'em, uma equa\c c\~ao para o
funcional caracter\'\i stico utilizando a transformada de Fourier dos campos
determin\'\i sticos, ou seja, efetuando as seguintes transforma\c c\~oes: 
\begin{eqnarray}
z_{i}\left( k,t\right) =\int d^{3}xy_{i}\left( x,t\right)
e^{-ik_{j}x_{j}}\quad \\
\quad y_{i}\left( x,t\right) =\left( 2\pi \right) ^{-3}\int
d^{3}kz_{i}\left( k,t\right) e^{ik_{j}x_{j}}
\end{eqnarray}
\begin{eqnarray}
m_{i}\left( k,t\right) =\int d^{3}xl_{i}\left( x,t\right)
e^{-ik_{j}x_{j}}\quad \\
\quad l_{i}\left( x,t\right) =\left( 2\pi \right) ^{-3}\int
d^{3}km_{i}\left( k,t\right) e^{ik_{j}x_{j}}
\end{eqnarray}
\begin{eqnarray}
g_{i}\left( k,t\right) =\left( 2\pi \right) ^{-3}\int d^{3}xf_{i}\left(
x,t\right) e^{ik_{j}x_{j}}\quad \\
\quad f_{i}\left( x,t\right) =\int d^{3}kg_{i}\left( k,t\right)
e^{-ik_{j}x_{j}} .
\end{eqnarray}
Como $y$ e $f$ s\~ao reais, segue da transforma\c c\~ao de Fourier que 
\begin{equation}
z_{i}^{\ast }\left( k,t\right) =z_{i}\left( -k,t\right) \quad e\quad
g_{i}^{\ast }\left( k,t\right) =g_{i}\left( -k,t\right),
\end{equation}
donde vem, tamb\'em, que a incompressibilidade de $l_{i}$ fica representada
por 
\begin{equation}
k_{i}m_{i}=0 .
\end{equation}
Esta transforma\c c\~ao no espa\c co das fun\c c\~oes implica que 
\begin{equation}
\frac{\delta F[y]}{\delta y_{\alpha }(x,t)}=\int d^{3}k\frac{\delta F[z]}{%
\delta z_{\beta }(k,t)}\frac{\delta z_{\beta }(k,t)}{\delta y_{\alpha }(x,t)}%
=\int d^{3}k\frac{\delta F[z]}{\delta z_{\beta }(k,t)}e^{-ik_{j}x_{j}}.
\end{equation}
Portanto, para a n-\'esima derivada teremos 
\begin{eqnarray}
\frac{\delta ^{n}F[y]}{\delta y_{\alpha _{1}}(x_{1},t_{1})...\delta
y_{\alpha _{n}}(x_{n},t_{n})}= \\
\int d^{3}k_{1}...d^{3}k_{n}\frac{\delta ^{n}F[z]}{\delta z_{\alpha
_{1}}(z_{1},t_{1})...\delta z_{\alpha _{n}}(z_{n},t_{n})}\exp \left\{ -i%
\left[ x_{1_{j}}k_{1_{j}}+...+x_{n_{j}}k_{n_{j}}\right] \right\} .
\end{eqnarray}
Substituindo na (\ref{eqt3}) vem\footnote{Para obtermos esta equa\c c\~ao, utilizamos o operador de proje\c c\~ao $\left[ \delta _{ij}-\frac{k_{i}k_{j}}{k^{2}}\right] $ para ressaltarmos que o campo de velocidades responde somente \`a parte solenoidal da for\c ca.}:
\begin{eqnarray}
\int dtd^{3}km_{i}\left( k,t\right) \left\{ \left[ \frac{\partial }{\partial
t}+vk^{2}\right] \frac{\delta F[z]}{\delta z_{i}(k,t)}-iF\left[ z\right] 
\left[ \delta _{ij}-\frac{k_{i}k_{j}}{k^{2}}\right] g_{j}\left( k,t\right)
\right\} =  \label{eqt4} \\
\int dtd^{3}k_{1}d^{3}k_{2}m_{i}\left( k_{1}+k_{2},t\right) \left[
k_{1_{j}}+k_{2_{j}}\right] \frac{\delta ^{2}F[z]}{\delta
z_{i}(k_{1},t)\delta z_{j}(k_{2},t)} . \nonumber
\end{eqnarray}
A equa\c c\~ao acima nos sugere que podemos generalizar nossa abordagem um pouco mais tratando a for\c ca como sendo um processo estoc\'astico.

Suponhamos que no instante inicial a velocidade e a for\c ca, sejam estatisticamente independentes e dados. Logo, ao inv\'es do funcional $F[y]$, consideraremos um novo funcional que descreva a probabilidade conjunta da for\c ca e da velocidade. Assim, definimos 
\begin{equation}
F\left[ y,q\right] \equiv \int dP\left( u,f\right) \exp i\left\{ \int
dtd^{3}x\left( y_{i}u_{i}+q_{i}f_{i}\right) \right\}.
\end{equation}
Agora, se tomamos 
\begin{eqnarray}
y_{i}\left( x,t\right) &=&\delta \left( t-0\right) r_{i}\left( x\right) \\
q_{i}\left( x,t\right) &=&\delta \left( t-0\right) s_{i}\left( x\right) .
\end{eqnarray}
substituindo na defini\c c\~ao de $F$ e utilizando a condi\c c\~ao de independ\^encia, vem: 
\begin{eqnarray}
F\left[ y,q\right] =\left\langle \exp i\int d^{3}x\left[ r_{i}\left(x\right) u_{i}\left( x,0\right) +s_{i}\left( x\right) f_{i}\left( x,0\right)\right] \right\rangle = \nonumber \\
\left[ \int dP\left( u\right) \exp i\int d^{3}xr_{i}\left( x\right) u\left(x,0\right) \right] \times \left[ \int dP\left( f\right) \exp i\int d^{3}xs_{i}\left( x\right) f_{i}\left( x,0\right) \right] \nonumber \\
=\chi \left[ r\right] \Psi \left[ s\right] ,  \nonumber
\end{eqnarray}
onde $\chi \left[ r\right] $ e$\Psi \left[ s\right] $ s\~ao respectivamente os funcionais caracter\'\i sticos para a velocidade
e para a for\c ca no instante inicial.

Fica f\'acil ver que a modifica\c c\~ao que incide sobre a (\ref{eq3}) \'e simplesmente 
\begin{equation}
if_{\alpha }F[y,q]\rightarrow \frac{\delta F\left[ y,q\right] }{\delta
q_{\alpha }\left( x,t\right) },
\end{equation}
levando-nos a reescrev\^e-la como: 
\begin{eqnarray}
\int dtd^{3}xl_{\alpha }\left[ \frac{\partial }{\partial t}\frac{\delta
F[y,q]}{\delta y_{\alpha }(x,t)}-i\frac{\partial }{\partial x_{\beta }}\frac{%
\delta ^{2}F[y,q]}{\delta y_{\alpha }(x,t)\delta y_{\beta }(x,t)}\right] =
\label{eqt5} \\
\int dtd^{3}xl_{\alpha }\left[ \nu \frac{\partial }{\partial x_{\beta }}%
\frac{\partial }{\partial x_{\beta }}\frac{\delta F[y,q]}{\delta y_{\alpha
}(x,t)}+\frac{\delta F\left[ y,q\right] }{\delta q_{\alpha }\left(
x,t\right) }\right],  \nonumber
\end{eqnarray}
sendo que a condi\c c\~ao de incompressibilidade n\~ao muda formalmente.

\section{Solu\c {c}\~ao para N\'umero de Reynolds Nulo}

Observar o comportamento da equa\c c\~ao para o n\'umero de Reynolds
nulo significa que estamos olhando para uma equa\c c\~ao onde os efeitos
de transporte de energia entre os turbilh\~oes s\~ao desprez\'\i veis,
ou seja, toda a energia recebida da for\c ca externa \'e dissipada
atrav\'es do termo viscoso. Em outras palavras, esperamos que esta solu\c c%
\~ao descreva o comportamento da regi\~ao dissipativa do espectro.

Comecemos pelo caso mais simples em que a for\c ca externa \'e
determin\'\i stica. Obviamente o termo que deve ser eliminado \'e o termo
na segunda derivada funcional de $F$, que na representa\c c\~ao espectral
\'e o termo essencialmente n\~ao local. Portanto, a equa\c c\~ao toma a seguinte forma: 
\begin{equation}
\int dtd^{3}km_{i}\left( k,t\right) \left\{ \left[ \frac{\partial }{\partial
t}+vk^{2}\right] \frac{\delta F[z]}{\delta z_{i}(k,t)}-iF\left[ z\right] %
\left[ \delta _{ij}-\frac{k_{i}k_{j}}{k^{2}}\right] g_{j}\left( k,t\right)
\right\} =0.
\end{equation}

Como a integral acima deve ser nula para todo campo $m$, necessariamente a
express\~ao entre par\^enteses deve ser nula, ou seja, 
\begin{equation}
\left[ \frac{\partial }{\partial t}+vk^{2}\right] \frac{\delta F[z]}{\delta
z_{i}(k,t)}-iF\left[ z\right] \left[ \delta _{ij}-\frac{k_{i}k_{j}}{k^{2}}%
\right] g_{j}\left( k,t\right) =0 .  \label{eqrn0}
\end{equation}
Suponhamos que o termo de for\c ca seja nulo. Logo, a equa\c c\~ao fica
reduzida a 
\begin{equation}
\left[ \frac{\partial }{\partial t}+vk^{2}\right] \frac{\delta F[z]}{\delta
z_{i}(k,t)}=0 .
\end{equation}
Como a equa\c c\~ao acima n\~ao tem derivadas funcionais em $z$, devemos
ter: 
\begin{equation}
\frac{\delta F[z]}{\delta z_{i}(k,t)}=R_{i}\left[ z\right] e^{-vk^{2}t}.
\end{equation}
Como neste ponto o funcional $R_{i}\left[ z\right] $ \'e arbitr\'ario,
tomemos $R_{i}\left[ z\right] =z_{i}(k,t)$. Mas, por ser $F$ um funcional
caracter\'\i stico de fluxo incompress\'\i vel, devemos ter: 
\begin{equation}
k_{i}z_{i}=0 .
\end{equation}
Agora, como queremos tomar $z$ como um campo arbitr\'ario, basta inserirmos
o operador de proje\c c\~ao solenoidal diante do campo $z$ na nossa solu%
\c c\~ao. E para descrevermos uma condi\c c\~ao inicial
arbitr\'aria, j\'a que $R_{i}\left[ z\right] $ deve ser essencialmente a
derivada funcional do funcional que \'e solu\c c\~ao, escrevemos a solu%
\c c\~ao como: 
\begin{equation}
F[z]=A\left[ \int dte^{-vk^{2}t}\left[ \delta _{ij}-\frac{k_{i}k_{j}}{k^{2}}%
\right] z_{j}(k,t)\right] .
\end{equation}
Para resolver (\ref{eqrn0}), podemos tentar multiplicar o funcional, que
j\'a sabemos ser solu\c c\~ao da equa\c c\~ao n\~ao for\c cada,
por um funcional que ao ser derivado cancele o termo for\c cado. Ou seja,
tentemos substituir 
\begin{equation}
F=A\left\{ \int dte^{-vk^{2}t}\left[ \delta _{ij}-\frac{k_{i}k_{j}}{k^{2}}%
\right] z_{j}(k,t)\right\} \exp \left[ i\int dtd^{3}kh_{j}z_{j}\right] ,
\label{eqrn1}
\end{equation}
na equa\c c\~ao for\c cada, levando a seguinte equa\c c\~ao: 
\begin{equation}
\left[ \frac{\partial }{\partial t}+vk^{2}\right] h_{i}\left( k,t\right) =%
\left[ \delta _{ij}-\frac{k_{i}k_{j}}{k^{2}}\right] g_{j}\left( k,t\right) ,
\end{equation}
equa\c c\~ao que tem como solu\c c\~ao 
\begin{equation}
h_{i}\left( k,t\right) =e^{-\nu k^{2}t}\int_{0}^{t}d\tau e^{\nu k^{2}\tau } 
\left[ \delta _{ij}-\frac{k_{i}k_{j}}{k^{2}}\right] g_{j}\left( k,\tau
\right) .
\end{equation}
Consideremos, neste momento, o caso em que a for\c ca \'e aleat\'oria e o
n\'umero de Reynolds nulo. A (\ref{eqt5}) fica 
\begin{equation}
\frac{\partial }{\partial t}\frac{\delta F[y,q]}{\delta y_{\alpha }(x,t)}%
-\nu \frac{\partial }{\partial x_{\beta }}\frac{\partial }{\partial x_{\beta
}}\frac{\delta F[y,q]}{\delta y_{\alpha }(x,t)}=\frac{\delta F\left[ y,q%
\right] }{\delta q_{\alpha }\left( x,t\right) }.  \nonumber
\end{equation}
Transformando para a representa\c c\~ao espectral, onde 
\begin{equation}
g_{i}\left( k,t\right) =\left( 2\pi \right) ^{-3}\int d^{3}xq_{i}\left(
x,t\right) e^{ik_{j}x_{j}},
\end{equation}
teremos 
\begin{equation}
\left[ \frac{\partial }{\partial t}+vk^{2}\right] \frac{\delta F[y,q]}{%
\delta z_{\alpha }(k,t)}=\left[ \delta _{\alpha \beta }-\frac{k_{\alpha
}k_{\beta }}{k^{2}}\right] \frac{\delta F[y,q]}{\delta g_{\beta }(k,t)} .
\end{equation}
Por analogia com (\ref{eqrn1}) e lembrando que a condi\c c\~ao inicial
\'e que o campo de for\c ca seja estatisticamente independente do campo
de velocidade, teremos como solu\c c\~ao 
\begin{eqnarray}
F &=&\chi \left\{ \int dte^{-vk^{2}t}\left[ \delta _{ij}-\frac{k_{i}k_{j}}{%
k^{2}}\right] z_{j}(k,t)\right\} \times \\
&&\times \Psi \left\{ \int dte^{-vk^{2}t}\left[ \delta _{ij}-\frac{k_{i}k_{j}%
}{k^{2}}\right] z_{j}(k,t)+g\left( k,t\right) \right\}.
\end{eqnarray}

\section{Abordagens Ing\^enuas de Solu\c {c}\~ao}

Possu\'\i mos uma equa\c c\~ao diferencial para o funcional
caracter\'\i stico que, \`a primeira vista, nos parece muito apraz\'\i vel,
pois partimos de uma equa\c c\~ao n\~ao-linear e inomog\^enea e
chegamos a uma equa\c c\~ao linear e homog\^enea(\ref{eqt5}).

Todavia, n\~ao existe um m\'etodo geral para resolver equa\c c\~oes
funcionais integro-diferenciais, nem tampouco teoremas de exist\^encia e
unicidade das solu\c c\~oes para o caso geral\footnote{Recentemente tem havido um grande esfor\c co no estudo de equa\c c\~oes
diferenciais funcionais em espa\c cos de Banach. Mas nenhum resultado significativo
\'e de nosso conhecimento(\textit{vide: Approximation Methods in Banach
Spaces, Mesure Theory in Infinite Dimension Spaces MIA, Kluwer }).}. Outra
dificuldade \'e que equa\c c\~oes deste tipo n\~ao s\~ao de
f\'acil implementa\c c\~ao computacional.

Portanto, nesta sec\c c\~ao, tentaremos mostrar algumas abordagens j\'a
realizadas e onde aparecem as suas dificuldades.

Reescrevendo a equa\c c\~ao para o funcional caracter\'\i stico como 
\begin{eqnarray}
\left( l,\frac{\partial }{\partial t}D_{y}F[y,q]-i\nabla
D_{y}D_{y}F[y,q]\right) =  \label{eqt6} \\
\left( l,\nu \Delta D_{y}F[y,q]+D_{q}F\left[ y,q\right] \right),  \nonumber
\end{eqnarray}
onde $\left( a,b\right) =\int dtd^{3}xa_{i}b_{i}$ \'e o produto interno do
espa\c co das solu\c c\~oes da equa\c c\~ao de Navier-Stokes, $D_{y}=%
\frac{\delta }{\delta y\left( x,t\right) }$ e $D_{q}=\frac{\delta }{\delta
q\left( x,t\right) }$ s\~ao ``gradientes funcionais'' em rela\c c\~ao
aos campos $y\left( x,t\right) $ e $q\left( x,t\right) $, respectivamente,
sendo que os outros s\'\i mbolos t\^em seu significado usual.

Em uma primeira vista, poder\'\i amos pensar em expandir $F$ como uma
s\'erie de funcionais {\it \`a la Taylor}, ou seja, 
\begin{equation}
F\left[ y,q\right] =F^{\left( 0,0\right) }+\sum_{n=1}^{\infty
}\sum_{m=1}^{\infty }F^{\left( n,m\right) },  \label{eqt7}
\end{equation}
onde 
\begin{equation}
F^{\left( n,m\right) }=\int dt_{1}d^{3}x_{1}...\int dt_{n}d^{3}x_{n}\int
dt_{n+1}d^{3}x_{n+1}...\int dt_{n+m}d^{3}x_{n+m}\times \nonumber\\
\end{equation}
\bigskip
\begin{equation}
\frac{\delta ^{n+m}F\left[ 0,0\right] }{\delta y_{\alpha _{1}}\left(
x_{1},t_{1}\right) ...\delta y_{\alpha _{n}}\left( x_{n},t_{n}\right) \delta
q_{\alpha _{n+1}}\left( x_{n+1},t_{n+1}\right) ...\delta q_{\alpha
_{n+m}}\left( x_{n+m},t_{n+m}\right) }\times \nonumber \\
\end{equation}
\bigskip
\begin{equation}
\times y_{\alpha _{1}}\left( x_{1},t_{1}\right) ...y_{\alpha _{n}}\left(
x_{n},t_{n}\right) \times q_{\alpha _{n+1}}\left( x_{n+1},t_{n+1}\right)
...q_{\alpha _{n+m}}\left( x_{n+m},t_{n+m}\right). \nonumber
\end{equation}
Assim, \'e evidente que 
\begin{equation}
F^{\left( 0,0\right) }=F\left[ 0,0\right] =1 .
\end{equation}

Mas desta forma, ganhamos o problema de converg\^encia da s\'erie,
j\'a que os momentos n\~ao necessariamente se anulam quando, pelo menos
dois pontos est\~ao separados por uma dist\^ancia infinita.

Se, por exemplo, ignoramos o termo de for\c ca de (\ref{eqt6}), fazemos $l=y$%
, utilizamos a expans\~ao proposta em (\ref{eqt7}) e agrupamos os
polin\^omios de mesma ordem em $y$, obtemos 
\begin{equation}
\left( y,\left[ \frac{\partial }{\partial t}-\nu \Delta \right]
D_{y}F^{\left( n\right) }\right) =\left( y,i\nabla D_{y}D_{y}F^{\left(
n+1\right) }\right),  \label{eqt8}
\end{equation}
que \'e exatamente a equa\c c\~ao para o momento da n-\'esima ordem,
sendo que este, como j\'a hav\'\i amos visto, envolve o c\'alculo do
momento de ordem subseq\"uente.

Portanto, a \'unica vantagem, neste caso, \'e est\'etica, j\'a que
deixa a equa\c c\~ao de uma forma mais compacta. O problema de
converg\^encia da s\'erie pode ser contornado, utilizando uma
expans\~ao no logaritmo de $F$, ou seja, uma expans\~ao nos
cumulantes,  garantindo assim, a converg\^encia das integrais, mas
obviamente n\~ao resolvendo o problema de fechamento.

Poder\'\i amos tentar utilizar uma expans\~ao do tipo Gram-Charlier\footnote{A conveni\^encia desta expans\~ao adv\'em do seguinte fato: se ignoramos os termos da segunda derivada funcional, ou seja, N\'umero de Reynolds nulo, e de for\c ca, a solu\c c\~ao ser\'a estacion\'aria, o que significa que, se tivermos um campo gaussiano, este permanecer\'a como tal para todo tempo posterior\cite{Lewis}.}, ou seja, 
\begin{equation}
F=e^{F^{\left( 1\right) }+F^{\left( 2\right) }}\left[ 1+\sum_{n=3}^{\infty
}F^{\left( n\right) }\right].
\end{equation}
Mas esta tamb\'em nos leva a equa\c c\~oes que precisam de fechamento,
muito embora a determina\c c\~ao destes $F^{\left( n\right) }$ seja
equivalente \`a determina\c c\~ao de um n\'umero infinito de momentos.

Neste ponto, poder\'\i amos aventar a possibilidade de truncar a expans\~ao
em uma ordem espec\'\i fica, para conseguirmos uma primeira aproxima\c c%
\~ao do problema. Mas, 
\begin{eqnarray}
\left| F\right| &=&\left| 1+\sum_{n=1}^{k}F^{\left( n\right)
}+\sum_{n=k+1}^{\infty }F^{\left( n\right) }\right| \leq 1\Rightarrow \\
&\Rightarrow &\left| F\right| <\left| 1+\sum_{n=1}^{k}F^{\left( n\right)
}\right| +\left| \sum_{n=k+1}^{\infty }F^{\left( n\right) }\right| .
\end{eqnarray}
Se supusermos que $\left| \sum_{n=k+1}^{\infty }F^{\left( n\right)}\right| =0$, ent\~ao, existir\'a pelo menos um $y$, a saber $y=0$, tal que $\left| 1+\sum_{n=1}^{k}F^{\left( n\right) }\right| >1$. Logo, a s\'erie truncada n\~ao pode ser Funcional Caracter\'\i stico de nenhum processo estoc\'astico. Ou seja, s\'o \'e v\'alida a expans\~ao de Taylor com resto, n\~ao nos fornecendo, portanto, um m\'etodo de aproxima\c c\~oes sucessivas.

Em outras palavras, mesmo que n\~ao saibamos exatamente todos os momentos, cumulantes ou desvios gaussianos(dependendo da expans\~ao utilizada), temos que ter uma estimativa para eles, se quisermos utilizar este m\'etodo. Contudo, estimativas normalmente vir\~ao de uma hip\'otese ou modelo, e n\~ao diretamente da equa\c c\~ao para o funcional.

Poder\'\i amos, ainda, tentar fazer uma expans\~ao da solu\c c\~ao com base no n\'umero de Reynolds mas este procedimento \'e desaconselhado pelos seguintes aspectos: (1) estamos interessados em Turbul\^encia Desenvolvida, logo devemos tomar o limite quando o n\'umero de Reynolds vai para o infinito. Este fato obriga-nos a fazer ressomas de s\'eries perturbativas, cuja converg\^encia n\~ao \'e bem estabelecida, podendo levar-nos a resultados errados em caso de converg\^encia\cite{Fournier}; (2) a invari\^ancia por transforma\c c\~ao aleat\'oria de Galileu \'e quebrada para qualquer aproxima\c c\~ao, pois a expans\~ao \'e feita sobre o termo n\~ao linear da equa\c c\~ao, sendo que a invari\^ancia \'e conseguida atrav\'es da anula\c c\~ao
do mesmo com o termo de derivada temporal; (3) todos os resultados conseguidos por m\'etodos diagram\'aticos foram conseguidos tamb\'em por m\'etodos mais simples.

Mesmo com estas dificuldades, alguns f\'\i sicos v\^em utilizando m\'etodos diagram\'aticos com um relativo sucesso( ver\cite{L'vov} e suas refer\^encias), muito embora nos pare\c ca improv\'avel que este m\'etodo consiga obter resultados que n\~ao foram obtidos por outras t\'ecnicas menos abstratas.

\chapter{A Equa\c c\~ao de Burgers}

Neste cap\'\i tulo trataremos da equa\c c\~ao de Burgers unidimensional, que \'e
uma redu\c c\~ao para uma dimens\~ao da equa\c c\~ao de Navier-Stokes, sem o
termo de press\~ao. Esta equa\c c\~ao, embora introduzida por J.M. Burgers\cite{Burgers} como um modelo para o estudo da turbul\^encia,  n\~ao apresenta caos, mesmo na vers\~ao for\c cada. Por outro lado, ela surge naturalmente em estado s\'olido no estudo de pol\'\i meros diretos e crescimento de superf\'\i cie, neste caso sendo mais conhecida como equa\c c\~ao {\bf KPZ}\cite{Kardar}.

O interesse neste modelo simples adv\'em do fato de ser integr\'avel(via transforma\c c\~ao de Hopf-Cole\cite{Hopf2}) e termos muitas informa\c c\~oes sobre as propriedades das solu\c c\~oes, no caso determin\'\i stico\cite{Bec, Falkovich, Fournier,Woyczynski} e uma formula\c c\~ao matem\'atica precisa para o caso estoc\'astico \cite{Funaki}. Isso faz com que este modelo seja um laborat\'orio ideal para o teste de t\'ecnicas que posteriormente devem ser aplicadas \`a equa\c c\~ao de Navier-Stokes.

Este cap\'\i tulo tentar\'a dar uma vis\~ao panor\^amica sobre a equa\c c\~ao determin\'\i stica, e mostrar os principais resultados conhecidos para o caso aleat\'orio.

\section{A Equa\c c\~ao Determin\'\i stica}

A equa\c c\~ao de Burgers \'e 
\begin{equation}
\frac{\partial }{\partial t}u=-u\frac{\partial }{\partial x}u+\nu \frac{\partial ^{2}}{\partial x^{2}}u ,
\end{equation}
tamb\'em podendo ser escrita na forma dita conservativa 
\begin{equation}
\frac{\partial }{\partial t}u=\frac{\partial }{\partial x}\left[ -\frac{1}{2}\left( u\right) ^{2}+ \nu \frac{\partial }{\partial x}u\right],
\end{equation}
quando adquire uma forma mais palat\'avel, se introduzimos a transforma\c c\~ao de Hopf-Cole, 
\begin{equation}
u=-\frac{\partial }{\partial x}\psi ,\qquad \psi =2\nu \ln w ,
\end{equation}
levando-nos, a menos de uma fase arbitr\'aria no tempo, \`a equa\c c\~ao de calor 
\begin{equation}
\frac{\partial }{\partial t}w= \nu \frac{\partial ^{2}}{\partial x^{2}}u .
\end{equation}
Se tomarmos como dom\'\i nio toda a reta real e tempos estritamente positivos, a solu\c c\~ao \'e 
\begin{equation}
w\left( x,t\right) =\frac{1}{\sqrt{\pi 4\nu t}}\int e^{-\frac{\left(
x-a\right) ^{2}}{4 \nu t}}w_{0}\left( a\right) da .
\end{equation}
Escrevendo a condi\c c\~ao inicial como 
\begin{equation}
w_{0}\left( x\right) =\lim_{t\rightarrow 0^{+}}w\left( x,t\right) ,
=e^{f\left( x\right) }
\end{equation}
vem, da transforma\c c\~ao de Hopf-Cole, que 
\begin{equation}
u_{0}\left( x\right) =-2\nu \frac{d}{dx}f
\end{equation}
\begin{equation}
u\left( x,t\right) =\frac{\int d\xi \frac{x-\xi }{t}e^{-\frac{1}{2\nu }\left[
\frac{\left( x-\xi \right) ^{2}}{2t}+\int_{0}^{\xi }d\zeta u_{0}\left( \zeta
\right) \right] }}{\int d\xi e^{-\frac{1}{2\nu }\left[ \frac{\left( x-\xi
\right) ^{2}}{2t}+\int_{0}^{\xi }d\zeta u_{0}\left( \zeta \right) \right] }} .
\label{b1}
\end{equation}
Importante observar que, para esta solu\c c\~ao existir, devemos necessariamente ter 
\begin{equation}
\lim_{\left| \xi \right| \rightarrow \infty }\frac{\int_{0}^{\xi }d\zeta
u_{0}\left( \zeta \right) }{\xi ^{2}}=0 .
\end{equation}

\section{Aproxima\c c\~ao Heur\'\i stica}

Para viscosidade nula, a equa\c c\~ao de Burgers, dada a exist\^encia do Teorema de Hopf\cite{Sinai}, pode ser escrita como: 
\begin{equation}
\frac{d}{dt}\left( u\right) =0,\qquad \frac{d}{dt}=\frac{\partial }{\partial
t}+u\frac{\partial }{\partial x} ,
\end{equation}
ou seja, as part\'\i culas lagrangeanas conservam suas velocidades. Assim, a
solu\c c\~ao desta equa\c c\~ao pode ser escrita como: 
\begin{eqnarray}
u\left( X\left( a,t\right) ,t\right) &=&u_{0}\left( a\right) \\
X\left( a,t\right) &=&a+tu_{0}\left( a\right),
\end{eqnarray}
onde utilizamos letras mai\'{u}sculas para coordenadas eulerianas,i.e. $%
u\left( X\left( a,t\right) ,t\right) $ e letras min\'{u}sculas latinas para
coordenadas lagrangeanas, i.e. $a$. Todavia, esta \'e uma solu\c c\~ao
impl\'\i cita da equa\c c\~ao de Burgers, chamada de lagrangeana, que
permanece \'{u}nica, invers\'\i vel e regular, at\'e o instante que o
jacobiano da transforma\c c\~ao euler-lagrangeana permanece
invers\'\i vel, ou seja, at\'e que 
\begin{equation}
J\left( a,t\right) =\frac{\partial }{\partial a}X\left( a,t\right) =1+t\dot{u}_{0}\left( a\right)
\end{equation}
seja positivo. O instante em que isto acontece \'e denotado por $t_{\ast }$, e o chamaremos de \textit{tempo de pr\'e-choque} , sendo definido por: 
\begin{equation}
t_{\ast }=\frac{1}{-\min_{a}\left[ \dot{u}_{0}\left( a\right) \right] }.
\end{equation}
Notemos que no caso em que $u_{0}\left( a\right) =ca \quad c>0$, o tempo de
pr\'e-choque \'e negativo, ou seja, se no instante inicial n\~ao
existe nenhuma part\'\i cula no fluido capaz de ultrapassar a sua vizinha, o
fluxo permanecer\'a sempre numa fase regular(onde todo o fluxo forma um \'unico choque). Isto pode ser melhor visto se substitu\'\i mos esta solu\c c\~ao na equa\c c\~ao para a
coordenada euleriana, i.e. $X\left( a,t\right) =\left( 1+tc\right) a$. \
Como solu\c c\~oes n\~ao decrescentes, no conjunto das solu\c c\~oes da equa\c c\~ao de Burgers, s\~ao raras, n\~ao precisamos concentrar nossa aten\c c\~ao sobre este caso. Concentraremos nossa aten\c c\~ao no conjunto de solu\c c\~oes cujos gradientes sejam n\~ao constantemente positivos.

Seja $a_{\ast }$ a posi\c c\~ao lagrangeana tal que $t_{\ast }=1/\left[-u_{0}\left( a_{\ast }\right) \right]$,  portanto a derivada espacial da velocidade euleriana em um instante $t$, e no ponto $X_{\ast }\left( a_{\ast
},t\right) $ ser\'a 
\begin{equation}
\frac{\partial }{\partial X}u=\frac{\partial u}{\partial a}\frac{\partial a}{\partial X}=\dot{u}_{0}\left( a_{\ast }\right) \frac{1}{1-\frac{t}{t_{\ast }}},
\end{equation}
que \'e infinito no tempo de pr\'e-choque. Por isso, chamamos esta
singularidade em $a_{\ast }$ de pr\'e-choque. Fazendo uma expans\~ao de
Taylor ao redor do pr\'e-choque, tomando $u_{0}\left( a_{\ast }\right) =0,a_{\ast }=0$, condi\c c\~ao
esta que pode ser conseguida por transforma\c c\~ao de coordenadas e transforma\c c\~ao de Galileu, vem: 
\begin{equation}
u_{0}\left( a\right) =-\frac{1}{t_{\ast }}a+\frac{b}{6}a^{3}+O\left(
a^{4}\right).
\end{equation}
Como o gradiente da velocidade \'e m\'\i nimo global em $a_{\ast }$,
devemos ter a segunda derivada nula e a terceira positiva, ou seja $b>0$. A
expans\~ao para a posi\c c\~ao euleriana fica 
\begin{equation}
X\left( a,t\right) =\left( 1-\frac{t}{t_{\ast }}\right) a+\frac{tb}{6}a^{3}+O\left( a^{4}\right) .
\end{equation}
Se desejamos inverter esta transforma\c c\~ao, teremos dois
comportamentos diferentes na vizinhan\c ca do pr\'e-choque, dependento do
tempo, ou seja, 
\begin{eqnarray}
a\left( X,t\right) &=&\frac{1}{\left( 1-\frac{t}{t_{\ast }}\right) }X+O\left( X^{3}\right) ,\quad t<t_{\ast } \\
a\left( X,t_{\ast }\right) &=&\left( \frac{6}{bt_{\ast }}X\right)
^{1/3}+O\left( x^{2/3}\right) ,\quad t=t_{\ast },
\end{eqnarray}
implicando, para a velocidade, na seguinte expans\~ao: 
\begin{eqnarray}
u\left( X,t\right) &=&-\frac{1}{\left( t_{\ast }-t\right) }X+O\left(X^{3}\right) ,\quad t<t_{\ast } \\
u\left( X,t\right) &=&-\frac{1}{t_{\ast }}\left( \frac{6}{bt_{\ast }}X\right) ^{1/3}+O\left( x^{2/3}\right) ,\quad t=t_{\ast } \: .
\end{eqnarray}
Assim, podemos visualizar o pr\'e-choque como uma mudan\c ca local de um
comportamento linear decrescente, $t<t_{\ast }$, para um comportamento tipo
raiz c\'{u}bica, $t=t_{\ast }$, com tangente vertical.
          
 Podemos ainda analisar o comportamento da enstropia que \'e definida como: 
\begin{equation}
\Omega \left( t\right) =\frac{1}{2}\int \left[ \frac{\partial }{\partial x}u\left( x,t\right) \right] ^{2}dx=\int \left[ \dot{u}_{0}\frac{1}{1+t\dot{u}_{0}}\right] ^{2}\left( 1+t\dot{u}_{0}\right) da ,
\end{equation}
obtendo 
\begin{eqnarray}
\Omega \left( t\right)= &\propto &\left( t_{\ast }-t\right) ^{-1/2},\quad
t\uparrow t_{\ast } \\
\Omega \left( t\right)= &\propto &\int_{\varepsilon \downarrow 0}^{\infty
}\left| x\right| ^{-4/3},\quad t=t_{\ast } \: .
\end{eqnarray}

Estando o pr\'e-choque caracterizado, vejamos como ele se
forma. Para tal, consideremos que o campo inicial \'e anal\'\i tico, tal
que possamos fazer a extens\~ao da solu\c c\~ao impl\'\i cita ao
dom\'\i nio complexo, denotando $U\left( Z,t\right) $ a velocidade euleriana
complexa, $A$ a coordenada lagrangeana. Assim, se a aplica\c c\~ao
complexa 
\begin{equation}
Z=A+tU_{0}\left( A\right)
\end{equation}
deixa de ser invers\'\i vel em um ponto $A_{\ast }\left( t\right) $, cuja
imagem pela aplica\c c\~ao \'e $Z_{\ast }\left( t\right) $, ent\~ao a
expans\~ao em torno deste ponto ser\'a 
\begin{eqnarray}
Z= &\propto &Z_{\ast }\left( t\right) +\left( A-A_{\ast }\left( t\right)
\right) ^{2} \\
\left| \frac{\partial Z}{\partial A}\right|= &\propto &\left| A-A_{\ast
}\left( t\right) \right| \propto \sqrt{\left| Z-Z_{\ast }\left( t\right)
\right| }
\end{eqnarray}
e, portanto, o comportamento local do gradiente da velocidade ser\'a 
\begin{eqnarray}
\left| \frac{\partial U}{\partial Z}\right| &=&\left| \frac{\partial U}{%
\partial A}\right| /\left| \frac{\partial Z}{\partial A}\right| \propto 1/%
\sqrt{\left| Z-Z_{\ast }\left( t\right) \right| } \\
U-U_{\ast }= &\propto &\sqrt{\left| Z-Z_{\ast }\left( t\right) \right| },
\end{eqnarray}
ou seja, a exist\^encia de singularidades complexas no campo inicial que
se aproximam da reta real, conforme $t$ se aproxima de $t_{\ast }$, d\'a
origem aos pr\'e-choques.

Para entendermos globalmente a solu\c c\~ao, comecemos por reescrever (\ref{b1}) como: 
\begin{equation}
u\left( x,t\right) =\frac{\int da\frac{x-a}{t}e^{-\frac{1}{2\nu }\phi \left(
x,t,a\right) }}{\int dae^{-\frac{1}{2\nu }\phi \left( x,t,a\right) }},
\label{b2}
\end{equation}
onde 
\begin{equation}
\phi \left( x,t,a\right) =\left[ \frac{\left( x-a\right) ^{2}}{2t}%
+\int_{0}^{a}ad\tilde{a}u_{0}\left( \tilde{a}\right) \right] .
\end{equation}
Em uma solu\c c\~ao para $\nu $ tendendo a zero, os \'unicos pontos $a$ que contribuem para (\ref{b2}), para $x,t$ fixos, s\~ao os pontos nos quais $\phi \left( x,t,a\right) $ atinge um m\'\i nimo global. Ou, de uma maneira mais formal, definimos 
\begin{equation}
\psi \left( x,t\right) =\min_{a}\phi \left( x,t,a\right),
\end{equation}
que pode ser reescrita como 
\begin{equation}
\psi \left( x,t\right) =\frac{x^{2}}{2t}+L_{t}\left( a\left( x,t\right),
\right)
\end{equation}
onde 
\begin{eqnarray}
L_{t}\left( a\left( x,t\right) \right) &=&\min_{a}\left\{ \omega \left(
x,a\right) -\frac{xa}{t}\right\} , \\
\omega \left( a,t\right) &=&\int_{0}^{a}d\tilde{a}\left[ u_{0}\left( \tilde{a%
}\right) +\frac{\tilde{a}}{t}\right] .
\end{eqnarray}
Definimos uma fun\c c\~ao $C_{\omega }\left( a\right) $, chamada de 
\textit{envelope convexo}, tal que goze da propriedade de ser a maior fun\c{c%
}\~ao convexa, tal que $C_{\omega }\leq \omega $, podendo ser descrita
geometricamente, por interm\'edio da seguinte constru\c c\~ao: fixando $x$ e $%
t $, constru\'\i mos uma reta no plano $\left( \omega ,a\right) $, que seja
descrita pela equa\c c\~ao $\omega =\frac{x}{t}a+c$. Para cada $x$
podemos encontrar $c_{0}\left( x,t\right) =c_{0}$ tal que, para todo $%
c<c_{0} $ a reta $\omega =\frac{x}{t}a+c$ n\~ao intercepte o gr\'afico
de $\omega $, enquanto que para todo $c>c_{0}$ a intercepta\c c\~ao
acontece. Mas para $c=c_{0}$ a reta $\omega =\frac{x}{t}a+c_{0}$ \'e
tangente ao gr\'afico de $\omega $, em pelo menos um ponto. Portanto, $\psi
\left( x,t\right) $ \'e, em geral, um conjunto dos pontos $a$, onde a reta $%
\omega =\frac{x}{t}a+c_{0}$ \'e tangente ao gr\'afico de $\omega $.
Definimos, ent\~ao, 
\begin{eqnarray}
a_{-}\left( x,t\right) &=&\min_{a}\left\{ a;a\in \psi \left( x,t\right)
\right\} \\
a_{+}\left( x,t\right) &=&\max_{a}\left\{ a;a\in \psi \left( x,t\right)
\right\},
\end{eqnarray}
ou seja, estes s\~ao o menor e o maior valor da aplica\c c\~ao
plur\'\i vuca $\psi \left( x,t\right) $.

Finalmente, definimos o envelope convexo como: 
\begin{equation}
C_{\omega }\left( a\right) =\left\{ 
\begin{array}{c}
\omega \left( a,t\right) ,\quad a_{-}\left( x,t\right) =a_{+}\left(
x,t\right) \\ 
xa+c_{0},\quad a_{-}\left( x,t\right) <a_{+}\left( x,t\right) .
\end{array}
\right.
\end{equation}
Consideremos, agora, sua derivada $F_{t}\left( a\right) =\frac{d}{da}C_{\omega
}\left( a\right) $, que em geral ser\'a uma fun\c c\~ao
n\~ao-decrescente, e sua inversa $F_{t}^{-1}\left( x\right) $, que
geralmente n\~ao \'e uma aplica\c c\~ao bem definida em raz\~ao dos
pontos $x$, para os quais existe um conjunto $\left\{ a;F_{t}\left( a\right)
=x\right\} $. Nos pontos em que $F_{t}^{-1}\left( x\right) $ for
descont\'\i nua,vale a pena considerar $F_{t}^{-1}\left( x\right) $ como uma
curva cont\'\i nua sobre o plano , que tenha segmentos verticais para estes
valores de $x$, j\'a que posteriormente estaremos interessados em uma
interpreta\c c\~ao mais pict\'orica. Agora, podemos formular o seguinte
teorema:

\begin{theorem}[Hopf]
 Sejam $x,t$ tal que $\psi \left( x,t\right) $ consista de um ponto $%
a\left( x,t\right) =F_{t}^{-1}\left( x\right) $. Ent\~ao $%
\lim_{v\rightarrow 0}u\left( x,t\right) =u_{0}\left( x\right) $ existe e 
\begin{equation}
u\left( x,t\right) =\frac{x-F_{t}^{-1}\left( x\right) }{t} ,
\end{equation}
se $F^{-1}\left( x\right) $ \'e um intervalo de comprimento positivo, ent\~ao existem os limites laterais 
\begin{eqnarray}
u_{-}\left( x,t\right) &=&\lim_{\tilde{x}\rightarrow 0^{-}}u_{0}\left( x+%
\tilde{x}\right) =\frac{x-a_{-}\left( x,t\right) }{t} \\
u_{+}\left( x,t\right) &=&\lim_{\tilde{x}\rightarrow 0^{+}}u_{0}\left( x+%
\tilde{x}\right) =\frac{x-a_{+}\left( x,t\right) }{t} .
\end{eqnarray}
\end{theorem}

Ao inv\'es de fazermos a demonstra\c c\~ao matem\'atica deste
teorema, vamos dar uma vis\~ao mais ''f\'\i sica'' dele.

Como $a\left( x,t\right) $ \'e a coordenada da part\'\i cula, onde $\phi
\left( x,t,a\right) $ atinge um m\'\i nimo global para um dado $\left(
x,t\right) $, \'e f\'acil ver que $a\left( x,t\right) $ \'e a
coordenada lagrangeana da qual emana a part\'\i cula do fluido, que
estar\'a na coordenada $\left( x,t\right) $ euleriana.

\begin{figure}[htb]
\epsfxsize=.6\textwidth
\begin{center}
\leavevmode
\epsfbox{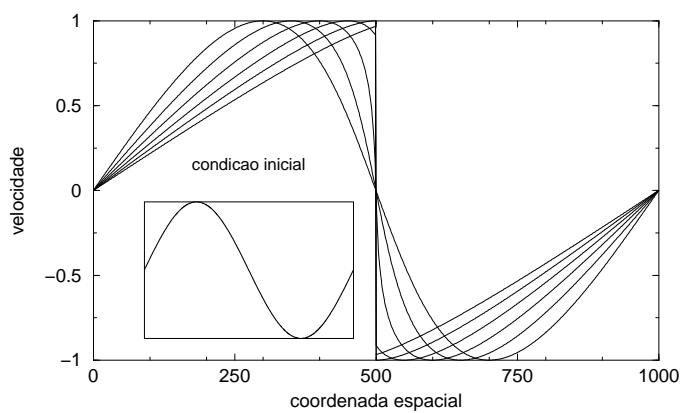}
\end{center}
\vskip 0.5 cm
\caption{Forma\c c\~ao do choque para condi\c c\~ao inicial peri\'odica.}  
\label{fig:formacao}
\end{figure}

\begin{figure}[htb]
\epsfxsize=.6\textwidth
\begin{center}
\leavevmode
\epsfbox{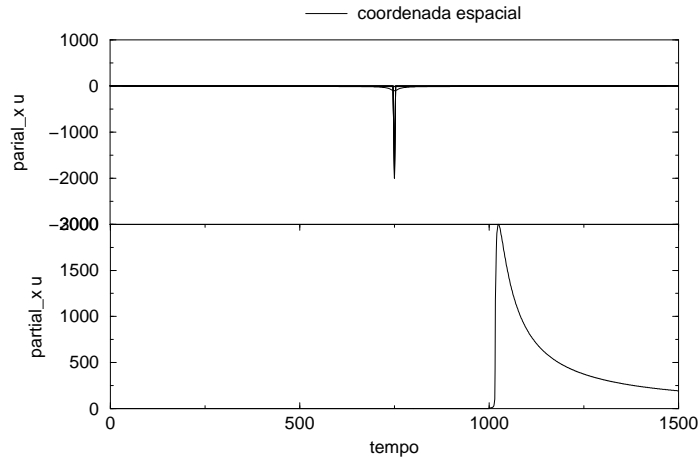}
\end{center}
\vskip 0.5 cm
\caption{Derivada espacial do campo de velocidade para condi\c c\~ao inicial peri\'odica.}  
\label{fig:derivada}
\end{figure}

Assim, podemos interpretar o resultado de Hopf e os resultados
heur\'\i sticos da seguinte forma: o choque entre part\'\i culas lagrangeanas
do fluxo, originadas da varia\c c\~ao espacial da velocidade,
gera a coalesc\^encia entre estas part\'\i culas, refletindo-se na perda
de inversibilidade da transforma\c c\~ao de coordenadas eulerianas para
lagrangeanas, forma\c c\~ao dos pr\'e-choques(fig: \ref{fig:derivada}), implicando, dessa forma, na
perda de continuidade do campo nestes pontos, sendo que estes pontos de
descontinuidade s\~ao chamados de choques(fig: \ref{fig:formacao})\footnote{ estas figuras t\^em um car\'acter apenas ilustrativo, j\'a que dispomos de resultados anal\'\i ticos.}. Assim, o perfil de velocidades
do campo euleriano \'e tipo serrote, sendo esta fase chamada de fase
dissipativa. Como ilustra\c c\~ao, consideremos o caso abaixo.

Suponha agora, que a solu\c c\~ao $u\left( x,t\right) $ apresente somente um choque, localizado em $X\left( t\right) $, e que 

\vskip .7cm  
    
\begin{figure}[htb]
\epsfxsize=.6\textwidth
\begin{center}
\leavevmode
\epsfbox{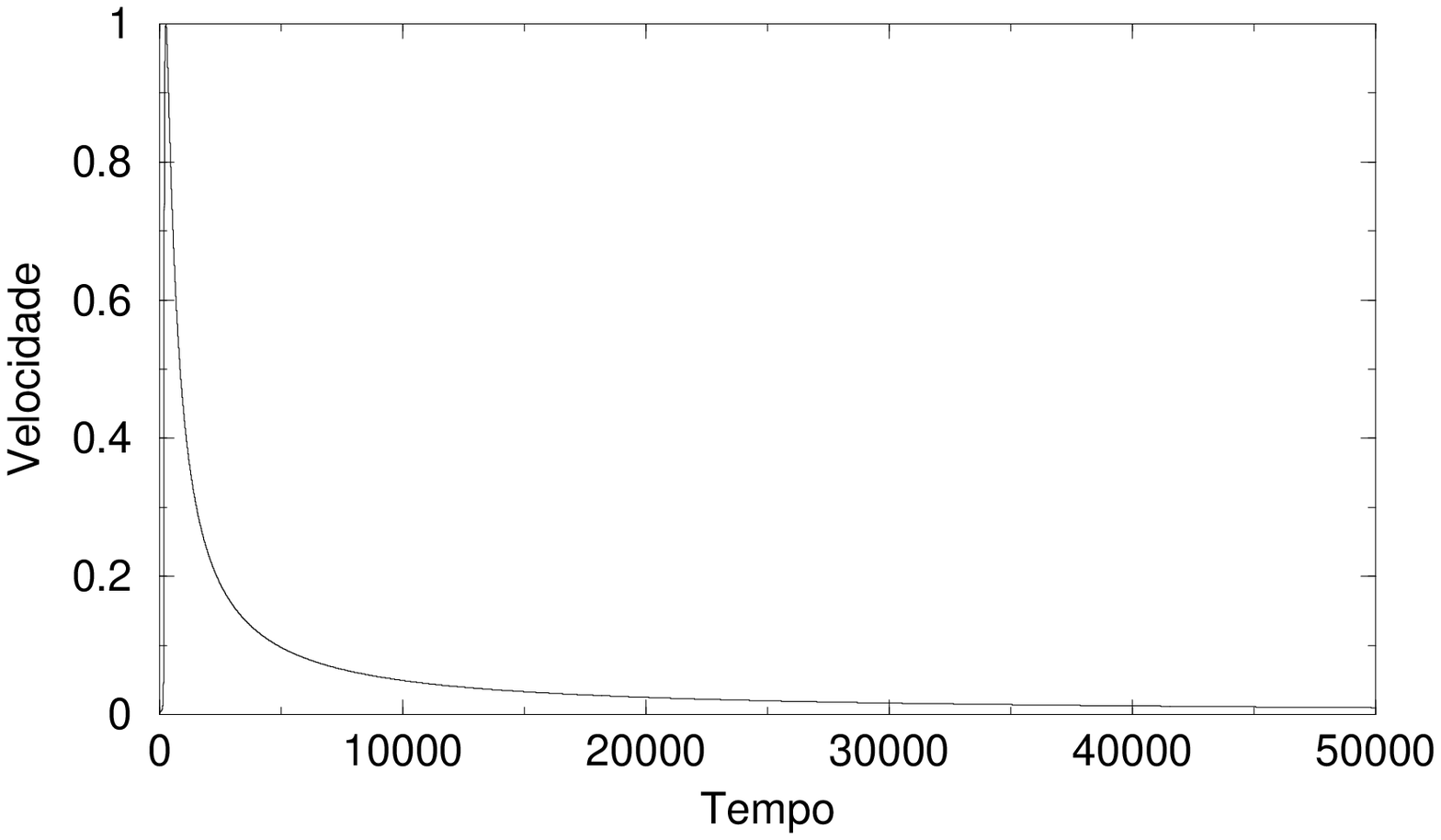}
\end{center}
\vskip 0.5 cm
\caption{Decaimento da Solu\c c\~ao}  
\label{fig:decaimento}
\end{figure}

\vskip .7cm  

\begin{equation}
\lim_{x\rightarrow X\left( t\right) ^{\pm }}u\left( x,t\right) =u_{\pm } ,
\end{equation}
onde $u_{\pm }$ \'e o valor da descontinuidade. Pela conserva\c c\~ao
do momento total, vem: 
\begin{eqnarray}
0=\frac{d}{dt} \int dxu\left( x,t\right) =\int_{-\infty }^{X\left( t\right)%
}dx\frac{\partial u}{\partial t}+\int_{X\left( t\right) }^{\infty }dx\frac{%
\partial u}{\partial t}+\frac{dX\left( t\right) }{dt}\left(
u_{-}-u_{+}\right) \Leftrightarrow \\
\Leftrightarrow \frac{1}{2}\int_{-\infty }^{X\left( t\right) }dx\frac{%
\partial }{\partial x}u^{2}\left( x,t\right) +\frac{1}{2}\int_{X\left(
t\right) }^{\infty }dx\frac{\partial }{\partial x}u^{2}\left( x,t\right) =%
\frac{dX\left( t\right) }{dt}\left( u_{-}-u_{+}\right) \Leftrightarrow \\
\Leftrightarrow \frac{dX\left( t\right) }{dt}=\frac{1}{2}\left(
u_{-}+u_{+}\right),
\end{eqnarray}
dando-nos a equa\c c\~ao para o movimento do choque(fig\ref{fig:decaimento}). Levando a cabo um
procedimento similar para a conserva\c c\~ao de energia, obtemos 
\begin{equation}
\frac{dE}{dt}=-\frac{1}{12}\left( u_{-}+u_{+}\right) ^{3} .
\end{equation}
Mas, como no caso de viscosidade positiva t\'\i nhamos dissipa\c c\~ao,
por consist\^encia, devemos ter no caso de viscosidade nula, $u_{-}>u_{+}$%
. Esta dissipa\c c\~ao \'e interpretada pela absor\c c\~ao do choque 
$X\left( t\right) $, das part\'\i culas inicialmente dentro do intervalo
lagrangeano $\left[ a_{-},a_{+}\right] $. Assim, uma vez absorvidas, estas
part\'\i culas t\^em a mesma velocidade do choque. Agora, pela solu\c c%
\~ao impl\'\i cita, devemos ter $u_{0}\left( a_{\pm }\right) =u_{\pm }$.
Portanto, a posi\c c\~ao euleriana de uma part\'\i cula, inicialmente em $%
a $ est\'a definida, sendo que chamaremos, doravante, a transforma\c c%
\~ao de coordenadas euler-lagrangeanas correta de $X_{e}\left( a,t\right) $ que, como vimos, \'e igual a $X\left( a,t\right) $, fora dos intervalos de
choque, e igual a $X\left( t\right) $, no interior destes intervalos.

\vskip .7cm            

\section{A Equa\c c\~ao Estat\'\i stica}

Nesta sec\c c\~ao, mais que nas anteriores, seguiremos os passos do
artigo de Fournier e Frisch, j\'a que este artigo \'e bastante intuitivo
sem, contudo, lan\c car m\~ao de id\'eias fenomenol\'ogicas
esp\'{u}rias.

Vamos supor um campo inicial homog\^eneo e erg\'odico, assim $%
\left\langle u_{0}\left( a\right) \right\rangle =0$, e a covari\^ancia
ser\'a 
\begin{equation}
U\left( x-\tilde{x},t\right) =\left\langle u\left( x,t\right) u\left( \tilde{%
x},t\right) \right\rangle
\end{equation}
e o espectro de energia que, \'e sua transformada de Fourier, ser\'a 
\begin{equation}
E\left( k,t\right) =\frac{1}{2\pi }\int dxe^{-ikx}U\left( x,t\right) .
\end{equation}
Consideremos a transformada de Fourier do campo de velocidade 
\begin{equation}
\hat{u}\left( k,t\right) =\frac{1}{2\pi }\int dxe^{-ikx}u\left( x,t\right) .
\end{equation}
Logo, teremos 
\begin{equation}
\left\langle \hat{u}\left( k,t\right) u\left( \tilde{k},t\right)
\right\rangle =\delta \left( k+\tilde{k}\right) E\left( k,t\right) .
\end{equation}
Podemos introduzir as chamadas RFL(Representa\c c\~oes
Fourier-Lagrangeanas) se escrevemos a transformada de Fourier do campo de
velocidades, tomando $x=X_{e}\left( a,t\right) $. Ou seja, 
\begin{equation}
\hat{u}\left( k,t\right) =\frac{1}{2\pi }\int dae^{-ikX_{e}}u_{0}\left(
a\right) \frac{\partial }{\partial a}X_{e} .
\end{equation}
Agora, por integra\c c\~ao por partes, v\^em as tr\^es RFL 
\begin{eqnarray}
\left\langle \hat{u}\left( k,t\right) u\left( \tilde{k},t\right) \right\rangle &=&\left( \frac{1}{2\pi }\right) ^{2}\int dad\tilde{a} \left\langle u_{0}\left( a\right) u_{0}\left( \tilde{a}\right) \frac{\partial }{\partial a}X_{e}\frac{\partial }{\partial \tilde{a}}\tilde{X}_{e}e^{-i\left( kX_{e}+\tilde{k}\tilde{X}_{e}\right) }\right\rangle \nonumber \\
E\left( k,t\right) &=&\frac{1}{2\pi }\left( \frac{1}{k}\right) ^{2}\int
dae^{-ika}\left\langle \dot{u}_{0}\left( a\right) \dot{u}_{0}\left( a\right)
e^{-ik\left[ \xi \left( a,t\right) -\xi \left( 0,t\right) \right]
}\right\rangle \\
E\left( k,t\right) &=&\frac{1}{2\pi }\left( \frac{1}{kt}\right) ^{2}\int
dae^{-ika}\left[ {\cal C}\left( t,k,a\right) -{\cal C}\left(t,k,0\right) \right] \nonumber,
\end{eqnarray}
onde 
\begin{equation}
\xi \left( a,t\right) =X_{e}\left( a,t\right) -a
\end{equation}
\'e o deslocamento lagrangeano, e 
\begin{equation}
{\cal C}\left( t,k,a\right) =\left\langle e^{-ik\left[ \xi \left(
a,t\right) -\xi \left( 0,t\right) \right] }\right\rangle .
\end{equation}

\subsection{Condi\c c\~ao Inicial Com Tempo de Pr\'e-Choque
Estat\'\i stico Positivo}

Muito embora cada realiza\c c\~ao tenha um tempo de pr\'e-choque, a
exist\^encia de um limite inferior, num sentido estat\'\i stico, para o
tempo de pr\'e-choque no caso aleat\'orio, depende da medida de
probabilidade sobre o espa\c co de probabilidade considerado. Em outras
palavras, depende da distribui\c c\~ao de probabilidade do gradiente da
velocidade. Por exemplo, se o campo inicial for gaussiano, h\'a uma
probabilidade finita, mesmo para argumentos tendendo a menos infinito, assim
sendo o tempo de pr\'e-choque estat\'\i stico \'e $T_{\ast }=0$.

Por outro lado, podemos tomar condi\c c\~oes iniciais tais que $P\left[ 
\dot{u}\left( a\right) <-M\right] =0$, ou seja, o conjunto das realiza\c c%
\~oes que apresenta $t_{\ast }$ menores que $1/M$ \'e nulo, de tal sorte
que $T_{\ast }=1/M$. Assim, podemos dizer que a distribui\c c\~ao \'e
inferiormente limitada. Ent\~ao, caracterizemos este caso.

Na fase regular$\left( t<T_{\ast }\right) $, podemos calcular a enstropia
pois, 
\begin{equation}
\Omega \left( t\right) =\frac{1}{2}\int da\frac{\dot{u}^{2}\left( a\right) }{%
1+t\dot{u}\left( a\right) }=\left\langle \frac{\dot{u}^{2}\left( a\right) }{%
1+t\dot{u}\left( a\right) }\right\rangle ,
\end{equation}
onde utilizamos a hip\'otese de ergodicidade, denotando, para facilitar, $%
\dot{u}^{2}\left( a\right) =w$, j\'a que a m\'edia n\~ao pode depender
do ponto calculado, reescrevemos a igualdade acima como: 
\begin{equation}
\Omega \left( t\right) =\int_{-1}^{\infty }dwP\left( w\right) \frac{w^{2}}{%
1+tw},
\end{equation}
sendo que aqui estamos assumindo que, $-M=T_{\ast }=1$.

Tamb\'em assumimos que a medida de probabilidade do gradiente \'e
absolutamente cont\'\i nua em rela\c c\~ao \`a Medida de Lebesgue, ou
seja, o conjunto dos pontos de choque eulerianos tem medida de probabilidade
nula. Hip\'otese que \'e v\'alida, no nosso entendimento, somente para
a fase regular e para tempos n\~ao muito pr\'oximos de $T_{\ast }$, pois
se uma parte substancial das realiza\c c\~oes tiver seu tempo de
pr\'e-choque pr\'oximo a $T_{\ast }$\footnote{Ou seja, se $P\left[ \frac{1}{-\min_{a}\dot{u}\left( a\right) }<T_{\ast
}+\varepsilon \right] =a+O\left( \varepsilon \right) $, onde $0<\varepsilon
<a<1$.}, como no caso gaussiano\cite{Sinai}, por exemplo(mais especificamente como o movimento browniano), o conjunto dos pontos regulares(fora do intervalo de
choque) \ tem dimens\~ao Hausdorff 1/2, fen\^omeno este vinculado ao
''aumento'' do n\'{u}mero de choques. Assim, que se h\'a um n\'{u}mero
grande de choques maduros num dado instante, obviamente a distribui\c c%
\~ao n\~ao pode ser absolutamente cont\'\i nua \`a de Lebesgue, pois
os choques trazem em si uma descontinuidade nos campos de velocidade e de
seu gradiente. A \'{u}nica exce\c c\~ao seria se os choques se tornassem
densos na reta e, portanto, restabelec\^essemos a condi\c c\~ao da
medida de probabilidade ser absolutamente cont\'\i nua com rela\c c\~ao
\`a Medida de Lebesgue, sendo o \'{u}nico detalhe que, em vez de
termos a distribui\c c\~ao de probabilidade regular(fora do intervalo de
choque), ter\'\i amos a probabilidade dos intervalos de choque.

Pela segunda RFL podemos escrever a enstropia como: 
\begin{equation}
\Omega \left(t\right) =\int_{-1}^{\infty }dkk^{2}E\left( k,t\right).
\end{equation}
Portanto, o comportamento cr\'\i tico de $\Omega $ \'e determinado pela
forma da distribui\c c\~ao $P\left( w\right) $, na vizinhan\c ca do
limite inferior da integral. Vamos supor que 
\begin{equation}
P\left( w\right) \propto \left( w+1\right) ^{\alpha }.
\end{equation}
A condi\c c\~ao de normaliza\c c\~ao nos diz que $\alpha >1$ e,
observando o comportamento, ainda na fase regular, teremos 
\begin{eqnarray}
\Omega (t) \approx 
\begin{array}{c}
cte,\quad \alpha >0 \\ 
-\ln ( T_{\ast }-t) ,\quad \alpha =0 \\ 
( T_{\ast }-t) ^{\alpha },\quad -1<\alpha <0 .
\end{array}
,\quad t\uparrow T_{\ast} 
\end{eqnarray}
Portanto, $\alpha $ regula a acumula\c c\~ao dos tempos de
pr\'e-choques das realiza\c c\~oes, logo a n\~ao diverg\^encia para 
$\alpha >0$ significa, em um senso amplo, que a maior parte das realiza\c c%
\~oes tem um tempo de pr\'e-choque bem acima do tempo de pr\'e-choque
estat\'\i stico. N\~ao obstante, a enstropia \'e infinita para tempos
posteriores a $T_{\ast }$. Seguindo um modelo simples, Fournier e
Frisch(Ap\^endice B) obtiveram o seguinte comportamento para o espectro de
energia: 
\begin{eqnarray}
E\left( k,t\right) &\approx &k^{-\left( 3+2\alpha /3\right) }F_{\alpha
} \left[ k/K_{1}\left( t\right) \right] ,\quad k\rightarrow \infty \\
K_{1}\left( t\right) &\approx &\left( T_{\ast }-t\right) ^{-\frac{3}{2}%
},\quad t\uparrow T_{\ast }
\end{eqnarray}
onde $F_{\alpha }$ \'e fun\c c\~ao positiva de decrescimento
assint\'otico r\'apido, ou seja, nos instantes imediatamente anteriores
ao tempo de pr\'e-choque estat\'\i stico, o espectro assint\'otico ganha
uma lei de pot\^encia, sendo que seu dom\'\i nio aumenta quando $t$ tende
ao tempo de pr\'e-choque estat\'\i stico. Isto nos leva a ver $\alpha $
como um tempo adimensional que regula a fra\c c\~ao de realiza\c c%
\~oes, que tem seu tempo de pr\'e-choque ao redor do tempo de
pr\'e-choque estat\'\i stico. Sendo $\alpha =-1/2$ a condi\c c\~ao de
todas as realiza\c c\~oes terem o mesmo tempo de de pr\'e-choque, sendo
que esta fra\c c\~ao vai se diluindo \`a medida que $\alpha $ aumenta, podemos observar, tamb\'em, que esta lei de pot\^encia \'e mais forte do que a da fase dissipativa $k^{-2}$, como veremos a seguir.

A terceira RFL na fase dissipativa se escreve 
\begin{equation}
E\left( k,t\right) =\left( \frac{1}{\sqrt{2\pi }kt}\right) ^{2}\int
da\left\langle e^{-ik\left[ X_{e}\left( a,t\right) -X_{e}\left( 0,t\right) %
\right] }\right\rangle .
\end{equation}
Lembrando que $X_{e}$ \'e constante dentro dos intervalos de choque,
podemos considerar somente a contribui\c c\~ao dos eventos, de forma que a
origem e $h$ estejam dentro do mesmo intervalo de choque pois, caso
contr\'ario, a exponencial oscilar\'a muito rapidamente, cancelando a
contribui\c c\~ao, j\'a que desejamos somente informa\c c\~oes
assint\'oticas. Portanto, podemos escrever em uma primeira aproxima\c c%
\~ao 
\begin{equation}
E\left( k,t\right) \approx I\left( t\right) k^{-2},~k\rightarrow \infty ,
\end{equation}
onde \ $I\left( t\right) $ \'e essencialmente o comprimento m\'edio dos
intervalos de choque, que depende das condi\c c\~oes iniciais.

\subsection{Distribui\c c\~ao de probabilidade do gradiente da velocidade}

A distribui\c c\~ao de probabilidade para o gradiente da velocidade pode
ser escrito como: 
\begin{equation}
p\left( x,\xi ,t\right) =\left\langle \delta \left( \xi -\frac{\partial }{%
\partial x}u\left( x,t\right) \right) \right\rangle .
\end{equation}
Se tomamos o fluxo como homog\^eneo e erg\'odico, podemos reescrever a
equa\c c\~ao acima como: 
\begin{equation}
p\left( \xi ,t\right) =\lim_{L\rightarrow \infty }\frac{1}{2L}%
\int_{-L}^{L}dx\delta \left( \xi -\frac{\partial }{\partial x}u\left(
x,t\right) \right).
\end{equation}
Tamb\'em, podemos escrever esta express\~ao na vers\~ao lagrangeana, que \'e 
\begin{equation}
p\left( \xi ,t\right) =\lim_{L\rightarrow \infty }\frac{1}{2L}\int_{-L}^{L}da\left| \frac{\partial X_{e}\left( a,t\right) }{\partial a}\right| \delta \left( \xi -\left( \frac{\partial X_{e}\left( a,t\right) }{\partial a}\right) ^{-1}\dot{u} \left( a\right) \right).
\end{equation}
Mas como no intervalo de choque $\frac{\partial X_{e}\left( a,t\right) }{%
\partial a}=0$, podemos nos concentrar sobre o conjunto dos pontos regulares
em um dado momento dentro do intervalo $\left[ -L,L\right] $, denotado por $%
R_{L}\left( t\right) $, para o qual o envelope convexo \'e igual a fun\c c%
\~ao, logo podemos reescrever a express\~ao acima como: 
\begin{equation}
p\left( \xi ,t\right) =\lim_{L\rightarrow \infty }\frac{1}{2L}%
\int_{R_{L}\left( t\right) }da\left| 1+t\dot{u}\left( a\right) \right|
\delta \left( \xi -\frac{\dot{u}\left( a\right) }{1+t\dot{u}\left( a\right) }%
\right) ,
\end{equation}
denotando por $b_{k}\left( \xi \right) =b_{k}$ os n\'{u}meros, de forma que: 
\begin{equation}
\xi -\frac{\dot{u}\left( b_{k}\right) }{1+t\dot{u}\left( b_{k}\right) }=0 .
\label{b4}
\end{equation}
Portanto, podemos escrever $p\left( \xi ,t\right) $ como: 
\begin{equation}
p\left( \xi ,t\right) =\frac{1}{\left| 1-t\xi \right| ^{3}}\sum_{k}\frac{1}{%
\left| \ddot{u}\left( b_{k}\right) \right| }\lim_{L\rightarrow \infty }\frac{%
1}{2L}\int_{R_{L}\left( t\right) }da\delta \left( a-b_{k}\right).  \label{b5}
\end{equation}
Se desejamos estudar o comportamento assint\'otico $\xi \rightarrow
-\infty $, \'e importante perceber que os pontos $b_{k}$ estar\~ao nas
proximidades dos pr\'e-choques $a_k^{\ast}$. Agora, fazendo uma
expans\~ao de (\ref{b4}) em torno do pr\'e-choque mais pr\'oximo, vem: 
\begin{equation}
(b_k - a_k^{\ast})^2 \simeq - \frac{2}{t u_0^{(3)} (a_k^{\ast})}
\left[ 1 + \frac{1}{t \xi} - \frac{t}{t_j^{\ast}} \right] ,
\end{equation}
onde $t_{j}^{\ast }=-1/\dot{u}_{0}\left( a_{k}^{\ast }\right) $. Logo, ou
temos ra\'\i zes duplas ou nenhuma raiz, denotadas por $b_{k}^{\pm }$.
Considerando tempos anteriores \ ao tempo do k-\'esimo pr\'e-choque$%
\left( t\leq t_{j}^{\ast }\right) $, e a exist\^encia de ra\'\i zes\ \ $1+%
\frac{1}{t\xi }-\frac{t}{t_{j}^{\ast }}\leq 0$, podendo estas duas
desigualdades serem unidas na forma 
\begin{equation}
-\frac{1}{t}\leq \dot{u}_{0}\left( a_{k}^{\ast }\right) <\frac{-1}{t}-\frac{1%
}{t^{2}\xi } ,
\end{equation}
analogamente para $t>t_{j}^{\ast }$, vem: 
\begin{equation}
\frac{-1}{t}-\frac{1}{2t^{2}\xi }<\dot{u}_{0}\left( a_{k}^{\ast }\right) <-%
\frac{1}{t} ,
\end{equation}
condi\c c\~oes estas que podem ser resumidas em: 
\begin{equation}
\frac{-1}{t}-\frac{1}{2t^{2}\xi }<\dot{u}_{0}\left( a_{k}^{\ast }\right) <%
\frac{-1}{t}-\frac{1}{t^{2}\xi }.  \label{b6}
\end{equation}
Seja agora $I_{L}\left[ a,t,\xi \right] $ a fun\c c\~ao indicador dos
conjuntos de pontos que respeitam a desigualdade acima, dentro de $%
R_{L}\left( t\right) $.\ Expandindo, $u_0^{(3)}\left( a_{k}^{\ast
}\right) $ em torno do pr\'e-choque, podemos escrever (\ref{b5})\ como: 
\begin{equation}
p\left( \xi ,t\right) \simeq \frac{\left( 2t\right) ^{1/2}}{\left| t\xi
\right| ^{3}}\lim_{L\rightarrow \infty }\frac{1}{2L}\int_{R_{L}\left(
t\right) }da\sum_{k}\frac{I_{L}\left[ a,t,\xi \right] }{\left| u_0^{(3)}\left( a\right) \left[ 1+\frac{1}{t\xi }+t\dot{u}_{0}\left( a\right) %
\right] \right| ^{1/2}}\delta \left( a-a_{k}^{\ast }\right),
\end{equation}
que pode ser reescrito como: 
\begin{equation}
p\left( \xi ,t\right) \simeq \frac{\left( 2t\right) ^{1/2}}{\left| t\xi
\right| ^{3}}\lim_{L\rightarrow \infty }\frac{1}{2L}\left\{ \int_{-L}^{L}%
\frac{I_{L}\left[ a,t,\xi \right] \times I_{R_{L}\left( t\right) }}{\left| 
u_0^{(3)}\left( a\right) \left[ 1+\frac{1}{t\xi }+t\dot{u}_{0}\left(
a\right) \right] \right| ^{1/2}}\right\}.
\end{equation}
Agora escrevemos esta express\~ao em termos da distribui\c c\~ao
conjunta das tr\^es primeiras derivadas, tal que estejamos em um conjunto
de pontos regulares sobre a reta $p_{3}\left( \dot{u},\ddot{u},u^(3), R_{\infty }\left( t\right) \right) $. Logo vem: 
\begin{equation}
p\left( \xi ,t\right) \simeq \frac{\left( 2t\right) ^{1/2}}{\left| t\xi
\right| ^{3}}\int_{0}^{\infty }d u^(3)\int_{\frac{-1}{t}-\frac{1}{%
2t^{2}\xi }}^{\frac{-1}{t}-\frac{1}{t^{2}\xi }}d\dot{u}\frac{u^(3)}%
p_{3}(\dot{u},0,u^{(3)},R_{\infty }(t)){\left[
1+\frac{1}{t\xi }+t\dot{u}\right] ^{1/2}},
\end{equation}
fazendo $\dot{u}\simeq $ $\frac{-1}{t}$, vem: 
\begin{equation}
p\left( \xi ,t\right) \simeq J\left( t\right) \left| \xi \right|
^{-7/2},~\xi \rightarrow -\infty,
\end{equation}
sendo esta uma caracter\'\i stica do fluxo v\'alida para todos os tempos e
independente das condi\c c\~oes iniciais, ou seja, uma caracter\'\i stica
da equa\c c\~ao de Burgers.

\section{Observa\c c\~oes}

Como este cap\'\i tulo nos mostrou, a exist\^encia de singularidades nas solu\c c\~oes \'e o fator preponderante na deforma\c c\~ao da estat\'\i stica inicial, no caso da equa\c c\~ao de Burgers. O mesmo tipo de lei foi provado recentemente\cite{Bec1}, para o caso com uma for\c ca impulsiva no tempo e suave no espa\c co, mostrando que a lei dos 7/2 \'e um tra\c co dos pr\'e-choques. 

Para o caso com for\c ca do tipo ru\'\i do braco no tempo e suave no espa\c co, a controv\'ersia ainda perdura, embora  E. {\it et all}\cite{E,E1,E2} sustentem a exist\^encia de pot\^encia em 7/2 e Kraichnan\cite{Gotoh}, usando uma equa\c c\~ao tipo Fokker-Plank, argumente em prol da pot\^encia em 3.Contudo, recentemente, o pr\'oprio Kraichnan\cite{Kraichnan1} favorece a lei dos 7/2, mas sem obter qualquer resultado definitivo.

A dificuldade est\'a em saber se os pr\'e-choques est\~ao presentes e separados, o que levaria \`a Lei dos 7/2, ou se existe alguma forma de agrupamento entre eles, o que levaria \`a uma pot\^encia em 3.

\chapter{Conclus\~oes}
 
\vskip +3.0 cm

 Este trabalho teve como objetivo descrever uma parte da pesquisa atual feita em
 turbul\^encia. Obviamente, n\~ao foi poss\'\i vel entrar nos detalhes de todas
 as t\'ecnicas utilizadas mas acreditamos ter sido cumprida a proposta
 b\'asica do projeto. Dada a generalidade da tese, podemos agora fazer
 incurs\~oes por diversos problemas na \'area.  Uma interessante linha de
 pesquisa \'e a possibilidade de utiliza\c c\~ao de redes neurais para solu\c
 c\~ao de EDP's, j\'a que isto poder\'a propiciar um m\'etodo computacionalmente
 barato de fazer estat\'\i stica no espa\c co de fase({\it ensemble}), ao inv\'es de lan\c carmos m\~ao da hip\'otese erg\'odica.


\end{document}